\documentclass{article}
\usepackage{flushend} 
\usepackage{balance}

\usepackage{array,multirow,graphicx}
\usepackage{microtype}
\usepackage{graphicx}
\usepackage{subfigure}
\usepackage{booktabs} % for professional tables

\usepackage{hyperref}

\usepackage[accepted]{sty}
\usepackage{graphicx}
\usepackage{pythonhighlight}

\usepackage{times}
\usepackage{helvet}
\usepackage{courier}
\usepackage{subfigure} 
\usepackage{cancel}
\RequirePackage{amsthm,amsmath}
\usepackage{graphicx, amssymb, mathrsfs,amsfonts,url, bm}

\newcommand{\real}{\mathbb{R}}

\newcommand{\gap}{\,\,\,\,\,\,\,\,}

\newcommand{\bSigma}{\boldsymbol\Sigma}

\newcommand{\bA}{\bm{A}}

\newcommand{\bB}{\bm{B}}

\newcommand{\bM}{\bm{M}}

\newcommand{\bp}{\bm{p}}

\newcommand{\br}{\bm{r}}

\newcommand{\bv}{\bm{v}}

\newcommand{\bx}{\bm{x}}
\newcommand{\bX}{\bm{X}}
\newcommand{\by}{\bm{y}}

\newcommand{\bz}{\bm{z}}

\usepackage[font=small,skip=2pt]{caption}
 
\newcommand{\titlethis}{A Hybrid Approach on Conditional GAN for Portfolio Analysis}

\icmltitlerunning{\titlethis}

\begin{document}

\twocolumn[
\icmltitle{\titlethis}

\begin{icmlauthorlist}
	\icmlauthor{Jun Lu}{te}
	\icmlauthorsingle{\gap \gap Danny Ding}
\end{icmlauthorlist}

\icmlaffiliation{te}{Correspondence to: Jun Lu $<$jun.lu.locky@gmail.com$>$. Copyright 2022 by the author(s)/owner(s). July 13th, 2022}

%\icmlaffiliation{t2}{}

%\icmlcorrespondingauthor{Jun Lu}{jun.lu.locky@gmail.com}

\vskip 0.3in
]

% this must go after the closing bracket ] following \twocolumn[ ...

% This command actually creates the footnote in the first column
% listing the affiliations and the copyright notice.
% The command takes one argument, which is text to display at the start of the footnote.
% The \icmlEqualContribution command is standard text for equal contribution.
% Remove it (just {}) if you do not need this facility.

\printAffiliationsAndNotice{}  % leave blank if no need to mention equal contribution
%\printAffiliationsAndNotice{\icmlEqualContribution} % otherwise use the standard text.

\begin{abstract}
Over the decades, the Markowitz framework has been used extensively in portfolio analysis though it puts too much emphasis on the analysis of the market uncertainty rather than on the trend prediction.
While generative adversarial network (GAN), conditional GAN (CGAN), and autoencoding CGAN (ACGAN) have been explored to generate financial time series and extract features that can help portfolio analysis. The limitation of the CGAN or ACGAN framework stands in putting too much emphasis on generating series and finding the internal trends of the series rather than predicting the future trends. 
In this paper, we introduce a hybrid approach on conditional GAN based on deep generative models that learns the internal trend of historical data while modeling market uncertainty and future trends. 
We evaluate the model on several real-world datasets from both the US and Europe markets, and show that the proposed HybridCGAN and HybridACGAN models lead to better portfolio allocation compared to the existing Markowitz, CGAN, and ACGAN approaches.

\end{abstract}
\section{Introduction}

Financial portfolio management is largely based on linear models and the Markowitz framework \citep{markowitz1968portfolio, markowitz1976markowitz} though the underlying data and information in today's market has increased countless times over that of many years ago.
The framework, often known as the modern portfolio theory (MPT), has become one of the cornerstones of quantitative finance.
The fundamental idea behind the MPT is to create portfolio diversification while reducing specific risks and assessing the risk-return trade-offs for each asset.
The MPT, on the other hand, has been criticized for making ideal assumptions about the financial system and data: the expected mean returns, and the covariance matrix of the return series are estimated from historical observations and assumed constant in the future. 
This is such a strong assumption, though, that it will be impossible for the market to actually achieve this demand.

As the evaluation outcomes of cross-section risk, the traditional portfolio assessment approach creates portfolio risk indicators based on asset price series over the previous period, such as variance, value at risk, and expected loss.
The conventional method where the classical mean-variance optimization approach employed in MPT, however, has two clear shortcomings.
First, historical data typically cannot be simply utilized to indicate the future due to the capital market's quick-change since financial returns are notoriously stochastic with an extremely low signal-to-noise ratio.
When a reliable long-term prognosis is made available in a highly efficient market, traders immediately act on this forecast, which directly affects the price at hand; while future price variations are unpredictable again
\citep{kallberg1981remarks, kallberg1984mis, green1992will, best1991sensitivity, timmermann2004efficient}. 
Second, the linear components in the historical series are typically all that are included in the risk measuring indicators evaluated using conventional methods, leaving out the nonlinear information. This causes a discrepancy between the evaluation results and the actual situation
\citep{tsay2005analysis}.

The financial sector, on the other hand, has been significantly impacted by advances in machine learning, deep learning, artificial intelligence.
Machine learning in general has been used in a variety of applications, including forecasting, series generation, risk management, customer service, and portfolio management \citep{huang2005forecasting, kara2011predicting, takahashi2019modeling}.
Especially among these advancements, the generative adversarial networks (GANs) are a sort of neural network architectures that have shown promise in image generation and are now being used extensively to produce time series and other financial data \citep{goodfellow2014generative, esteban2017real, eckerli2021generative}. 
While models of the ARCH and GARCH families use traditional statistics to explain the change in variance over time in a time series by describing the variance of the current error component as a function of prior errors
\citep{engle1982autoregressive, bollerslev1986generalized, lu2022reducing}.
GANs are being used to address the problem of paucity of real data, as well as to optimize portfolios and trading methods which achieve better results \citep{takahashi2019modeling, mariani2019pagan}. 
However, due to its highly stochastic, noisy, and chaotic nature, market price forecasting is still a major topic in the time series literature. 
While previous work have tried to generate financial data based on historical trend, the generation still lacks guidance on the potential trend of the future series \citep{mariani2019pagan, lu2022autoencoding}.

In this light, we focus on GANs for better portfolio allocation that can both capture historical trends and generate series based on past data. 
We present a novel framework about portfolio analysis based on conditional GAN (CGAN) that incorporates a proposer providing a potential mean value of future series for data normalization to achieve stable strategies, hence the name \textit{HybridCGAN}.
% to overcome the issues and challenges encountered in portfolio management tasks. 
Similar to the CGAN and ACGAN models for portfolio analysis \citep{mariani2019pagan, lu2022autoencoding}, HybridCGAN can also directly model the market uncertainty via its complex multidimensional form, which is the primary driver of future price trends, such that the nonlinear interactions between different portfolios can be embedded effectively. 
We evaluate the proposed HybridCGAN method on two separate portfolios representing different markets (the US and the European markets) and industrial segments (e.g., Technology, Healthcare, Basic materials, and Industrials sectors). 
The empirical results show that the proposed approach is capable of realizing the risk-return trade-off and outperforms the classic MPT, CGAN- and ACGAN-based methodologies considerably.

\section{Related Work}
As aforementioned, there are several methods delving with portfolio allocation, including the Markowitz framework, the CGAN and ACGAN methodologies \citep{markowitz1968portfolio, mariani2019pagan, lu2022autoencoding}. 
The Markowitz framework relies on the assumption that the past trend can be applied in the future. While the CGAN and ACGAN methodologies partly solve the drawback in the Markowitz framework by simulating future data based on historical trends, it still lacks full ability to capture the information and features behind the past data. The proposed HybridCGAN (and HybridACGAN) model introduces an extra \textit{proposer} that can help the constructed networks to capture historical features and propose future trends.
\subsection{Markowitz Framework}
Portfolio allocation is a kind of investment portfolio where the market portfolio has the highest Sharpe ratio (SR \citep{sharpe1966mutual}) given the composition of assets \citep{markowitz1968portfolio}. For simplicity, we here only consider long-only portfolio. Denote $\br$ as the return on assets vector, $\bSigma$ as the asset covariance matrix, $\bv$ as the weight vector of each asset, and $r_f$ as the risk-free interest rate. If we measure portfolio risk by variance (or standard deviation),
then the overall return and risk of the portfolio are:
\begin{equation}
\begin{aligned}
		r = \bv^\top \br;\gap \sigma^2 = \bv^\top\bSigma\bv.
\end{aligned}
\end{equation}
The Sharpe ratio is simply calculated by 
\begin{equation}
\text{SR} =\frac{r - r_f}{\sigma} = \frac{\bv^\top\br - r_f}{\sqrt{\bv^\top\bSigma\bv}}.
\end{equation}
According to the definition of asset allocation, the weight of each asset in the market portfolio is the solution to the following optimization problem:
\begin{equation}
\begin{aligned}
&\gap \gap \gap \gap \mathop{\arg\max}_{\bv} \frac{\bv^\top\br - r_f}{\sqrt{\bv^\top\bSigma\bv}};\\	
&\text{s.t. } \sum_{i=1}^{N} v_i=1;\,\, 0\leq v_i\leq 1, \forall \, i\in \{1,2,\ldots,N \},
\end{aligned}
\end{equation}
where $N$ is the number of assets, and $v_i$ is the $i$-th element of the weight vector $\bv$.

\begin{figure}[t!]
\centering
\includegraphics[width=0.24\textwidth]{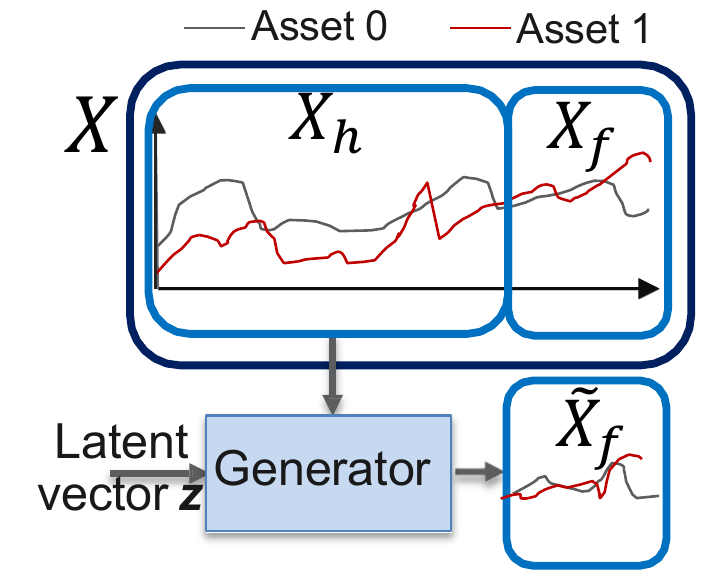}
\caption{A conceptual overview of the CGAN, ACGAN, and the proposed HybridCGAN and HybridACGAn generators' inputs and outputs.}
\label{fig:overview_of_GAN-series}
\end{figure}

\subsection{Portfolio Analysis with GAN}

Given the number of assets $N$,
we consider the matrix $\bX$ to span the whole analysis length: $w=h+f$ and $\bX\in \real^{N\times w}$. The matrix $\bX$ contains two components, the known historical series $\bX_h\in \real^{N\times h}$, and the unknown future series $\bX_f\in \real^{N\times f}$.
%Given the number of assets $N$, the matrix $\bX$ is of size $N\times w$; $\bX_h$ has shape $N\times h$; and $\bX_f$ is of shape $N\times f$.

Considering the (known) historical series $\bX_h\in \real^{N\times h}$ and a prior distribution of a random latent vector $\bz\in \real^m$, we use a generative deep-neural network $G$ to learn the probability distribution of future price trends $\bX_f$ within the target future horizon $f$.
Figure~\ref{fig:overview_of_GAN-series} provides a conceptual representation of the matrix $\bX$, and the inputs and outputs of the generator $G$.
Formally the generative model simulates a fake future matrix $\widetilde{\bX}_f$ by 
\begin{equation}
	\widetilde{\bX}_f = G( \bz,\bX_h),
\end{equation}
where $\bz\in \real^m$ is the latent vector sampled from a prior distribution (e.g., from a normal distribution). 
%In practice, the latent vector $\bz$ represents the unforeseeable future occurrences and phenomena that will have an impact on the marketplace. 
In practice, The unanticipated future events and phenomena that will affect the market are represented by the latent vector $\bz$.
Based on the most recent market conditions, the known historical series $\bX_h$ is used to extract features and condition the probability distribution of the future $\bX_f$.
Given the historical observation $\bX_h$ and following the Wasserstein GAN-GP (WGAN-GP) by \citet{gulrajani2017improved}, the generative $G$ is trained in adversarial mode against a discriminator network $D$ with the goal of minimizing the Wasserstein distance between the real future series $\bX_f$ and the fake series $\widetilde{\bX}_f$. Formally, the procedure is described by the following optimization problem:
\begin{equation}\label{equation:cgan_losses}
\begin{aligned}
&\mathop{\max}_{D} &\,&\mathbb{E}_{\bx\sim p(\text{data})}
\bigg\{D(\bx) - \mathbb{E}_{\bz\sim p(\bz)}\big[D(G(\bz,\bx_h))\big] \bigg\}
-\\
&& &\,\, \,\lambda_1 \cdot \mathbb{E}_{\overline{\bx}\sim p(\epsilon\, \text{data} + (1-\epsilon) G(\bz))}
\big[ ||\nabla_{\overline{\bx}} D(\overline{\bx}) ||_2 -1 \big]^2;\\
&\mathop{\max}_{G} &\, &\mathbb{E}_{\bx\sim p(\text{data})} \bigg\{\mathbb{E}_{\bz\sim p(\bz)}\big[ D(G(\bz, \bx_h))  \big]\bigg\},\\
\end{aligned}
\end{equation}
where $\bx_h$ contains the historical parts of the data $\bx$ ($\bx_h\in \bx$),  $G(\bz,\bx_h)$ indicates that the generator depends on the (historical) data $\bx_h$, and $\lambda_1$ controls the gradient penalty.
Theoretically, the optimization process finds the surrogate posterior probability distribution $p(\widetilde{\bX}_f | \bX_h)$ that approximates the real posterior probability distribution $p({\bX}_f | \bX_h)$.

\begin{figure}[h]
\centering  
%\vspace{-0.35cm} 
\subfigtopskip=2pt 
\subfigbottomskip=2pt 
\subfigcapskip=-2pt 
\subfigure[Generator]{\includegraphics[width=0.425\textwidth]{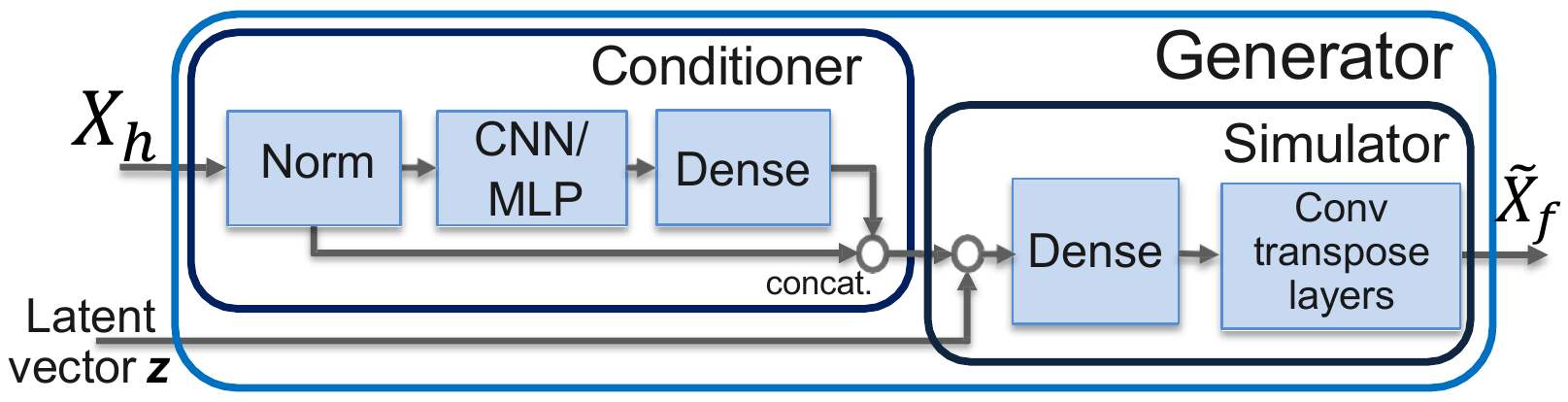} \label{fig:generator_cgan}}
\subfigure[Discriminator]{\includegraphics[width=0.325\textwidth]{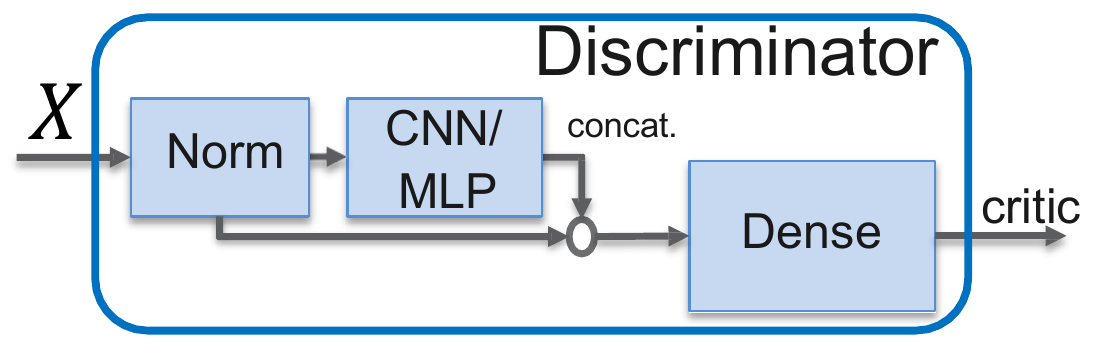} \label{fig:discriminator_cgan}}
\caption{Architectures of the CGAN generative and discriminative models for portfolio analysis.}
\label{fig:cgan_structure}
\end{figure}

\paragraph{Discriminator} The discriminator shown in Figure~\ref{fig:discriminator_cgan} (for both CGAN and ACGAN) takes as input either the real data matrix $\bX=[\bX_h, \bX_f]\in\real^{N\times w}$ or the synthetic data matrix $\widetilde{\bX}=[\bX_h, \widetilde{\bX}_f]\in \real^{N\times w}$.

The main drawback of the CGAN methodology is in that it puts too much emphasis on the \textit{conditioner} to extract features that can ``deceive" the discriminator (Figure~\ref{fig:generator_cgan}). When the discriminator is perfectly trained, this issue is not a big problem. 
However, in most cases, especially due to the scarcity of financial data, the discriminator works imperfectly such that the conditioner may lose important information of the historical data.

\begin{figure}[t!]
	\centering
	\includegraphics[width=0.425\textwidth]{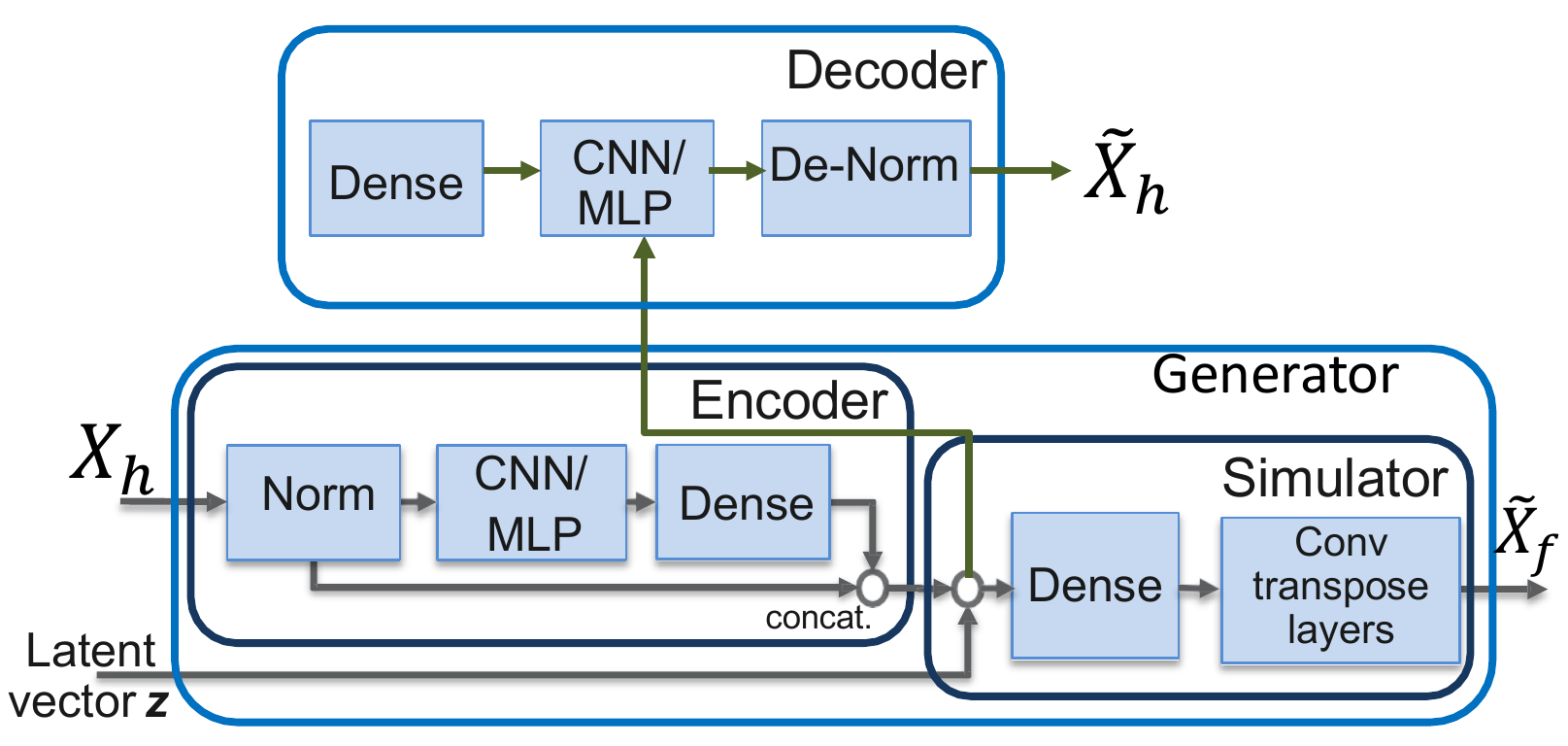}
	\caption{Architecture of the ACGAN generative model for portfolio analysis.}
	\label{fig:acgan_structure}
\end{figure}

\subsection{Autoencoding CGAN}

The Autoencoding CGAN (ACGAN) model partly solves the problem in the CGAN methodology, in which case it finds a balance between the information extraction and generation for cheating the discriminator via an embedded autoencoder providing the capability of keeping the intrinsic information of historical data  \citep{lu2022autoencoding}.
The ACGAN model has the same discriminator structure as the CGAN. However, it contains an extra \textit{decoder} in the generator as shown in Figure~\ref{fig:acgan_structure}. And therefore we call the conditioner an \textit{encoder} in the ACGAN context.

We use an \textit{encoding} deep-neural network $E$ to learn the features that can help the generator cheat the discriminator and can find the internal information itself; and a \textit{decoding} deep-neural network $F$ to reconstruct the historical series so as to force the encoder to do so. Formally the encoding and decoding models reconstruct the historical matrix by
$$
\begin{aligned}
	\by = E(\bX_h), \gap \widetilde{\bX}_h =F(\by).
\end{aligned}
$$ 
This process is known as \textit{autoencoding}, hence the name autoencoding conditional GAN (ACGAN). In a non-GAN context, the autoencoder is typically done by nonnegative matrix factorization or general matrix decomposition via alternative least squares or Bayesian inference \citep{lee1999learning, lu2021numerical, lu2022matrix, lu2022flexible}. Since we need to use the encoding part of the autoencoder to help trick the discriminator as well, the autoencoder is then constructed by deep-neural networks instead. Formally the process is described by the following optimization problem:
\begin{equation}\label{equation:acgan_propose_losses}
\begin{aligned}
&\mathop{\max}_{D} &\,&\mathbb{E}_{\bx\sim p(\text{data})}
\bigg\{D(\bx) - \mathbb{E}_{\bz\sim p(\bz)}\big[D(G(\bz,\bx_h))\big] \bigg\}
-\\
&& &\,\,\, \lambda_1 \cdot \mathbb{E}_{\overline{\bx}\sim p(\epsilon\, \text{data} + (1-\epsilon) G(\bz))}
\big[ ||\nabla_{\overline{\bx}} D(\overline{\bx}) ||_2 -1 \big]^2;\\
& \mathop{\max}_{G, E, F} &\, &\mathbb{E}_{\bx\sim p(\text{data})} \bigg\{\mathbb{E}_{\bz\sim p(\bz)}\big[ D(G(\bz, \bx_h))  \big]-\\
&&& \,\,\, \textcolor{black}{\lambda_2 \cdot g\big(\underbrace{F(E(\bx_h))}_{\widetilde{\bx}_h},\,\, \bx_h\big)}\bigg\}
,\\
\end{aligned}
\end{equation}
where $g(\cdot)$ denotes the error function; the authors apply the mean squared error as the loss function in \citet{lu2022autoencoding}. 
The parameter $\lambda_2$ controls how large the penalization by the autoencoder, and the term is thus known as the \textit{autoencoding penalty (AP)}.
In the original CGAN methodology, the conditioner is used to extract features that can help the generator to cheat the discriminator; however, it may lose some important information that captures the internal features of the market trend. 
The ACGAN then finds a balance between cheating the discriminator and keeping its market information.

\subsection{Data Normalization}
Following \citet{mariani2019pagan, lu2022autoencoding}, Given the frame window of $w=h+f$ days \footnote{$h$ for the historical length, $f$ for the future length. The historical series is denoted by $\bp_{1:h}\in \real^h$, and the real future series can be obtained by $\bp_{h+1:w}\in \real^f$.}, we consider the \textit{adjusted closing price} $\bp\in \real^w$ series for each asset. Then we unit-normalize the price series $\bp$ for each asset to fill in the range $[-1,1]$ for the initial $h$ days. This normalization procedure can help us to expose the values of neural networks limited within a suitable range that removes price-variability over multiple assets within the specified window.
In practice, the unit-normalization can be done by \textit{3-sigma normalization}: given the mean $\mu$ and standard deviation $\sigma$ of $\bp_{1:h}\in \real^h$, the normalization is done by
\begin{equation}\label{equation:acgan_datanormalization}
\widetilde{\bp} = (\bp-\mu)/(3\sigma).
\end{equation}
After generating the surrogate future series $\widetilde{\bp}_{h+1:w}$, we apply again a de-normalization procedure:
\begin{equation}\label{equation:denormalization-acgan}
\widehat{\bp}_{h+1:w} = \widetilde{\bp}_{h+1:w} \times 3\sigma +\mu.
\end{equation}

\section{Hybrid Methods}
\subsection{CGAN with Eavesdropping}
\begin{figure}[h]
\centering  
%	\vspace{-0.35cm} 
\subfigtopskip=2pt 
\subfigbottomskip=2pt 
\subfigcapskip=-5pt 
\subfigure[US, rebalance every 15 days]{\includegraphics[width=0.225\textwidth]{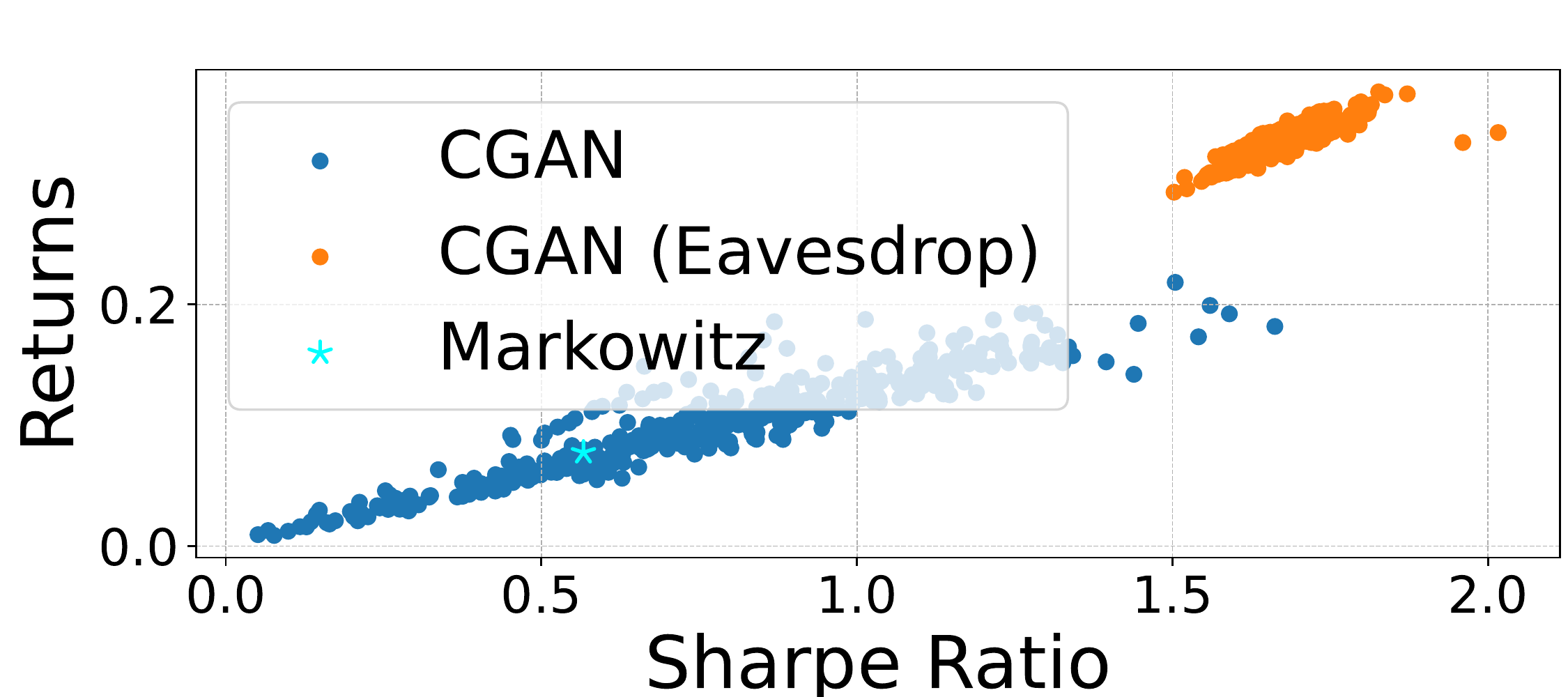} \label{fig:return_sharpe_us1_sample}}
\subfigure[US, rebalance every 20 days]{\includegraphics[width=0.225\textwidth]{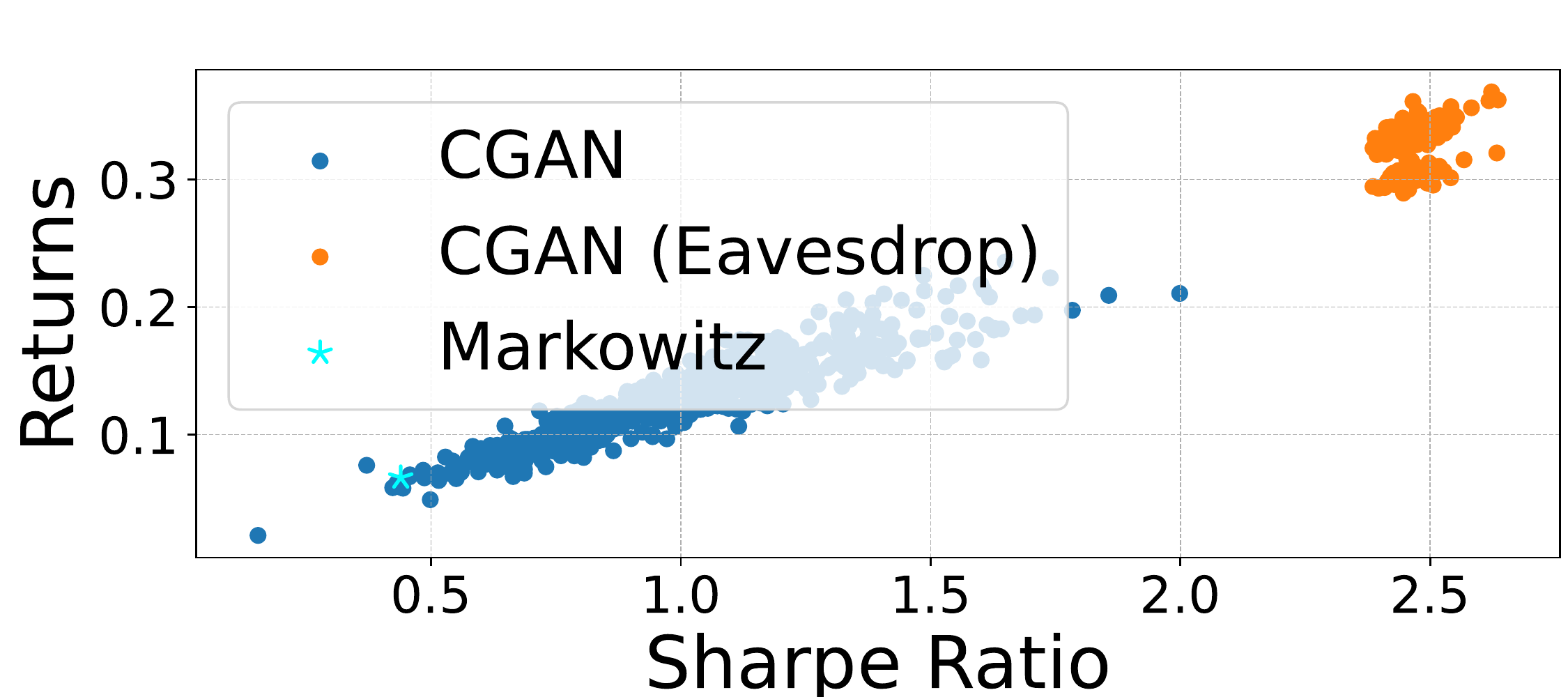} \label{fig:return_sharpe_us2_sample}}
\caption{\textbf{Eavesdroping:} (Annual) return-SR measured on the test period in US region by randomly sampling 1000 series for CGAN and \textbf{CGAN with Eavesdropping}. Similar results can be observed for ACGAN and ACGAN with Eavesdroppoing models.}
\label{fig:risk-sharperation_hybrid_eavesdrop}
\end{figure}

The proposed hybrid approaches highly rely on the data normalization procedure. 
Suppose in Eq.~\eqref{equation:acgan_datanormalization}, we obtain the mean $m$ of the whole asset $\bp\in \real^w$ in the window $w$ rather than the historical one $\bp_{1:h}\in \real^h$, and we apply the data normalization by this value:
\begin{equation}\label{equation:acgan_datanormalization_eavesdrop}
	\widetilde{\bp} = (\bp-\textcolor{blue}{m})/(3{\sigma}).
\end{equation}
Since the $\bp_{h+1:w}$ in $\bp$ cannot be obtained in practice and we thus call this method \textit{eavesdropping}.
A simple experiment on this approach, comparing the CGAN and CGAN with Eavesdropping, the return-SR (Sharpe ratio) plots in Figure~\ref{fig:risk-sharperation_hybrid_eavesdrop} shows that this eavesdropping procedure can increase the performance to a large extent, enforcing the mean Sharpe ratio from about 1.0 to 2.5. And the random draws are more clustered with small deviations such that the end strategy is more stable.
This is reasonable since the GAN finds the future means of the assets from this normalization and generates the series whose means are closer to these values.

\subsection{Proposed Methodology}
\begin{figure}[h]
	\centering  
	%\vspace{-0.35cm} 
	\subfigtopskip=2pt 
	\subfigbottomskip=2pt 
	\subfigcapskip=-2pt 
\subfigure[Proposer]{\includegraphics[width=0.285\textwidth]{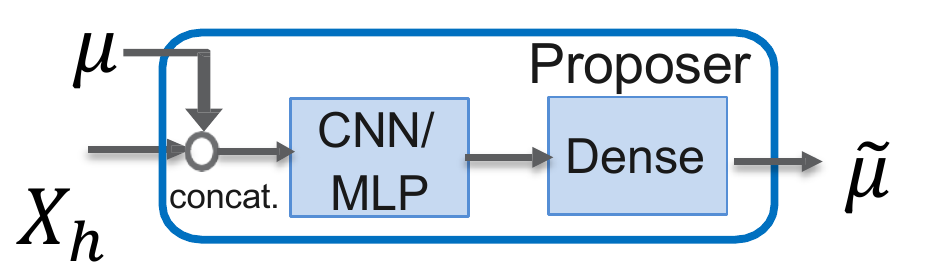} \label{fig:generator_hybridcgan_proposer}}
	\subfigure[Generator]{\includegraphics[width=0.425\textwidth]{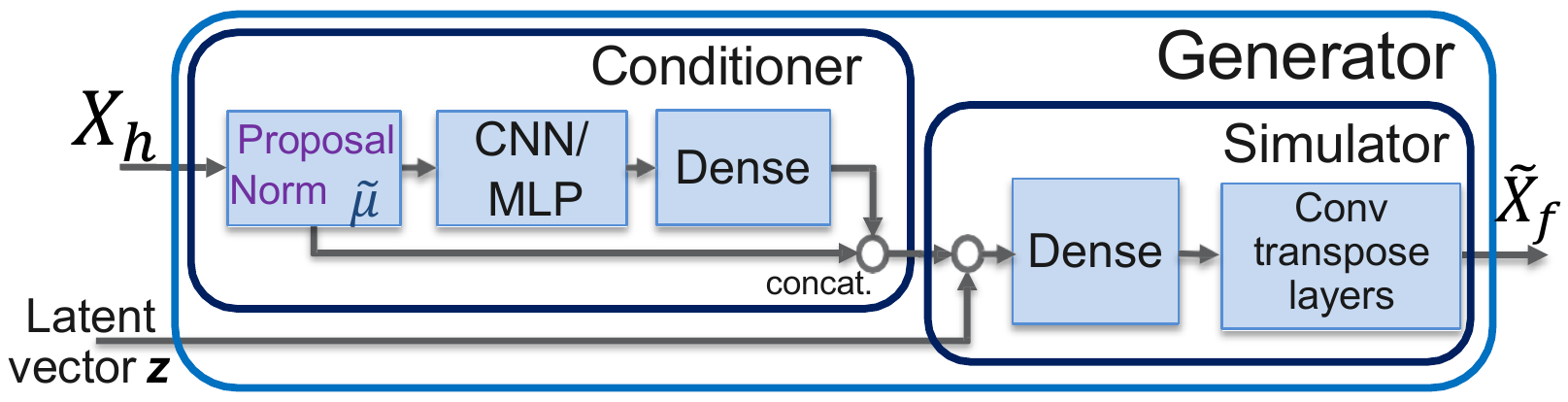} \label{fig:generator_hybridcgan}}
	\subfigure[Discriminator]{\includegraphics[width=0.325\textwidth]{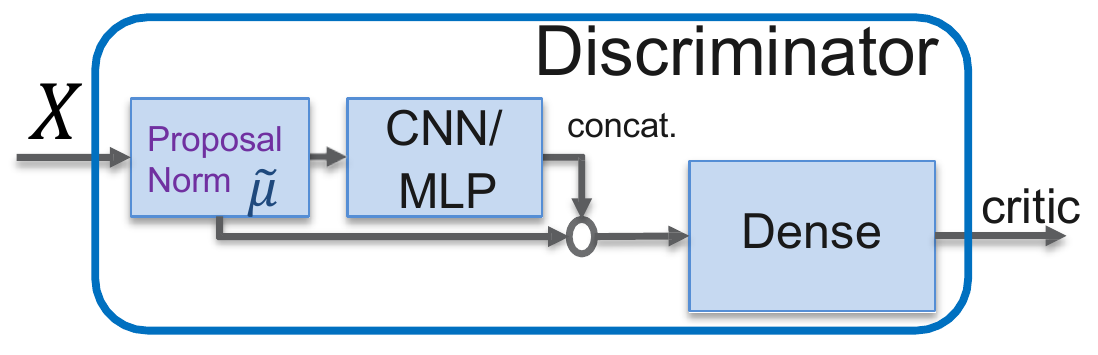} \label{figdiscriminator_hybridcgan}}
	\caption{Architectures of the HybridCGAN generative and discriminative models for portfolio analysis.}
	\label{fig:hybridcgan_structure}
\end{figure}

Though the eavesdropping methods are not practical since we use future data,
% and this has the problem of \textit{forward bias}, 
the idea can be applied to the proposed hybrid methods. 
This is known as the \textit{forward bias} issue in quantitative finance.
We incorporate an extra proposal network $P$ to find the surrogate of the mean values for each asset in corresponding windows as shown in Figure~\ref{fig:hybridcgan_structure}.
The proposer $P$ is trained by minimizing the mean squared error between the real mean value $m$ of the whole period and the prediction $\widetilde{\mu}$. 
Note that we incorporate the historical mean into the input of the proposal network to make the optimization easier:
\begin{equation}\label{equation:hybrid-meaninput}
\widetilde{\mu} = P(\bX_h, \mu),
\end{equation}
i.e., we concatenate $\bX_h$ and $\mu$ to predict $\widetilde{\mu}$.
After finding the surrogate mean value $\widetilde{\mu}$, the normalization follows:
\begin{equation}\label{equation:acgan_datanormalization_hybrid}
	\widetilde{\bp} = (\bp-\textcolor{blue}{\widetilde{\mu}})/(3{\sigma}).
\end{equation}
And the de-normalization:
\begin{equation}\label{equation:denormalization-hybriudacgan}
	\widehat{\bp}_{h+1:w} = \widetilde{\bp}_{h+1:w} \times 3\sigma +\textcolor{blue}{\widetilde{\mu}}.
\end{equation}
Since this approach combines the CGAN and deep neural network regression, we call it the \textit{HybridCGAN} model.
This hybrid approach can be easily extended into the ACGAN methodology in a similar way, termed the \textit{HybridACGAN} model.

\begin{table}
%	\parbox{.47\linewidth}{
		\centering
\scriptsize
\begin{tabular}{l|lllll}
	\hline
	& Ticker & Type & Sector & Company & Curr. \\ \hline
\parbox[t]{0.3mm}{\multirow{10}{*}{\rotatebox[origin=c]{90}{US Region}  }} 
& MSFT   & Share & IT               & Microsoft                & USD   \\
& GOOG   & Share & IT               & Alphabet                 & USD   \\
& XOM    & Share & Energy           & Exxon Mobil              & USD   \\
& HES    & Share & Energy           & Hess                     & USD   \\
& PFE    & Share & Healthcare       & Pfizer                   & USD   \\
& WBA    & Share & Consumer staples & Walgreens  Alliance & USD   \\
& KR     & Share & Consumer staples & The Kroger               & USD   \\
& IYR    & ETF   & Real estate      & iShares US Real Estate   & USD   \\
& IYY    & ETF   & Dow Jones        & iShares Dow Jones        & USD   \\
& SHY      & ETF   & US treasury bond & iShares Treasury Bond    & USD   \\
	\hline
	\hline 
	\parbox[t]{0.3mm}{\multirow{10}{*}{\rotatebox[origin=c]{90}{EU Region}  }} 
& VOW3.DE  & Share & Automotive       & Volkswagen               & EUR   \\
& BMW.DE   & Share & Automotive       & BMW                      & EUR   \\
& VK.PA    & Share & Industrials      & Vallourec S.A.           & EUR   \\
& SOI.PA   & Share & Industrials      & Soitec S.A.              & EUR   \\
& DTE.DE   & Share & Technology       & Deutsche Telekom AG      & EUR   \\
& SAP.DE   & Share & Technology       & SAP SE                   & EUR   \\
& BAS.DE   & Share & Basic materials  & BASF SE                  & EUR   \\
& BAYN.DE  & Share & Healthcare       & Bayer AG & EUR   \\
& $~^\wedge$FCHI  & Index & French market    & CAC 40                   & EUR   \\
& $~^\wedge$GDAXI & Index & German market    & DAX                      & EUR   \\
	\hline
\end{tabular}
\caption{Summary of the underlying portfolios in the US and EU markets, 10 assets for each market respectively. In each region, we include assets from various sectors to favor a somehow sector-neutral strategy.}
\label{table:us_eu_data_summary}		
\end{table}

\begin{figure}[h]
	\centering  
	%\vspace{-0.35cm} 
	\subfigtopskip=2pt 
	\subfigbottomskip=2pt 
	\subfigcapskip=-5pt 
	\subfigure{\includegraphics[width=0.425\textwidth]{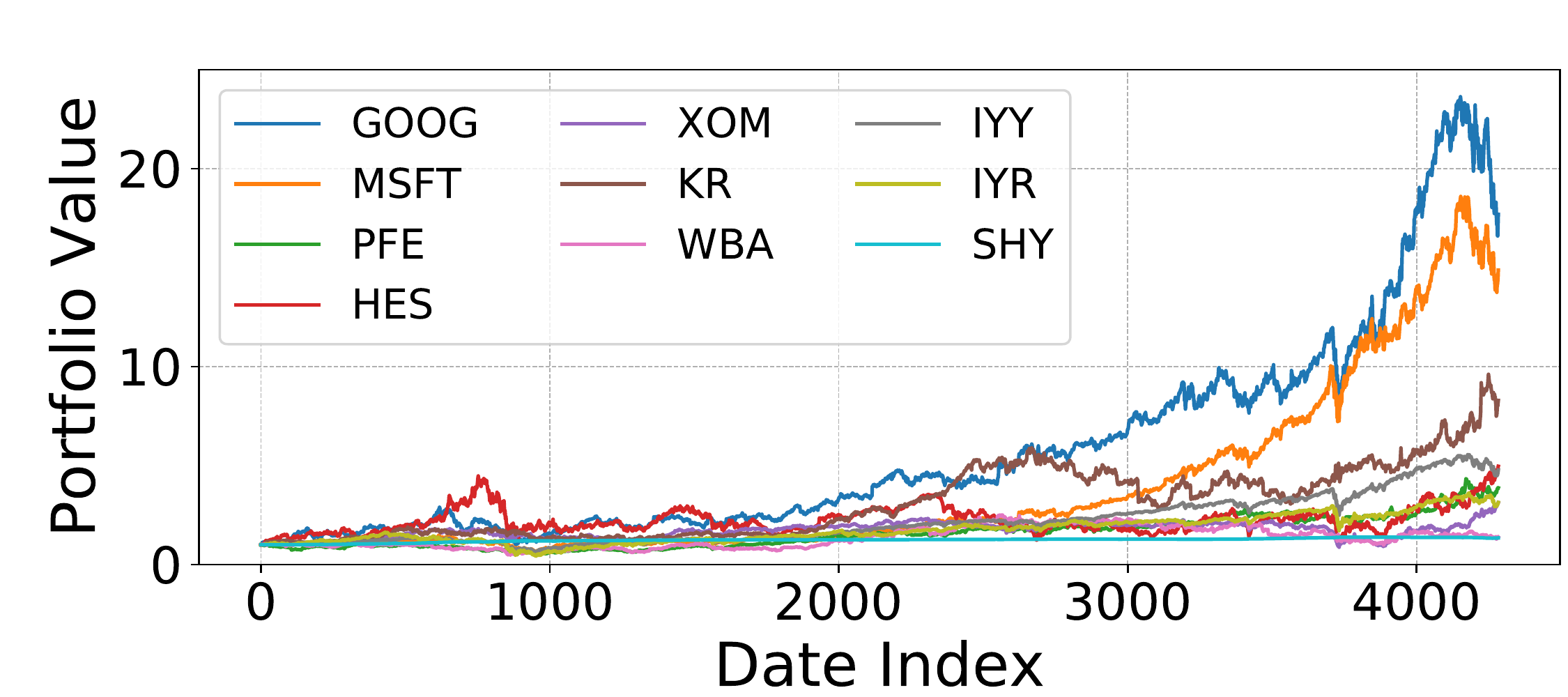} \label{fig:acgan_dataset_us}}
	\subfigure{\includegraphics[width=0.425\textwidth]{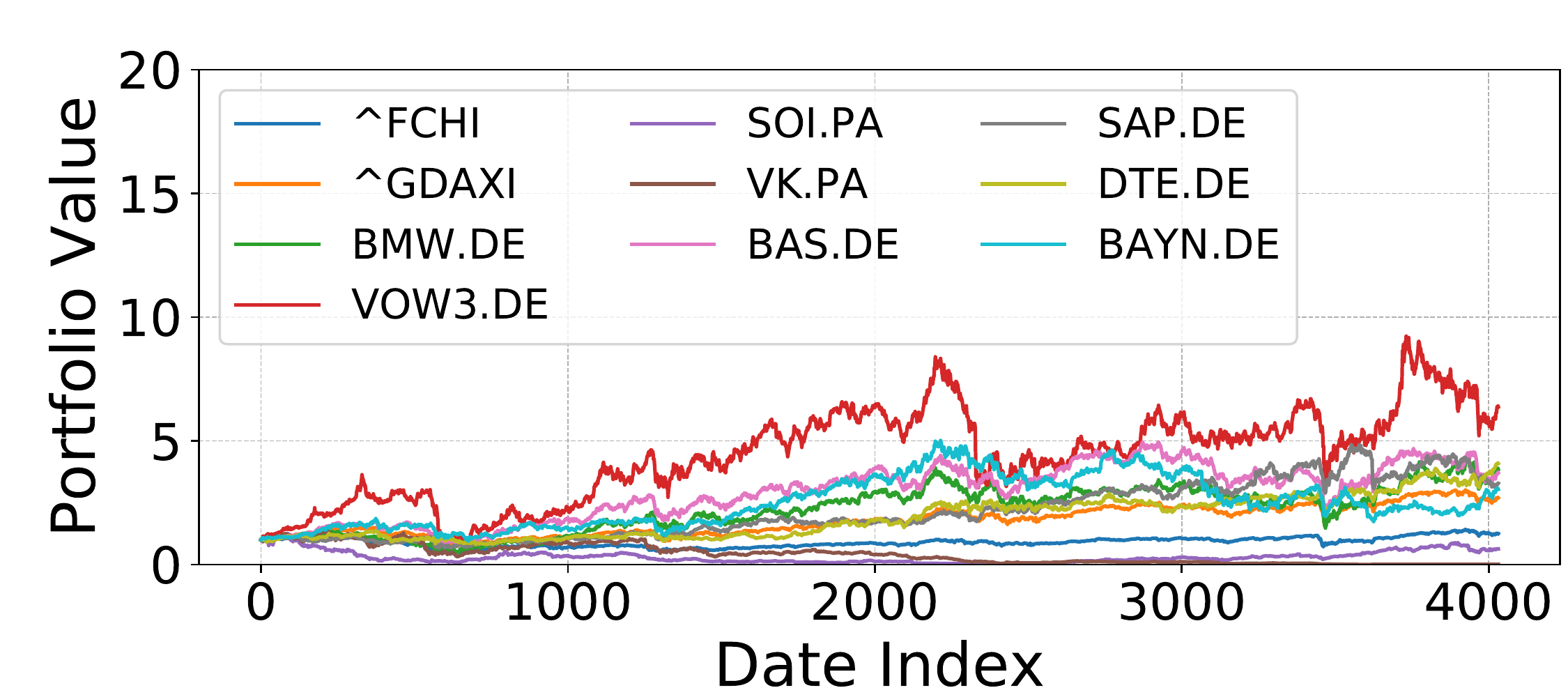} \label{fig:acgan_dataset_eu}}
	\caption{Different portfolios for the US (upper) and EU (lower) markets with a unit initial value.}
	\label{fig:acgan_dataset_us_eu}
\end{figure}

\section{Experiments}\label{section:ader_experiments}

\begin{algorithm}[!htb]
	\caption{Training and testing process for the HybridCGAN, HybridACGAN, CGAN, and ACGAN models.} 
	\label{alg:acgan_training}  
	\begin{algorithmic}[1] 
		\STATE {\bfseries General Input:} Choose parameters $w=h+f$; number of assets $N$; number of epoches $T$; latent dimension $m$;
		\STATE {\bfseries Training Input: } Training data matrix $\bM\in \real^{N\times D}$;
		\STATE Decide index set $\mathcal{S}_1=\{1,2,\ldots, D-w+1\}$ and \textbf{draw without replacement};
		\FOR{$t=1$ to $T$}
\FOR{$i$ $\in$ $\text{\textcolor{blue}{random}}(\mathcal{S}_1)$} 
\STATE $\bX = [\bX_h, \bX_f]=\bM[:,i:i+w-1]\in \real^{N\times w}$;
\STATE Randomly sample latent vector $\bz\in \real^m$;
\STATE Backpropatation for generator in Eq.~\eqref{equation:acgan_propose_losses} or \eqref{equation:cgan_losses};
\STATE Generate surrogate $\widetilde{\bX}_f = G( \bz,\bX_h)\in \real^{N\times f}$;
\STATE Backpropatation for discriminator in Eq.~\eqref{equation:acgan_propose_losses} or \eqref{equation:cgan_losses};
\ENDFOR
		\ENDFOR
		\STATE {\bfseries Inference Input: } Testing data matrix  $\bA\in \real^{N\times K}$;
		\STATE {\bfseries Inference Output: } Testing data matrix $\bB\in \real^{N\times K}$;
		\STATE Decide index set $\mathcal{S}_2=\{\textcolor{blue}{h+1},h+f+1,\ldots\}$;
		\STATE Copy the first $h$ days data $\bB[:,1:h]=\bA[:,1:h]$;
		\FOR{$i$ $\in$ $\text{\textcolor{blue}{ordered}}(\mathcal{S}_2)$} 
		\STATE $\bX = [\bX_h, \bX_f]=\bA[:,i:i+w-1]\in \real^{N\times w}$;
		\STATE Randomly sample latent vector $\bz\in \real^m$;
		\STATE Generate $\bB[:,i:i+f-1] = G(\bz,\bX_h)\in \real^{N\times f}$ with de-normalization in Eq.~\eqref{equation:denormalization-acgan};
		\ENDFOR
		\STATE Output the synthetic series $\bB$;
	\end{algorithmic} 
\end{algorithm}

\begin{figure*}[h]
\centering  
%\vspace{-0.35cm} 
\subfigtopskip=2pt 
\subfigbottomskip=2pt 
\subfigcapskip=-2pt 
\subfigure{\includegraphics[width=1\textwidth]{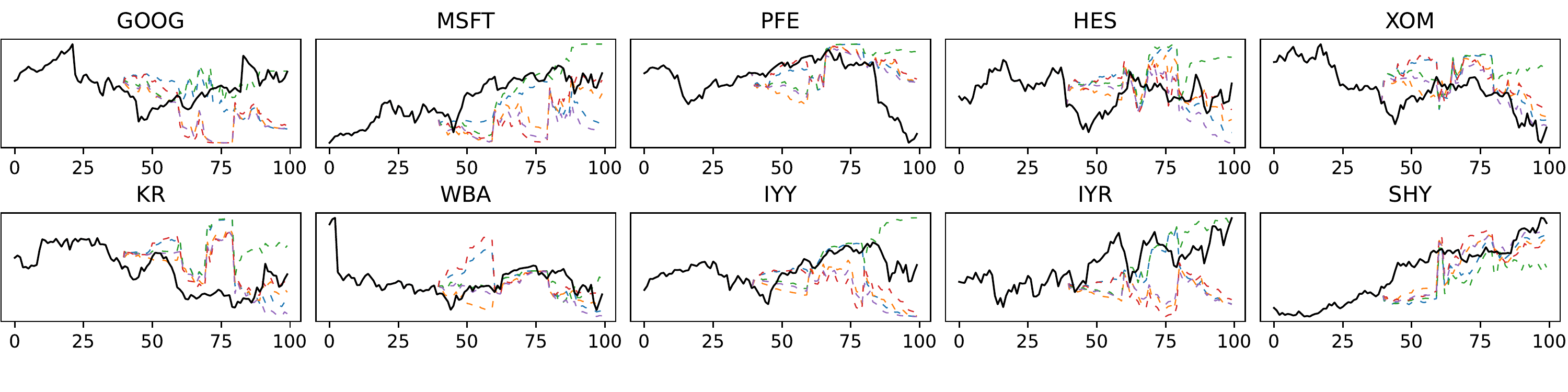}}
\subfigure{\includegraphics[width=1\textwidth]{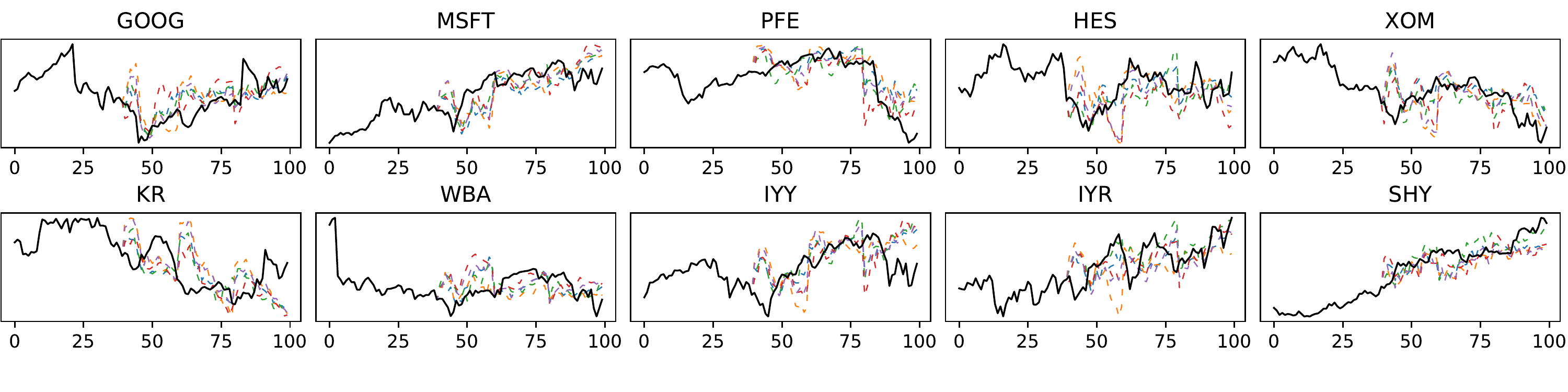}}
\caption{Actual price trend (black solid line) of the US assets, five representative simulations (colored dashed lines) generated by HybridCGAN (upper two rows), and five representative simulations generated by CGAN (lower two rows) for the first 100 trading days in the test set. 
%	The simulated price series of the whole period (800 trading days) and simulated price series of the CGAN model (100 and all trading days) can be found in Figure~\ref{fig:diversity_us_eu_acgan_800}, \ref{fig:diversity_us_eu_cgan}, and \ref{fig:diversity_us_eu_cgan_800} respectively.
}
\label{fig:diversity_us_eu_ecgan}
\end{figure*}
\begin{figure*}[!htb]
	\centering  
	\vspace{-0.35cm} 
	\subfigtopskip=2pt 
	\subfigbottomskip=2pt 
	\subfigcapskip=-5pt 
	\subfigure[US, rebalance every 10 days]{\includegraphics[width=0.325\textwidth]{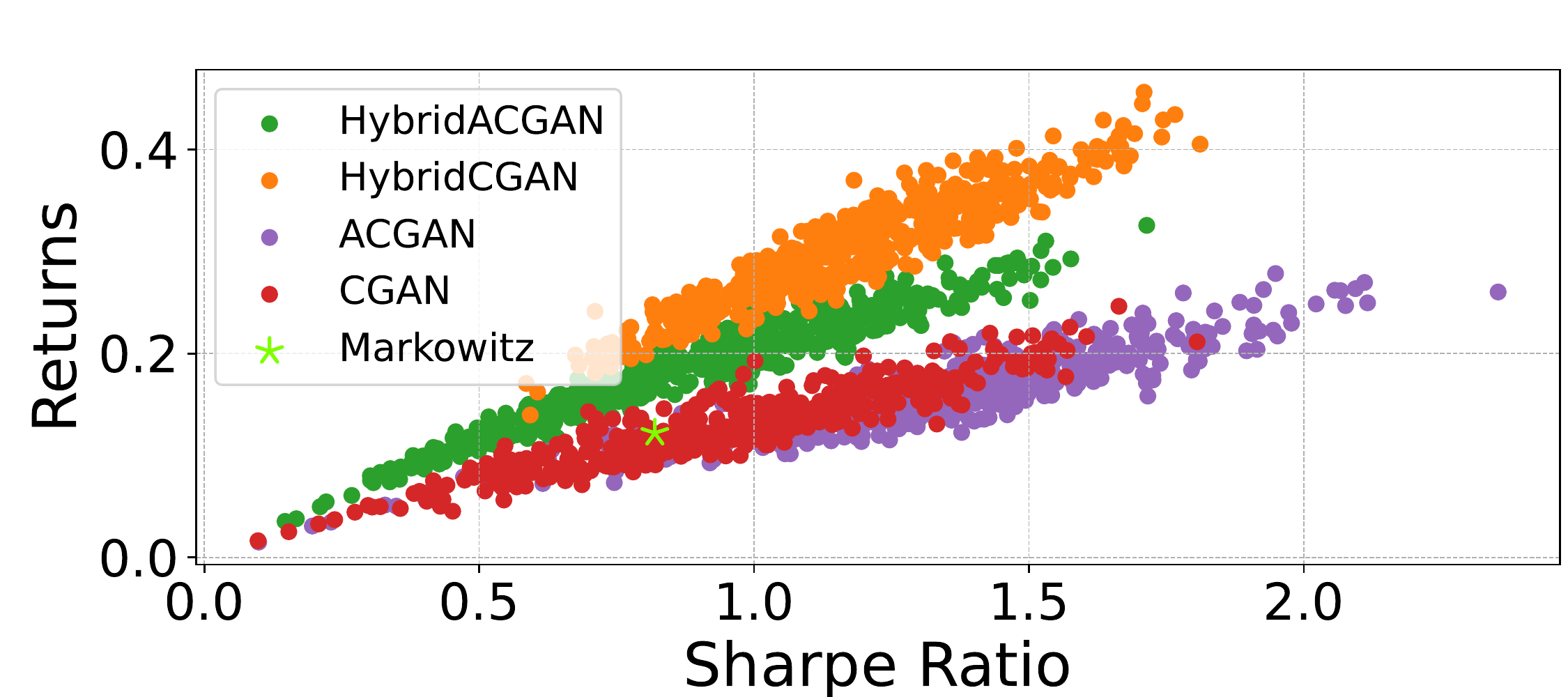} \label{fig:return_sharpe_us1}}
	\subfigure[US, rebalance every 15 days]{\includegraphics[width=0.325\textwidth]{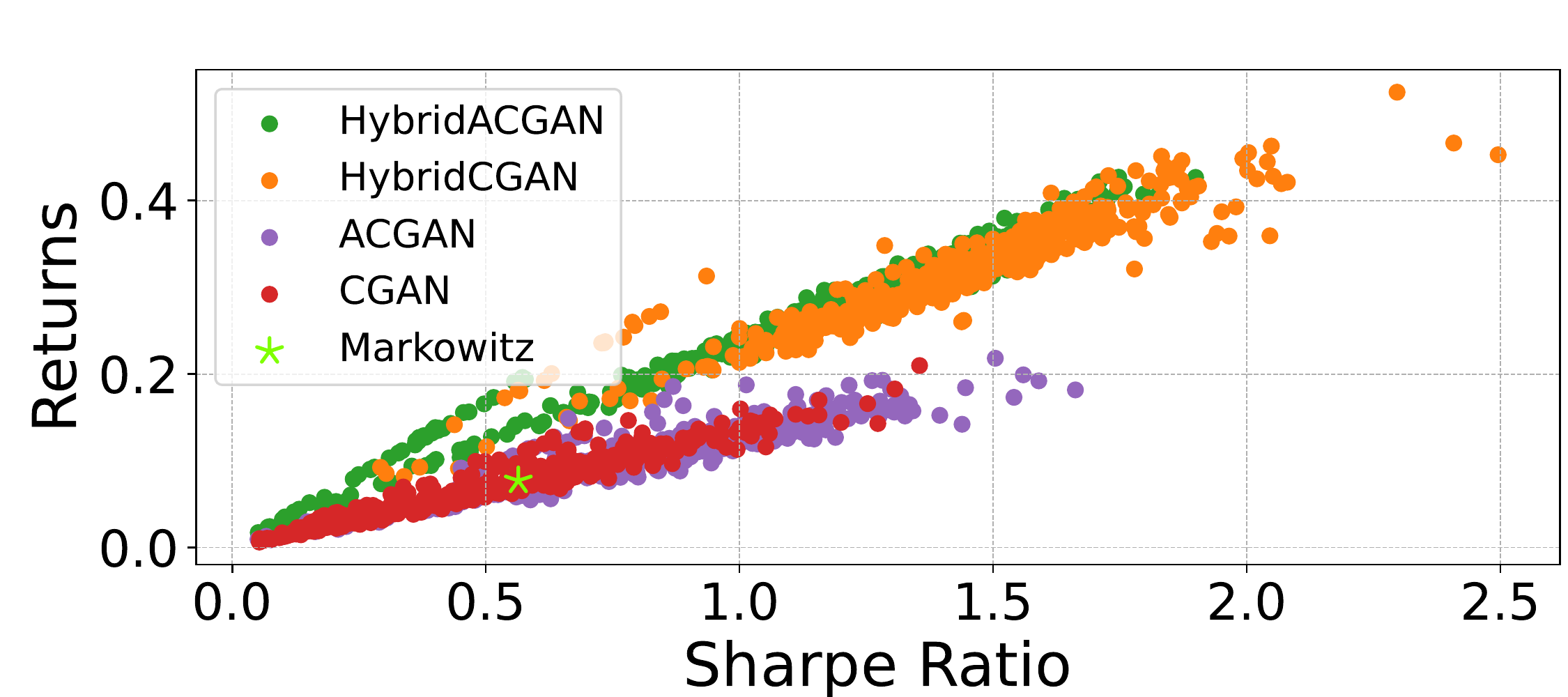} \label{fig:return_sharpe_us2}}
	\subfigure[US, rebalance every 20 days]{\includegraphics[width=0.325\textwidth]{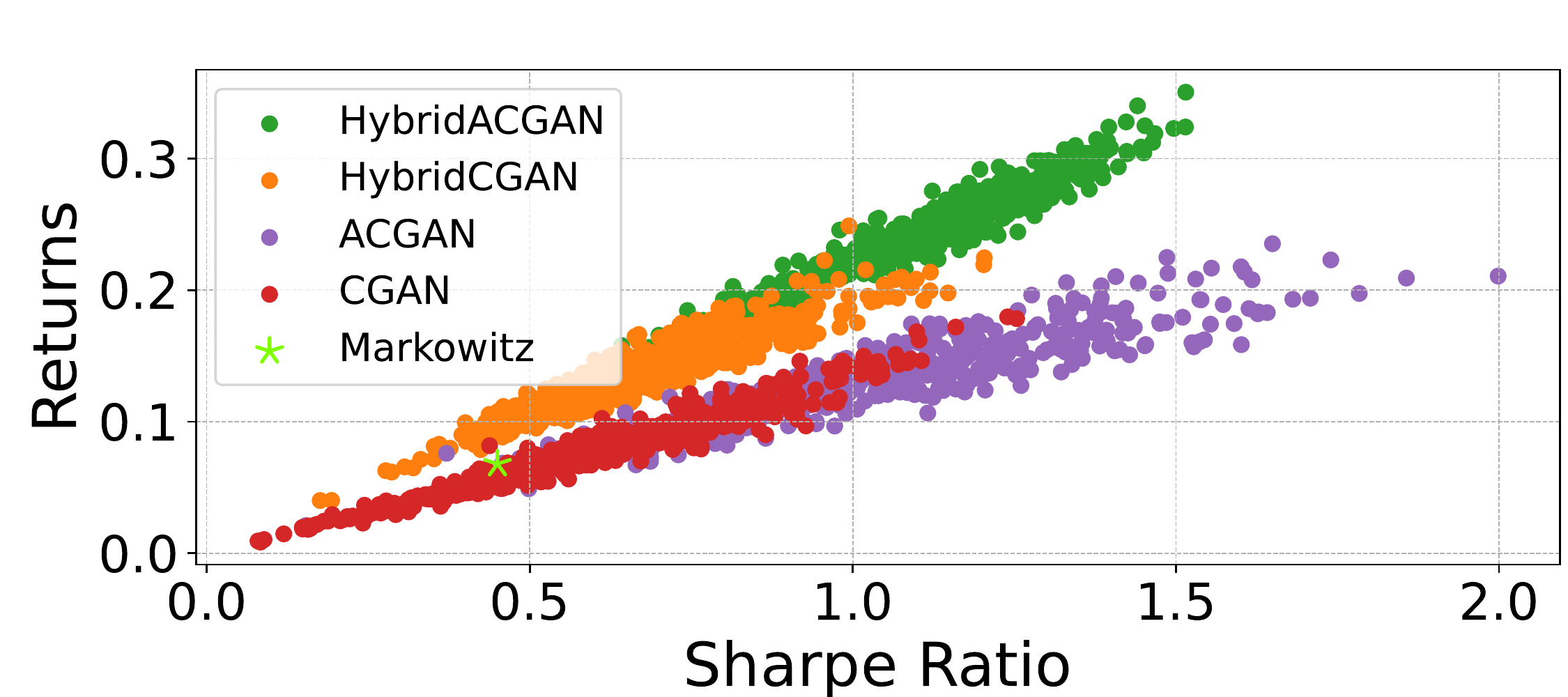} \label{fig:return_sharpe_us3}}
	\subfigure[EU, rebalance every 10 days]{\includegraphics[width=0.325\textwidth]{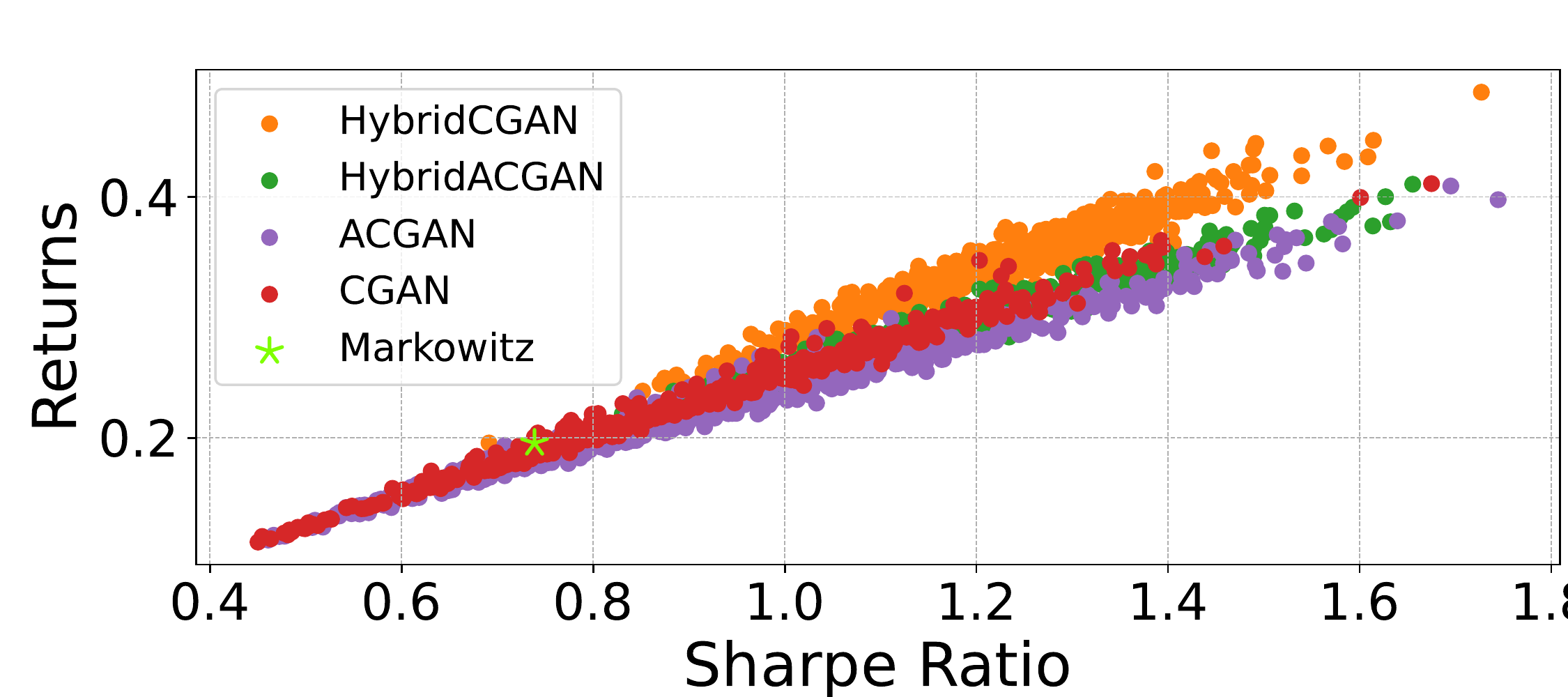} \label{fig:return_sharpe_eu1}}
	\subfigure[EU, rebalance every 15 days]{\includegraphics[width=0.325\textwidth]{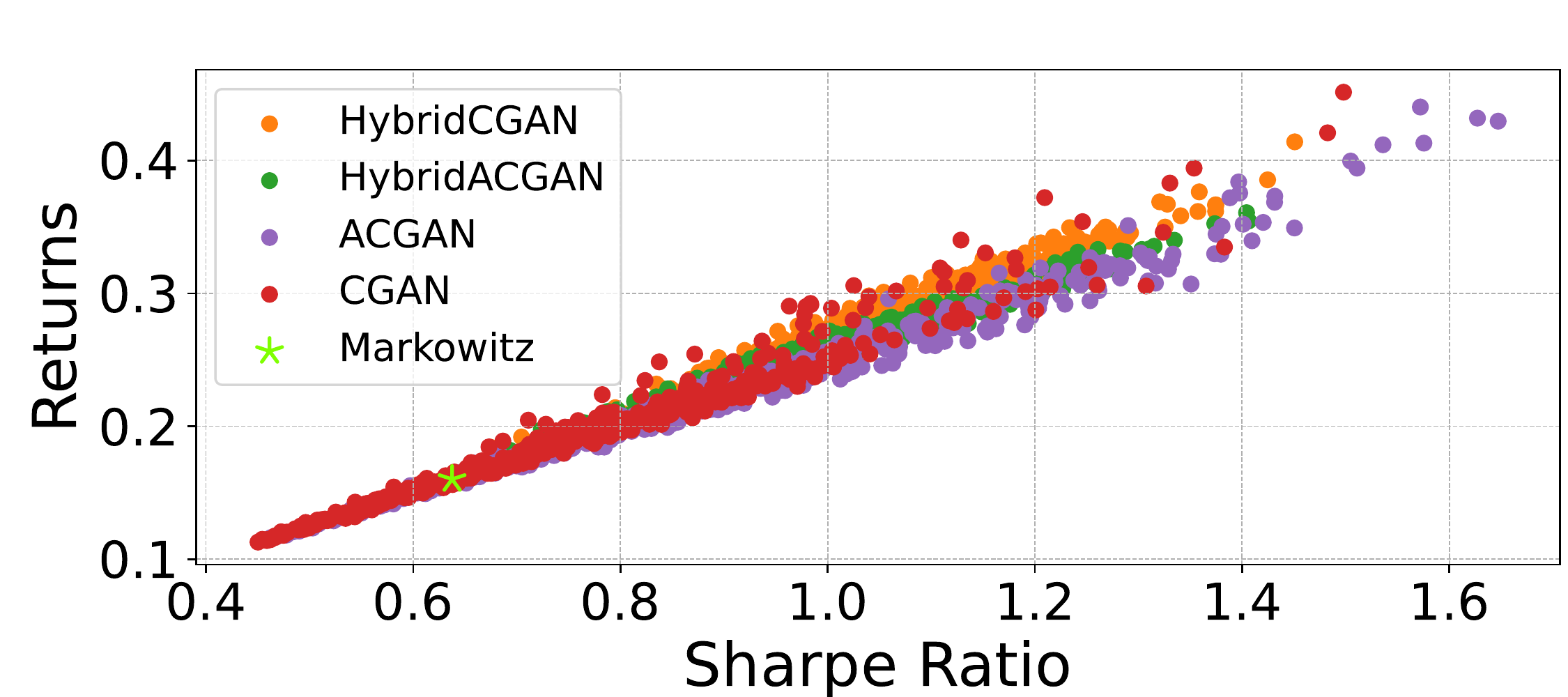} \label{fig:return_sharpe_eu2}}
	\subfigure[EU, rebalance every 20 days]{\includegraphics[width=0.325\textwidth]{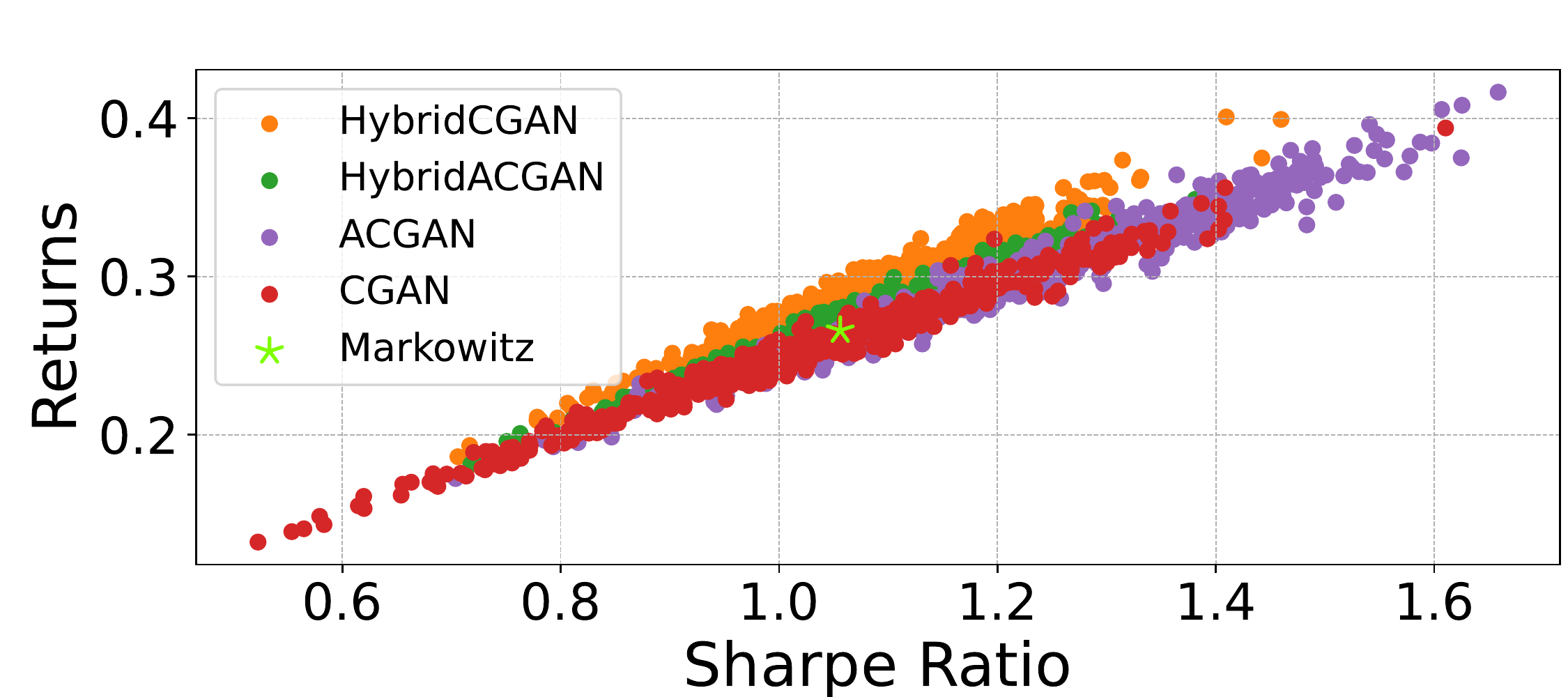} \label{fig:return_sharpe_eu3}}
	\caption{(Annual) return-SR measured on the test period by randomly sampling 1000 series.}
	\label{fig:risk-sharperation_acgan_compare}
\end{figure*}

To evaluate the strategy and demonstrate the main advantages of the proposed HybridCGAN and HybridACGAN methods, 
we conduct experiments with different analysis tasks; datasets from different geopolitical markets including the US and the European (EU) markets, and various industrial segments including Healthcare, Automotive, Energy and so on.
We obtain publicly available data from Yahoo Finance \footnote{\url{https://finance.yahoo.com/}.}. 
For the US market, we obtain data for a 17-year period, i.e., from 2005-05-24 to 2022-05-27, where the data between  2005-05-24 and 2019-03-28 is considered training data; 
while data between 2019-03-28 and 2022-05-27 is taken as the test set (800 trading days).
For the EU market, we obtain data for a 16-year period, i.e., from 2006-07-18 to 2022-06-07, where the data between 2006-07-18 and 2019-04-09 is considered training data;
while data between 2019-04-10 and 2022-06-07 is taken as the test set (800 trading days). 
The underlying portfolios are summarized in Table~\ref{table:us_eu_data_summary}:
\begin{itemize}
\item \textit{US market}: 10 assets of US companies from different industrial segments, i.e., GOOG and MSFT (from IT sector), PFE (from Healthcare sector), XOM and HES (from Energy sector), WBA and KR (from Consumer staples sector), and three ETFs (IYY, IYR, SHY).
\item \textit{EU market}: 10 portfolios of EU companies from different industrial segments, i.e., VOW3.DE and BMW.DE (from Automotive sector), VK.PA and SOI.PA (from Industrials sector),  DTE.DE and SAP.DE (from Technology sector), BAS.DE (from Basic materials sector), BAYN.DE (from Healthcare sector), and two indices, $~^\wedge$FCHI and $~^\wedge$GDAXI, that track the German and French stock markets respectively. 
\end{itemize}
The specific time periods and assets are chosen by following the four criteria. \textit{1). Data diversity}: in each market, we include companies from different sectors so that the final strategies are somewhat sector-neutral with fewer risks; \textit{2). Data availability}: we cover as a longer period as possible to make a decent prediction; the periods are selected to make all the assets have same frame length; \textit{3). Data correctness}: given the Yahoo Finance data source, we only include the data that do not have NaN values; \textit{4). Currency homogeneity}: in each region, the traded currencies are the same. Figure~\ref{fig:acgan_dataset_us_eu} shows the series of different assets where we initialize each portfolio with a unitary value for clarity.
In all experiments, the same parameter initialization is adopted when conducting different tasks. 
We compare the results in terms of performance of portfolio allocation and diversification of the assets. In a wide range of scenarios across various tasks, HybridCGAN and HybridACGAN improve portfolio evaluations, and lead to return-risks performances that are as good or better than the existing Markowitz framework, CGAN, and ACGAN methodologies.

\begin{figure*}[!htb]
	\centering  
	\vspace{-0.35cm} 
	\subfigtopskip=2pt 
	\subfigbottomskip=2pt 
	\subfigcapskip=-5pt 
	\subfigure[US, rebalance every 10 days,
	Sharpe ratios of HybridCGAN, HybridACGAN, CGAN, ACGAN, and Markowitz are 1.32, 0.95, 1.22, \textbf{1.72}, and 0.82 respectively.
	]{\includegraphics[width=0.325\textwidth]{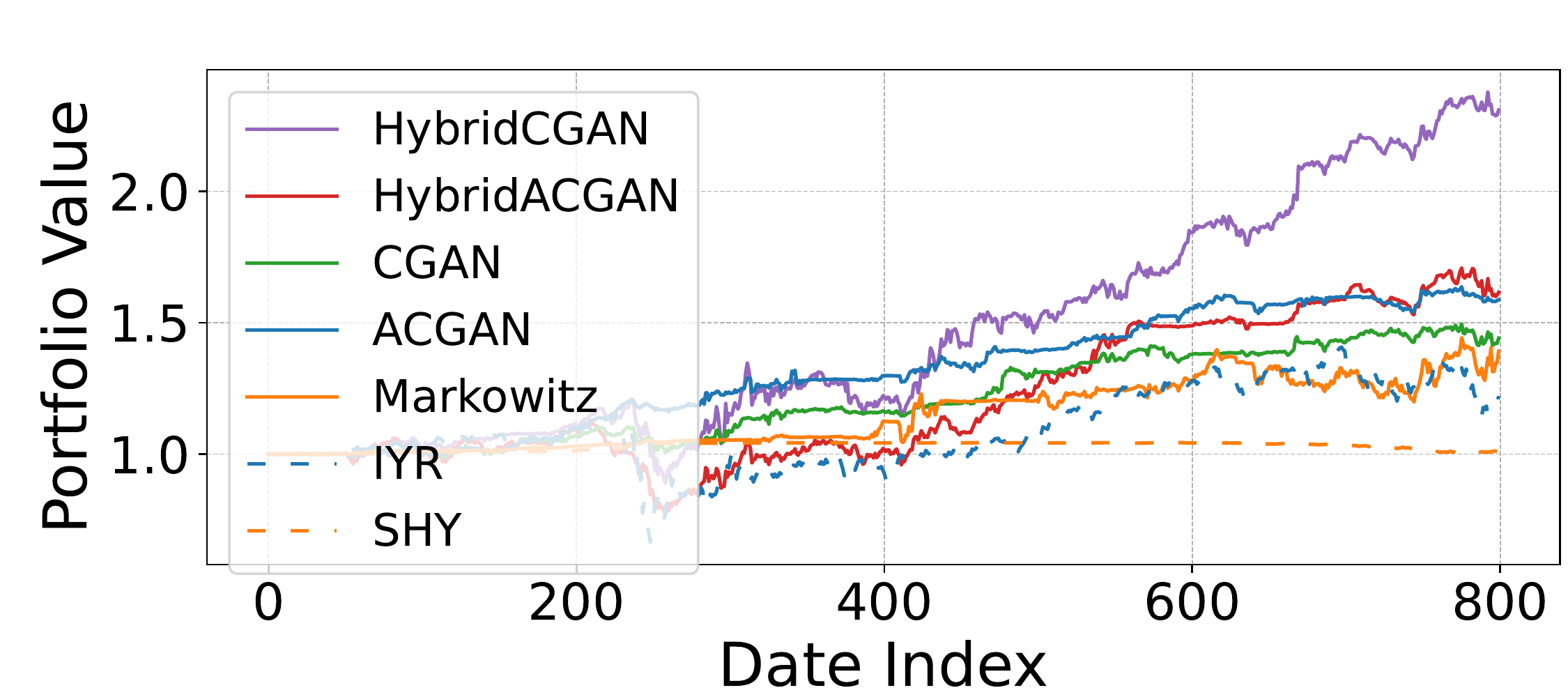} \label{fig:portfolio_value_us1}}
	\subfigure[US, rebalance every 15 days. Sharpe ratios of HybridCGAN, HybridACGAN, CGAN, ACGAN, and Markowitz are \textbf{1.47}, 1.18, 0.49, 0.82, and 0.56 respectively.]{\includegraphics[width=0.325\textwidth]{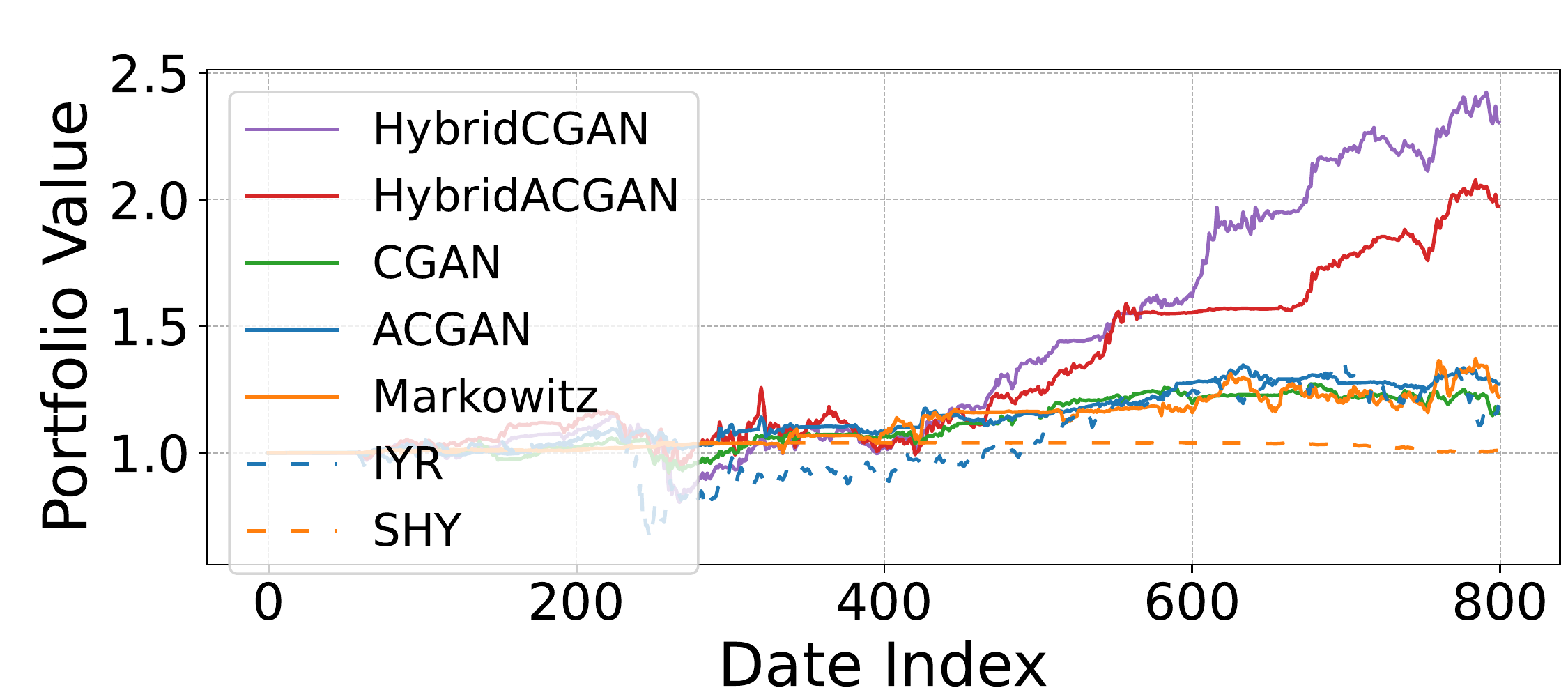} \label{fig:portfolio_value_us2}}
	\subfigure[US, rebalance every 20 days.
	Sharpe ratios of HybridCGAN, HybridACGAN, CGAN, ACGAN, and Markowitz are 0.74, \textbf{1.20}, 0.71, 1.17, and 0.45 respectively. ]{\includegraphics[width=0.325\textwidth]{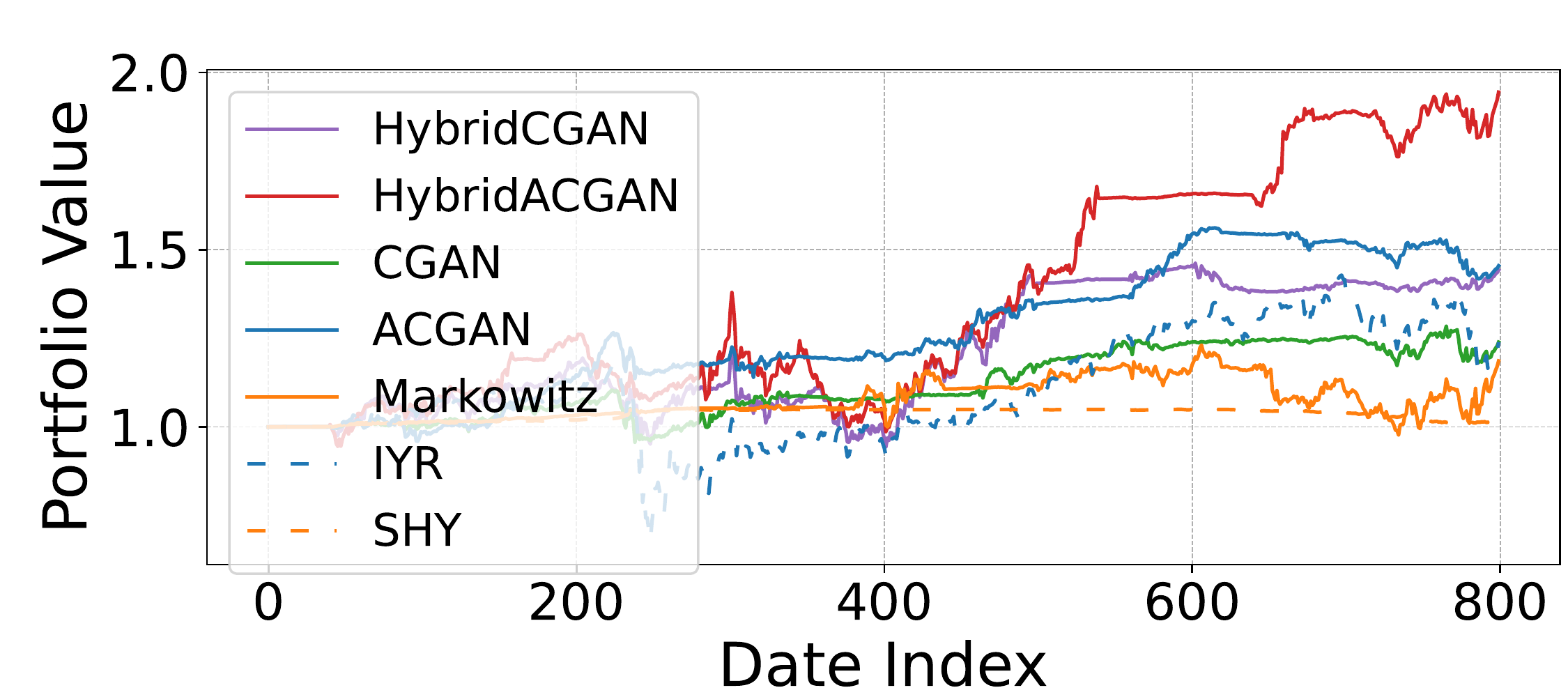} \label{fig:portfolio_value_us3}}
	\subfigure[EU, rebalance every 10 days, 
	Sharpe ratios of HybridCGAN, HybridACGAN, CGAN, ACGAN, and Markowitz are \textbf{1.33}, 1.22, 0.95, 1.02, and 0.74 respectively.
	]{\includegraphics[width=0.325\textwidth]{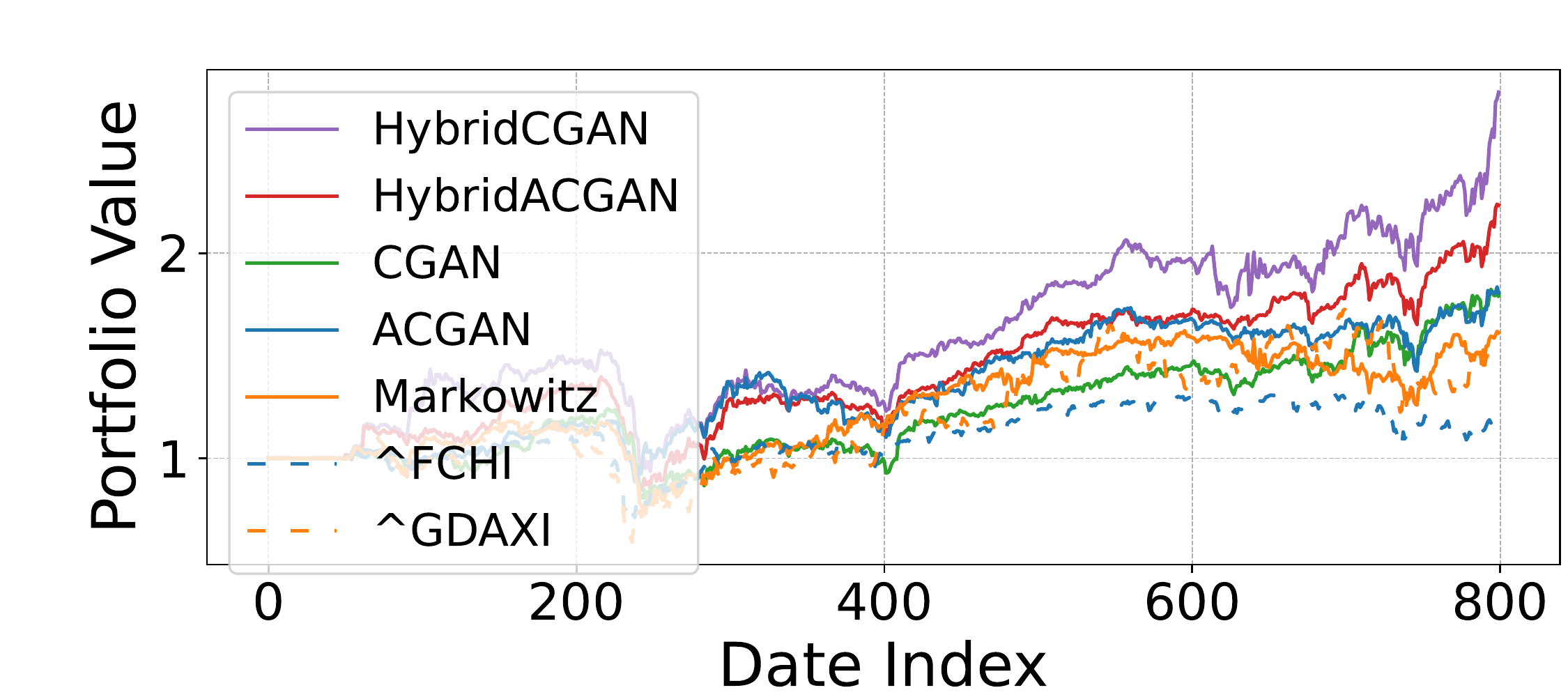} \label{fig:portfolio_value_eu1}}
	\subfigure[EU, rebalance every 15 days. Sharpe ratios of HybridCGAN, HybridACGAN, CGAN, ACGAN, and Markowitz are 0.96, \textbf{1.01}, 0.76, 0.96, and 0.64 respectively.]{\includegraphics[width=0.325\textwidth]{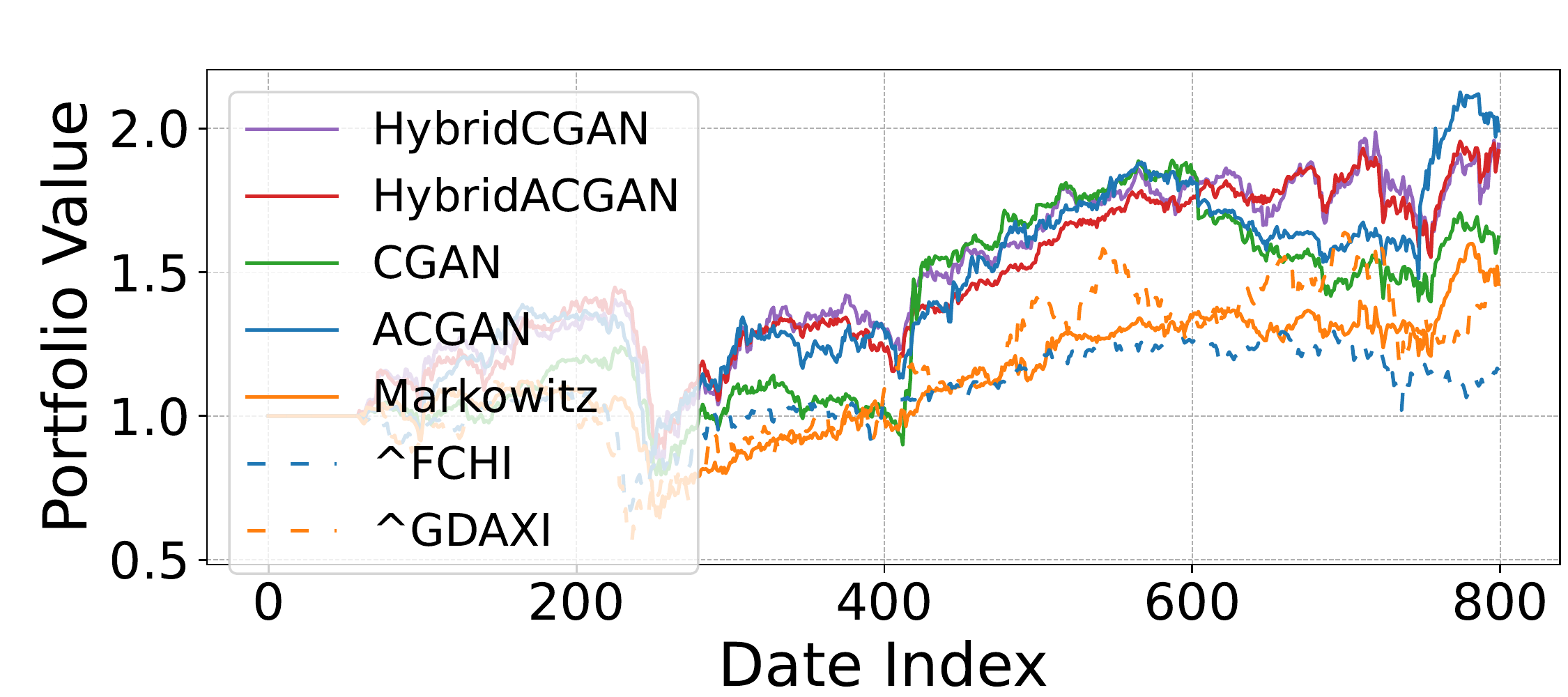} \label{fig:portfolio_value_eu2}}
	\subfigure[EU, rebalance every 20 days. 
	Sharpe ratios of HybridCGAN, HybridACGAN, CGAN, ACGAN, and Markowitz are 1.09, 1.06, 1.05, \textbf{1.27}, and 1.06 respectively.
	]{\includegraphics[width=0.325\textwidth]{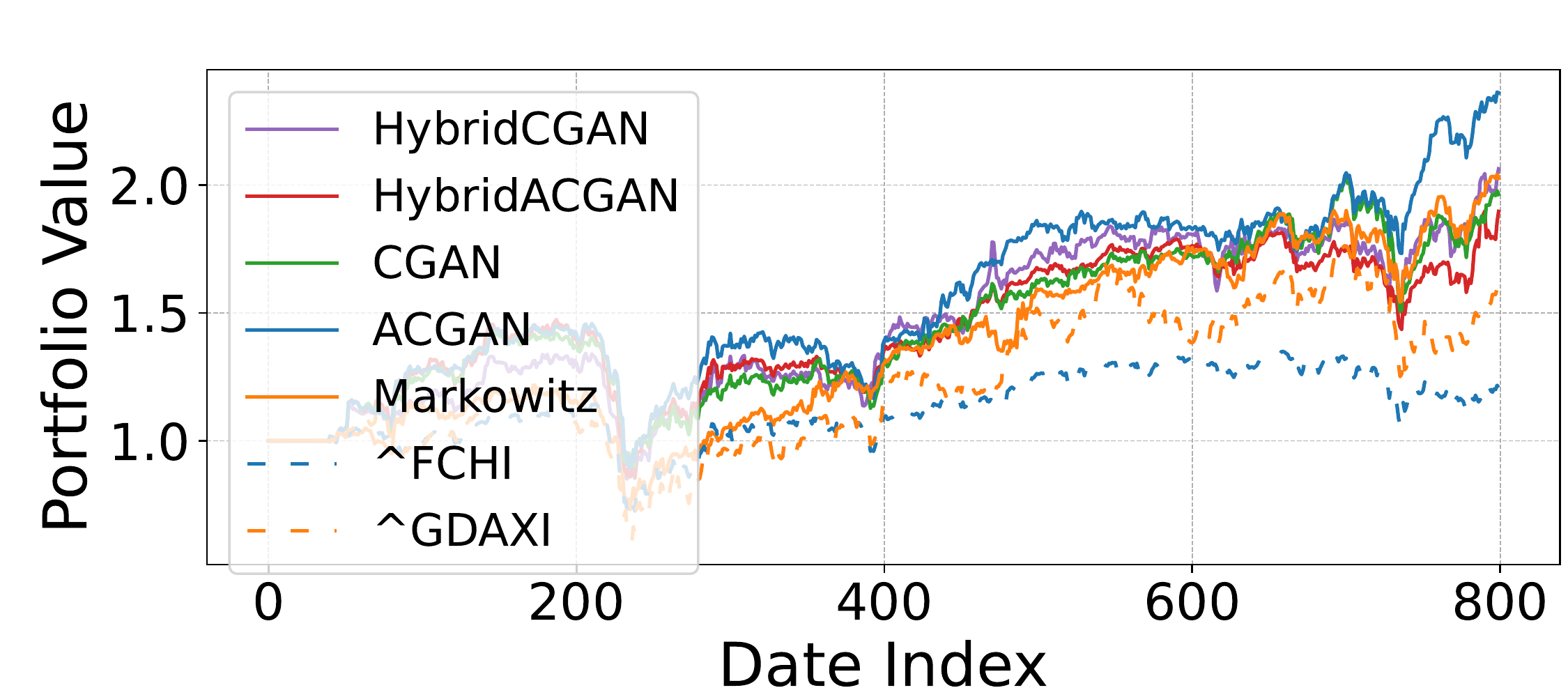} \label{fig:portfolio_value_eu3}}
	\caption{Portfolio values for different diversification risk settings. Reference benchmarks are shown with dashed lines (Index or ETF assets). HybridCGA, HybridACGAN, CGAN, ACGAN, and Markowitz with solid lines.}
	\label{fig:portfolio_value_acgan_compare}
\end{figure*}

\begin{figure}[!htb]
\centering  
\vspace{-0.35cm} 
\subfigtopskip=2pt 
\subfigbottomskip=2pt 
\subfigcapskip=-5pt 
\subfigure[US, $\eta=10$
]{\includegraphics[width=0.15\textwidth]{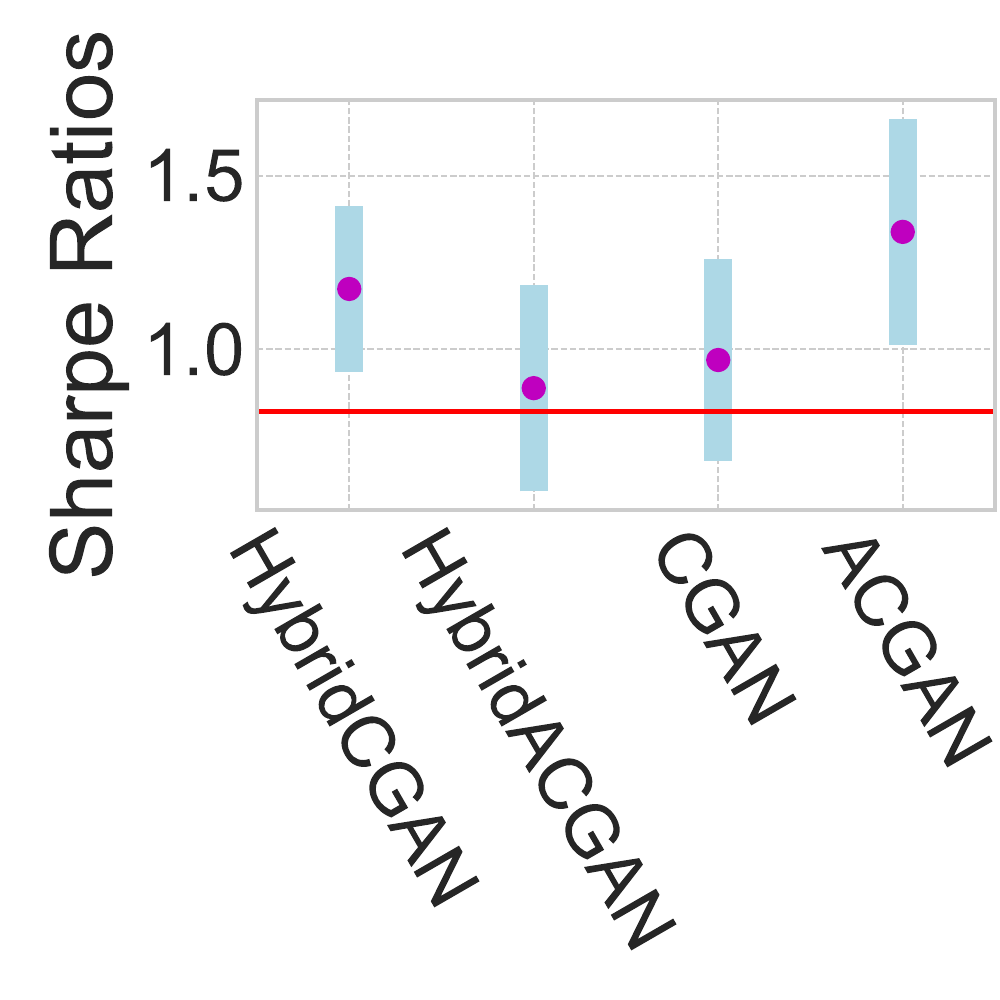} \label{fig:errorbarportfolio_value_us1}}
\subfigure[US, $\eta=15$]{\includegraphics[width=0.15\textwidth]{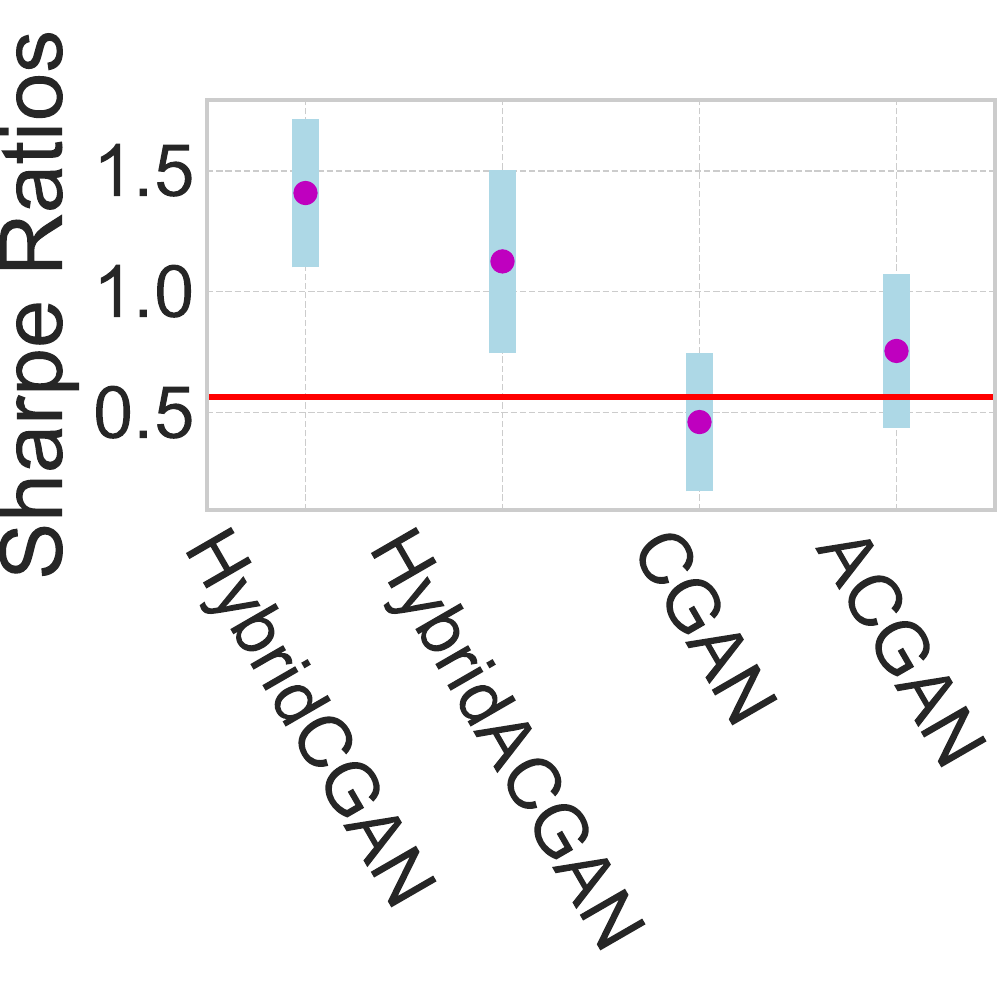} \label{fig:errorbarportfolio_value_us2}}
\subfigure[US, $\eta=20$ ]{\includegraphics[width=0.15\textwidth]{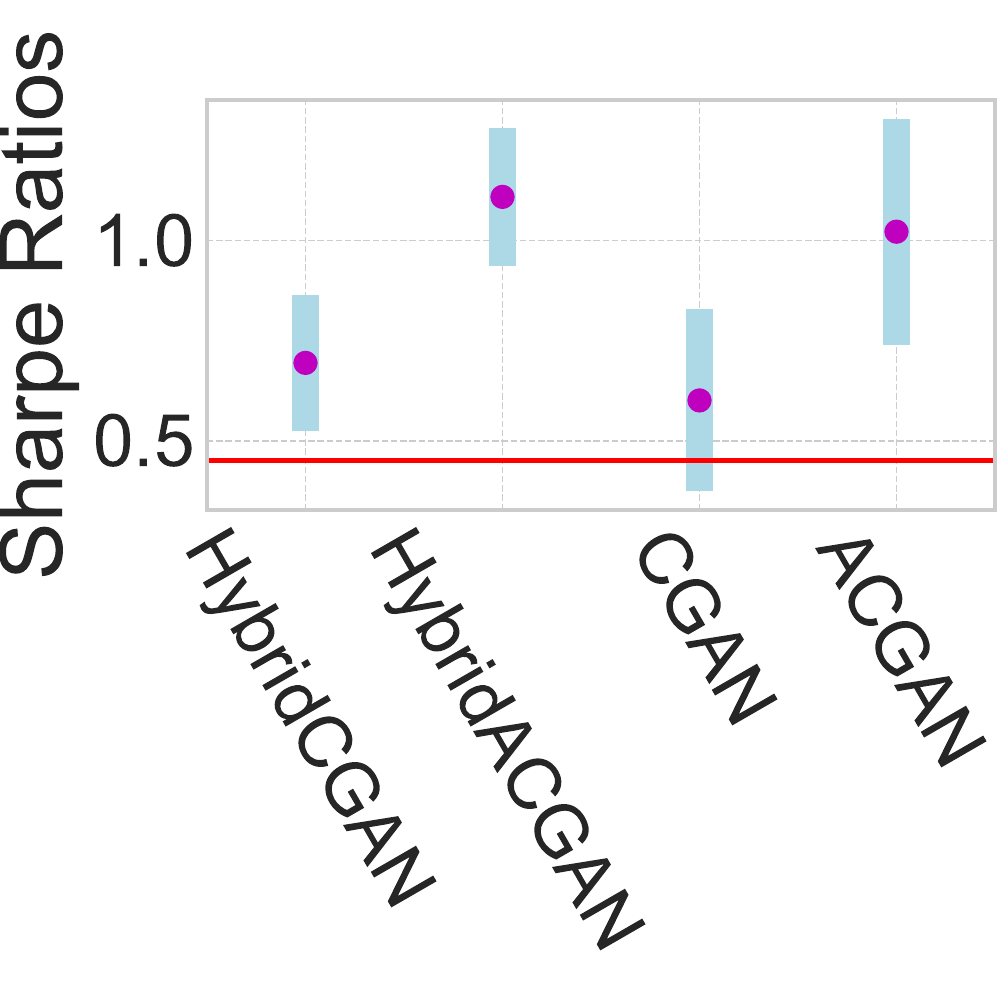} \label{fig:errorbarportfolio_value_us3}}
\subfigure[EU, $\eta=10$
]{\includegraphics[width=0.15\textwidth]{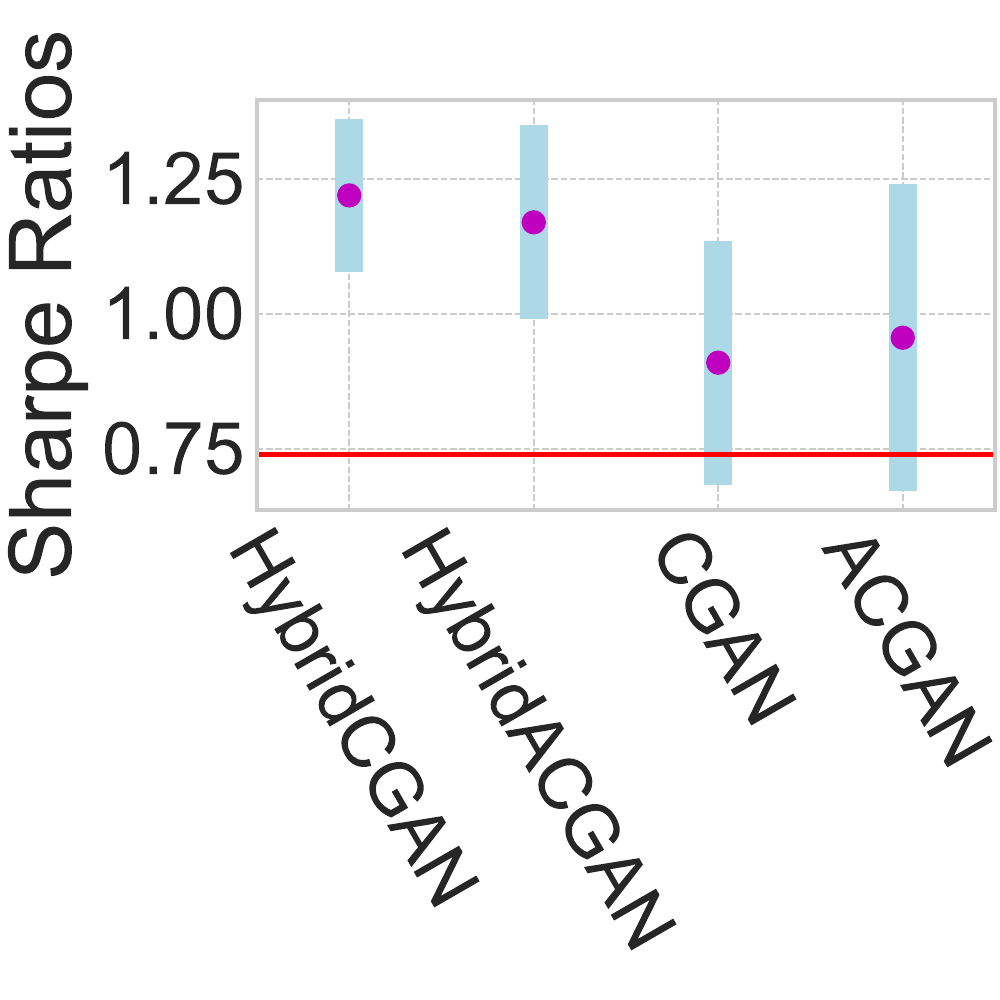} \label{fig:errorbarportfolio_value_eu1}}
\subfigure[EU, $\eta=15$]{\includegraphics[width=0.15\textwidth]{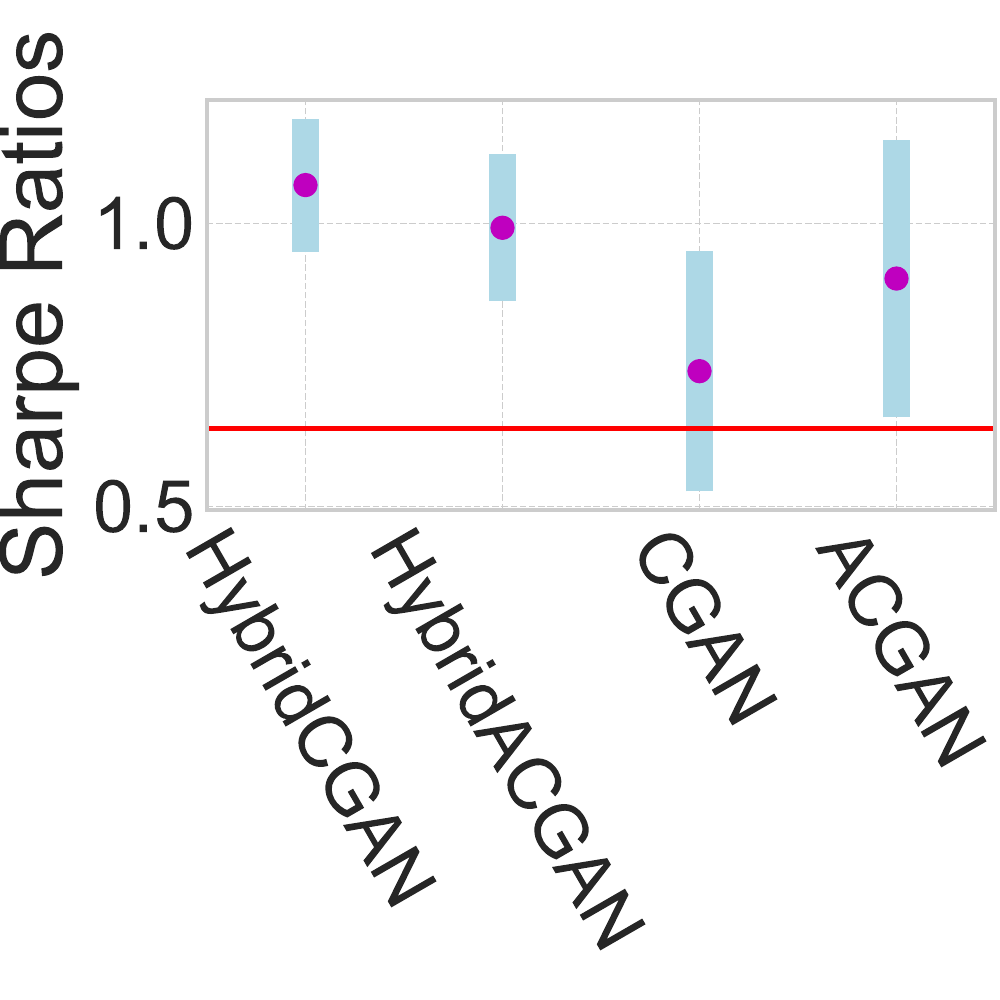} \label{fig:errorbarportfolio_value_eu2}}
\subfigure[EU, $\eta=20$]{\includegraphics[width=0.15\textwidth]{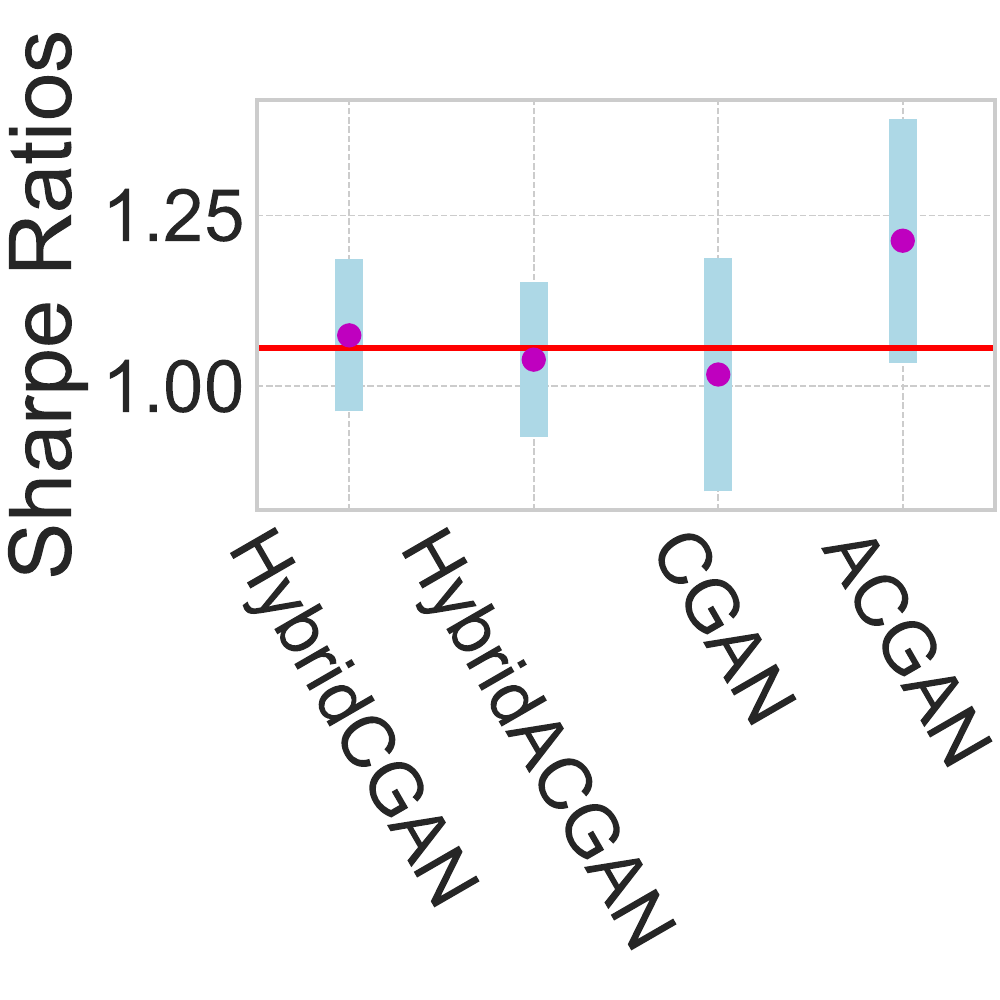} \label{fig:errorbarportfolio_value_eu3}}
\caption{The hybrid approaches surpass the non-hybrid alternatives reference approach on financial performance where the shaded bars are the standard deviation over means. Red horizontal lines are the relative Markowitz results.}
\label{fig:errorbar_hybrid_acgan_compare}
\end{figure}

Network structures for the conditioner (in HybridCGAN and CGAN), encoder, decoder (in HybridACGAN and ACGAN), generator, discriminator (in HybridCGAN, HybridACGAN, CGAN, and ACGAN), and proposer (in HybridCGAN and HybridACGAN) are provided in Appendix~\ref{appendix:hybrid_acgan_net_structures}. In all experiments, we train the network with 1,000 epochs. For simplicity, we set the risk-free interest $r_f=0$ to assess the Sharpe ratio evaluations.

\begin{figure}[!htb]
	\centering  
	%\vspace{-0.35cm} 
	\subfigtopskip=2pt 
	\subfigbottomskip=2pt 
	\subfigcapskip=10pt 
	\subfigure{\includegraphics[width=0.45\textwidth]{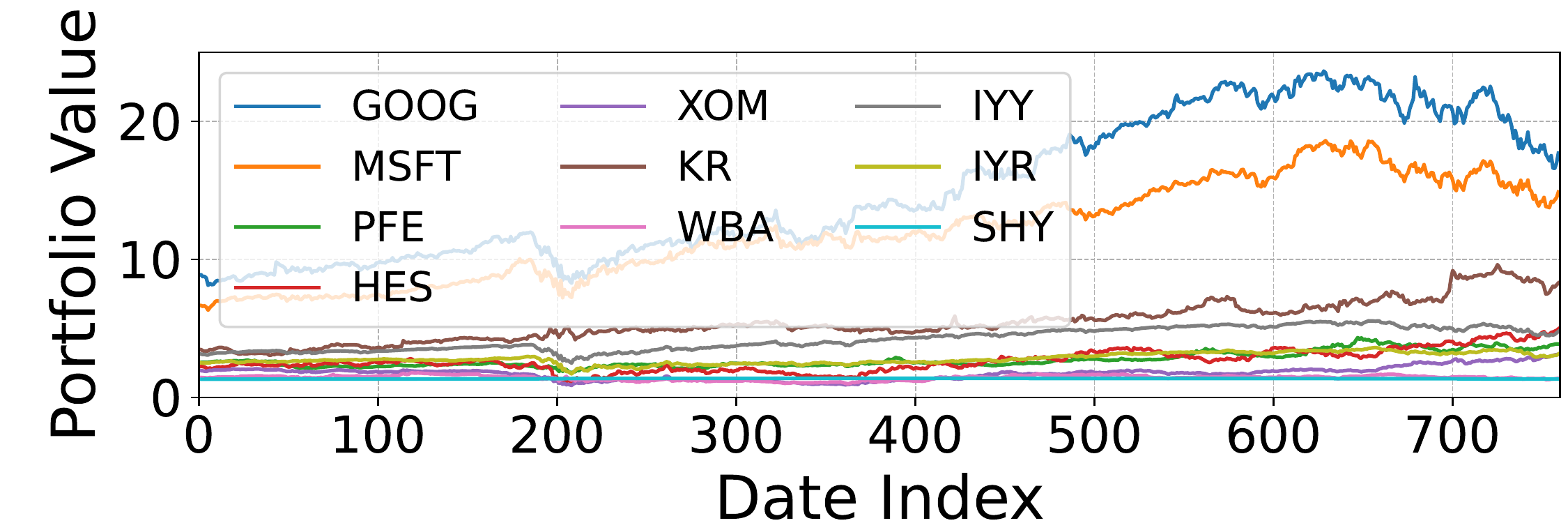} \label{fig:weight_diverse_us1}}\vspace{-0.4em}
	\subfigure{\includegraphics[width=0.45\textwidth]{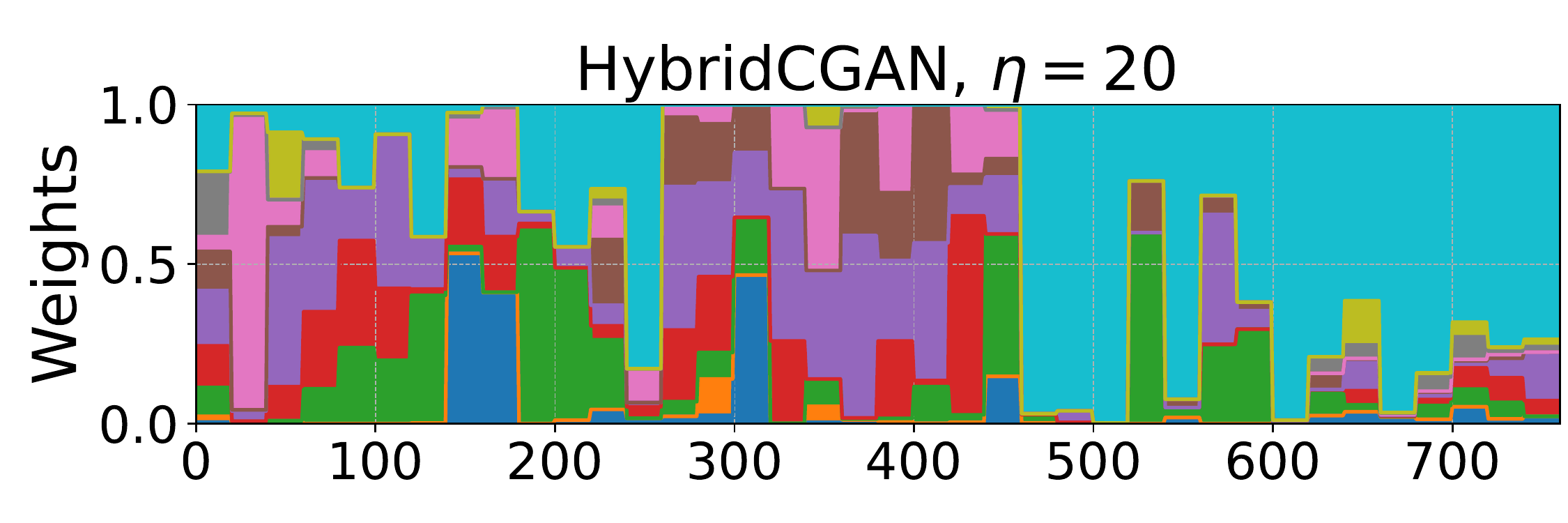} \label{fig:weight_diverse_us2}}\vspace{-1.15em}
	\subfigure{\includegraphics[width=0.45\textwidth]{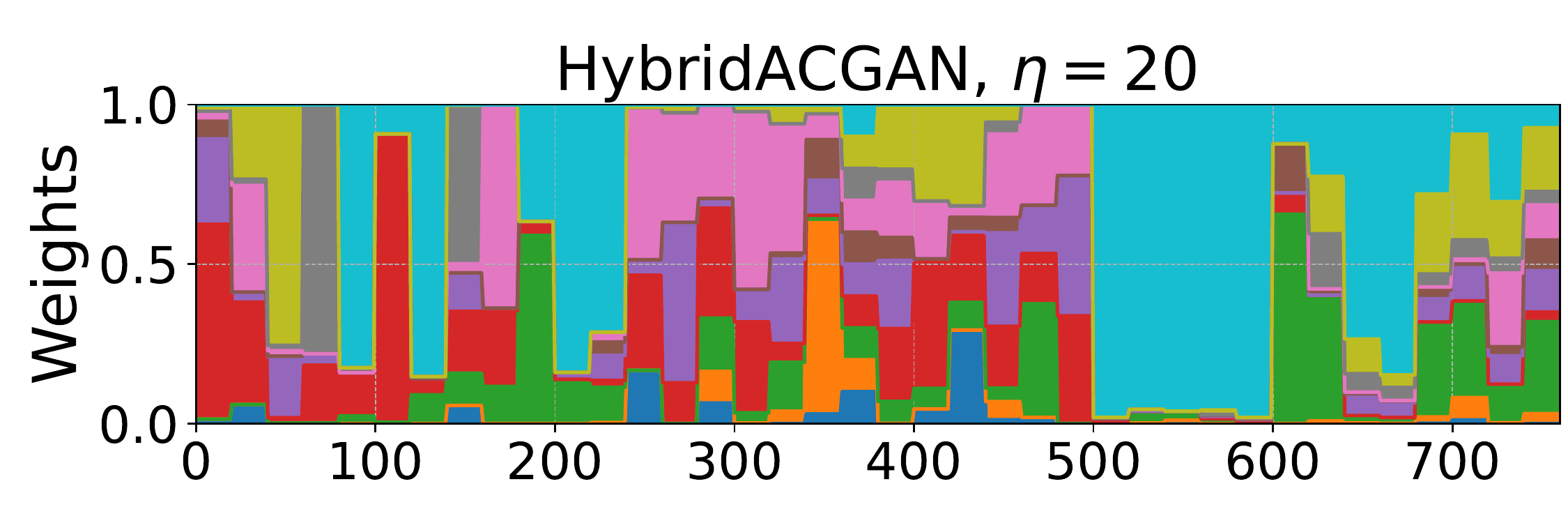} \label{fig:weight_diverse_us3}}\vspace{-1.15em}
	\subfigure{\includegraphics[width=0.45\textwidth]{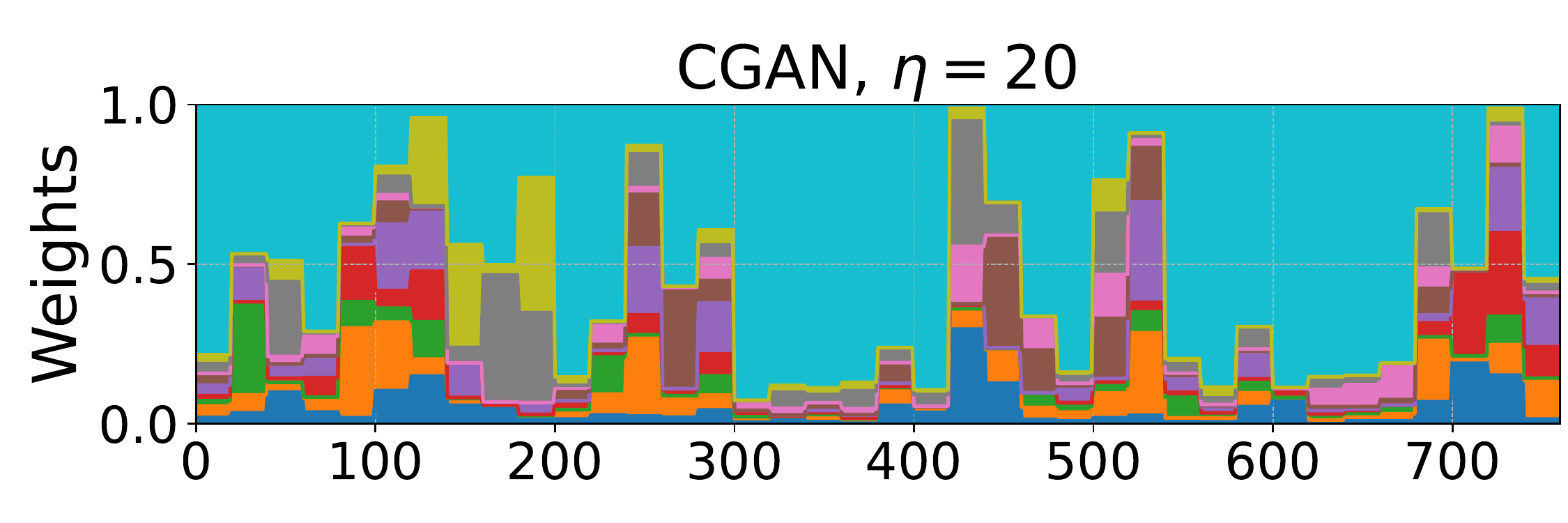} \label{fig:weight_diverse_us4}}\vspace{-1.14em}
	\subfigure{\includegraphics[width=0.45\textwidth]{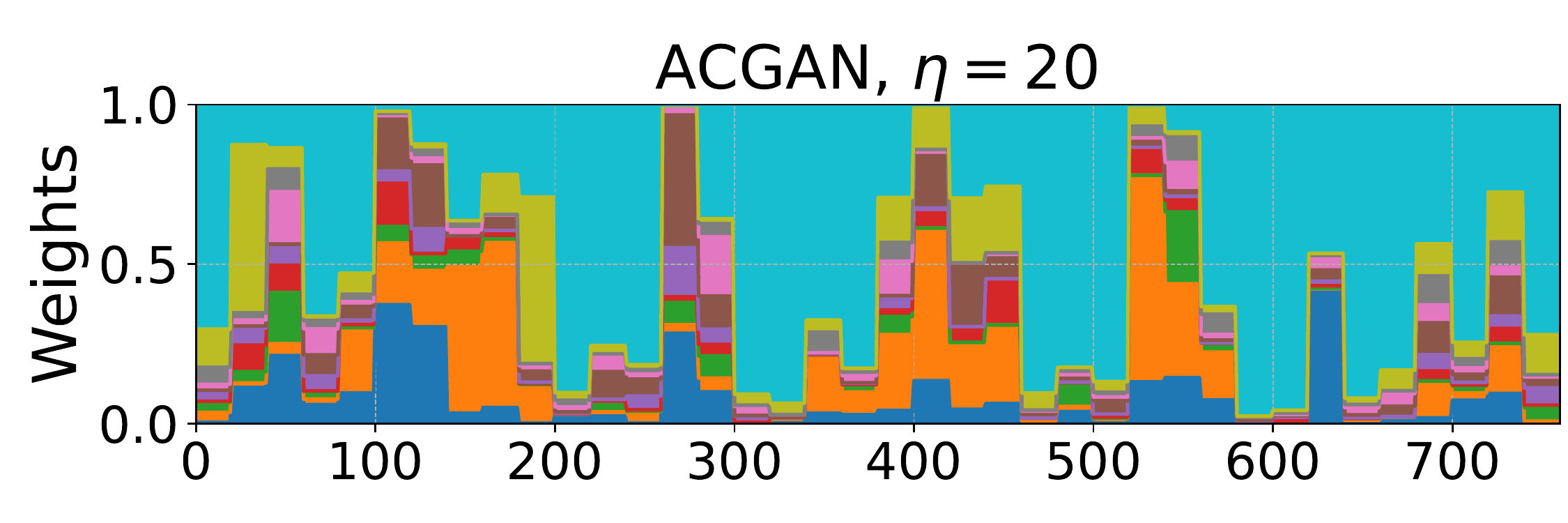} \label{fig:weight_diverse_us5}}\vspace{-1.14em}
	\subfigure{\includegraphics[width=0.45\textwidth]{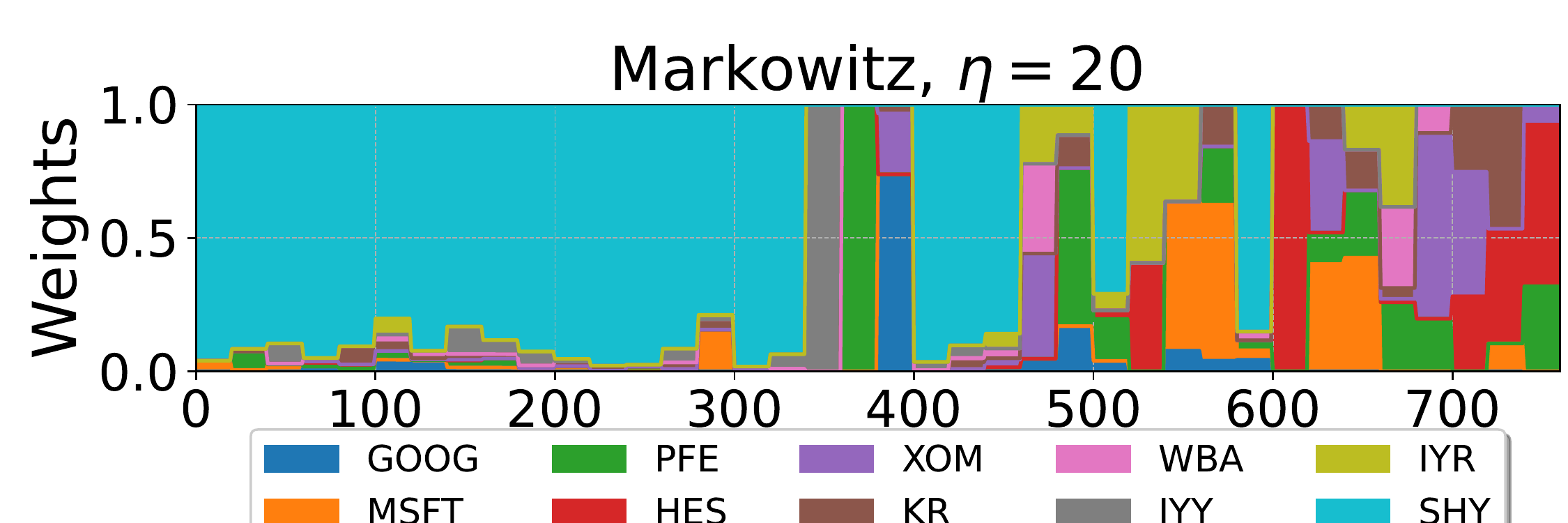} \label{fig:weight_diverse_us6}}
	\caption{Portfolio values of 10 assets, and weights distribution over time on the test period for HybridCGAN, HybridACGAN, CGAN, ACGAN, and Markowitz models for the US region with rebalancing every 20 days. Portfolios from the HybridCGAN and HybridACGAN are more diversified than those from CGAN, ACGAN, and Markowitz.}
	\label{fig:weight_diverse_bydays_us20}
\end{figure}

\subsection{Generating Analysis}
We follow the training and testing procedures in Algorithm~\ref{alg:acgan_training}. Given the training matrix $\bM$ of size $N\times D$ (where $N$ is the number of assets and $D$ is the number of days in the daily analysis context) and the window size $w$ ($w=h+f$ where $h$ is the length of the historical window and $f$ is the length of the future window), we define the index set $\mathcal{S}_1=\{1,2,\ldots, D-w+1\}$ so that $D-w$ samples can be extracted for each training epoch. While at the testing stage, given the testing matrix $\bA\in \real^{N\times K}$, the index set is obtained by $\mathcal{S}_2=\{\textcolor{black}{h+1},h+f+1,\ldots\}$ so that $(K-h)/f$ samples can be obtained (supposed here $(K-h)$ can be divided by $f$). 
The output $\bB\in \real^{N\times K}$ of Algorithm~\ref{alg:acgan_training} is the financial market simulation of the $N$ assets in $K$ days (here $N=10$ and $K=800$ in our datasets for both US and EU regions).
To be more concrete, the first $h$ days of $\bB$ are just copies of $\bA$, while the next $f$ days are the synthetic series based on the data of the first $h$ days; the next $f$ days are the synthetic series based on the data between the $f$-th and $(f+h)$-th days; and so on.

We set window size $h=40, f=20$ and $w=60$ in all experiments.
Figure~\ref{fig:diversity_us_eu_ecgan} shows the actual price trend (black solid line) of the US assets for the first 100 trading days in the test set, and five representative simulations generated by HybridCGAN and CGAN models (colored dashed lines). The proposed hybrid methods are not seeking simulations that are closer to the real series, but find the typical trends of the series, e.g., there is a big drawdown for GOOG, MSFT, PFE, HES, XOM, WBA, and IYY around 80-th day; and an increase for SHY around 80-th day.

\subsection{Portfolio Analysis}

After generating the synthetic series for each asset, we optimize over the fake series to generate minimal Sharpe ratio weight allocations (for HybridCGAN, HybridACGAN, CGAN, and ACGAN). For Markowitz framework, the optimization is done over the past data (here we use $h$ days).  
We consider three rebalance settings: a \textit{defensive setting} with rebalancing every $\eta=10$ days; a \textit{balanced setting} with $\eta=15$;
and an \textit{aggressive setting} with $\eta=20$.
Figure~\ref{fig:risk-sharperation_acgan_compare} presents the distribution of return-SR (Sharpe ratio) scatters with 1,000 draws from HybridCGAN, HybridACGAN, CGAN, and ACGAN models, and the one from Markowitz framework. The points in the upper-right corner are the better ones.

Figure~\ref{fig:errorbar_hybrid_acgan_compare} shows the distribution of Sharpe ratios over 1,000 draws where the shaded bars are the standard deviation over means.
We observe that all the HybridCGAN models surpass the alternative CGAN approaches;
 and in most cases, the HybridACGAN models perform better than the ACGAN model except when $\eta=10$ in the US region and $\eta=20$ in the EU region.
The hybrid approach has a smaller variance such that the final strategies are more stable.

The end strategies from HybridCGAN, HybridACGAN, CGAN, and ACGAN are the ones by taking average weight from these 1,000 draws on each rebalancing date (we call it \textit{mean strategy}).
%The \textit{ACGAN (Mean)} and \textit{CGAN (Mean)} are strategies by taking the average weight from these 1,000 draws on each rebalancing date.
Figure~\ref{fig:portfolio_value_acgan_compare} shows the portfolio value series of mean strategies for HybridCGAN, HybridACGAN, CGAN, ACGAN, and the one from Markowitz framework along the test period where we initialize each portfolio with a unitary value. The hybrid approaches dominate the other approaches in terms of the final portfolio values and Sharpe ratios.

%In all scenarios, the mean strategies of the proposed ACGAN model have better returns and Sharpe ratios than those of the Markowitz model and the CGAN model. 
When we apply the mean strategy in the US region with $\eta=15,20$, the hybrid versions of CGAN and ACGAN achieve both better return and Sharpe ratio evaluations compared to the mean strategies of non-hybrid approaches.
Similar results are observed in the EU region with $\eta=10,15$. 
The Sharpe ratio of ACGAN model for US region with $\eta=10$ obtains best performance among other results (SR=1.72); however, its final portfolio value is not the best where the HybridCGAN obtains the best final portfolio value with a decent Sharpe ratio of 1.32.

%The return results of the ACGAN and CGAN models are close in the defensive setting for the EU region; while ACGAN has better Sharpe ratio such that it takes smaller risk.

\subsection{Weight Distribution}

In Figure~\ref{fig:weight_diverse_bydays_us20}, we present the distributions of weights over time on the test period for the US region with rebalancing every 20 days and the corresponding portfolio values of the 10 assets. 
Since we use the first 40 trading days as the historical series, weights of only 760 days are shown in the figure. 
We observe that the Markowitz model puts a large weight on the SHY asset for the first 300 trading days; while in this period, companies in IT sector (GOOG and MSFT) receive large positive returns making the Markowitz result less competitive.

Moreover, we find that there is a big drawdown for the IT companies since the 600-th day and 700-th day. The HybridCGAN and HybridACGAN perform well compared to their non-hybrid versions in that they put less weight on these companies during the drawdown periods. Further, one can easily observe that the portfolios from the HybridCGAN and HybridACGAN are more diversified than those from CGAN, ACGAN, and Markowitz. 
Hybrid methods are able to
systematically improve the returns achievable.

For other rebalancing days and weights distribution for the EU region, results are provided in Figure~\ref{fig:weight_diverse_bydays_us_10}, \ref{fig:weight_diverse_bydays_us_15}, \ref{fig:weight_diverse_bydays_eu_10}, \ref{fig:weight_diverse_bydays_eu_15}, and \ref{fig:weight_diverse_bydays_eu_20}; and we shall not repeat the details.

\section{Conclusion}
The paper aims to solve the issue of poor prediction ability in the CGAN and ACGAN methodology for portfolio analysis. We propose a simple and computationally efficient algorithm that incorporates a deep neural regression model and requires little extra computation. A potential future work on the HybridCGAN and HybridACGAN models is to further reduce the variance of different draws such that the end strategy will be more stable and consistent.

%\paragraph{Acknowledgments}
%We greatly appreciate insightful discussions with Giovanni Mariani on the data normalization and the framework of the PAGAN (CGAN) methodology. 

%\pagebreak
%\small

\bibliography{bib}
\bibliographystyle{myplainnat}
%\pagebreak
\balance

\onecolumn
\appendix

\section{Network Structures}\label{appendix:hybrid_acgan_net_structures}
We provide detailed structures for the neural network architectures we used in our experiments in this section.
Given the number of assets $N$, historical length $h$, future length $f$ ($w=h+f$), and latent dimension $m$ for the prior distribution vector $\bz$, we consider multi-layer perceptron (MLP) structures, 
the detailed architecture 
%for each convolutional layer is described by C$(\langle\textit{kernel size}\rangle:\langle\textit{num outputs}\rangle:\langle\textit{activation function}\rangle)$; 
for each fully connected
layer is described by F$(\langle \textit{num inputs} \rangle :\langle \textit{num outputs} \rangle:\langle \textit{activation function} \rangle)$; 
for an activation function of LeakyRelu with parameter $p$ is described by $\text{LR}(\langle p \rangle )$; 
%for a max pooling layer is described by MP$(\langle \textit{kernel size} \rangle:\langle \textit{stride number} \rangle)$; 
and for a dropout layer is described by
DP$(\langle \textit{rate} \rangle)$. The \textit{conditioner} in CGAN shares the same structure as the \textit{encoder} in the ACGAN model (see Figure~\ref{fig:cgan_structure} and Figure~\ref{fig:acgan_structure}).
Then the network structures we use can be described as follows:
\begin{align}
&\textbf{Conditioner}=\textbf{Encoder}=
\text{F\big($N\cdot h$: 512: LR(0.2)\big)} \cdot 
\text{F\big(512:512:LR(0.2)\big)} \cdot 
\text{DP(0.4)}\cdot
\text{F(512:16)}\\
&\textbf{Decoder}=
\text{F\big(16:512:LR(0.2)\big)} \cdot 
\text{F\big(512:512:LR(0.2)\big)} \cdot 
\text{DP(0.4)}\cdot
\text{F(512:$N\cdot h$)}\\
&\textbf{Simulator (in CGAN or ACGAN)}=
\text{F\big($m$+16:128:LR(0.2)\big)} \cdot 
\text{F\big(128:256:LR(0.2)\big)} \cdot \nonumber\\
&\gap\gap\gap\gap\gap\gap\gap\gap \text{  F\big(256:512:LR(0.2)\big)} \cdot
\text{F\big(512:1024:LR(0.2)\big)} \cdot
\text{F\big(1024:$N\cdot f$:TanH\big)} \\
&\textbf{Simulator (in HybridCGAN or HybridACGAN)}=
\text{F\big($m$+16:128:LR(0.2)\big)} \cdot 
\text{F\big(128:256:LR(0.2)\big)} \cdot \nonumber\\
&\gap\gap\gap\gap\gap\gap\gap\gap \text{  F\big(256:512:LR(0.2)\big)} \cdot
\text{F\big(512:1024:LR(0.2)\big)} \cdot
\text{F\big(1024:$N\cdot f$:TanH\big)} \cdot \textcolor{blue}{100} \\
&\textbf{Discriminator}=
\text{F\big($N\cdot (h$+$f)$:512:LR(0.2)\big)} \cdot 
\text{F\big(512:512:LR(0.2)\big)} \cdot \text{  DP(0.4)} \cdot 
\text{F\big(256:512:LR(0.2)\big)} \cdot
\text{F(512:1)}\\
&\textbf{Proposer}=
\text{F\big($N\cdot (h+\textcolor{blue}{1})$: 512: LR(0.2)\big)} \cdot 
\text{F\big(512:512:LR(0.2)\big)} \cdot 
\text{DP(0.4)}\cdot
\text{F(512:16)},
\end{align}
where the highlighted 1 in proposer network is the input of the historical mean values (Eq.~\eqref{equation:hybrid-meaninput}). Due to effect of the proposal network, the outputs of the generator (or simulator) are not constrained into the range of [-1,1], we also multiply the result by 100 in the simulator of HybridCGAN or HybridACGAN.
We trained networks using Adam's optimizer with learning rate $2\times 10^{-5}$, $\beta_1 = 0.5$, and $\beta_2=0.999$. We set the penalization parameters $\lambda_1=10$ and $\lambda_2=3$. The latent dimension is $m=100$.
And we trained models for 1,000 epochs.

\section{Weight Distributions under Different Settings}
\begin{figure*}[!htb]
\centering  
%\vspace{-0.35cm} 
\subfigtopskip=2pt 
\subfigbottomskip=2pt 
\subfigcapskip=10pt 
\subfigure{\includegraphics[width=0.45\textwidth]{imgs/acgan_dataset_us_test.pdf} \label{fig:weight_diverse_us110}}\vspace{-0.2em}
\subfigure{\includegraphics[width=0.45\textwidth]{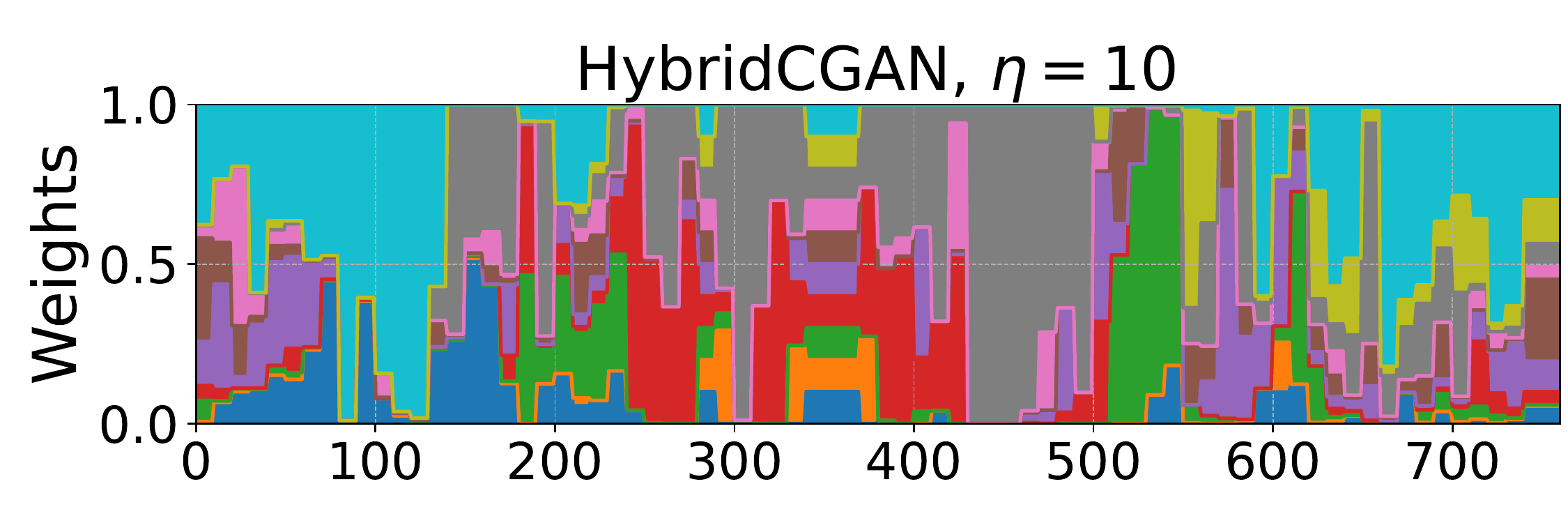} \label{fig:weight_diverse_us210}}\vspace{-0.2em}
\subfigure{\includegraphics[width=0.45\textwidth]{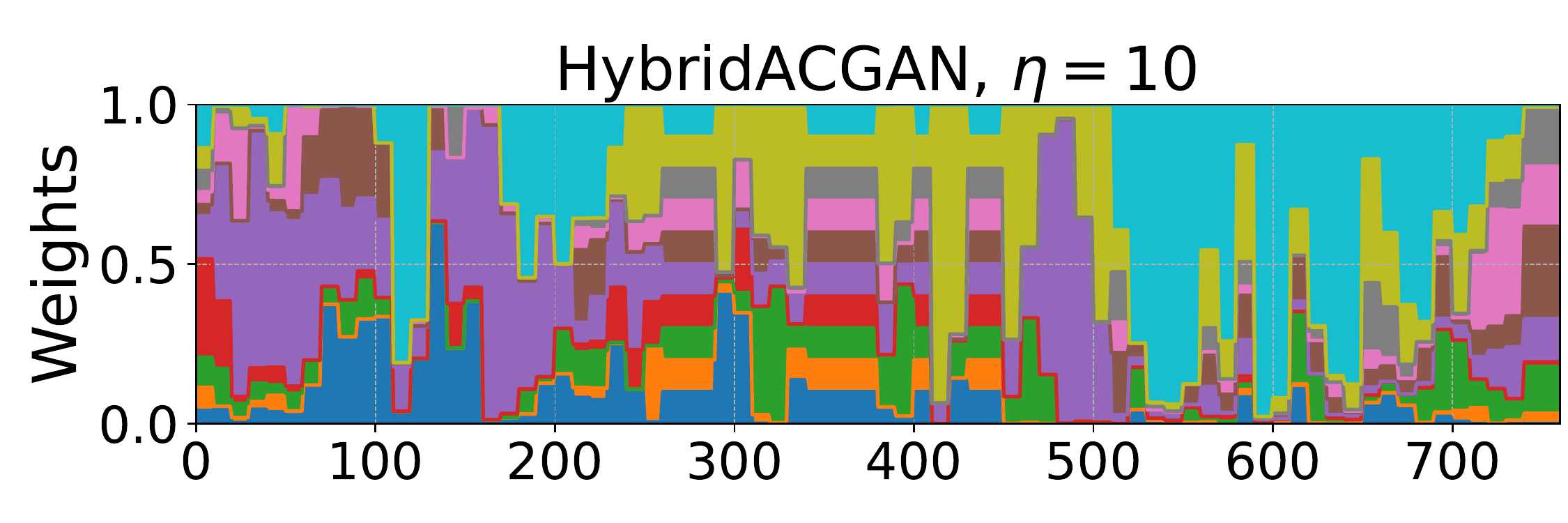} \label{fig:weight_diverse_us310}}\vspace{-0.3em}
\subfigure{\includegraphics[width=0.45\textwidth]{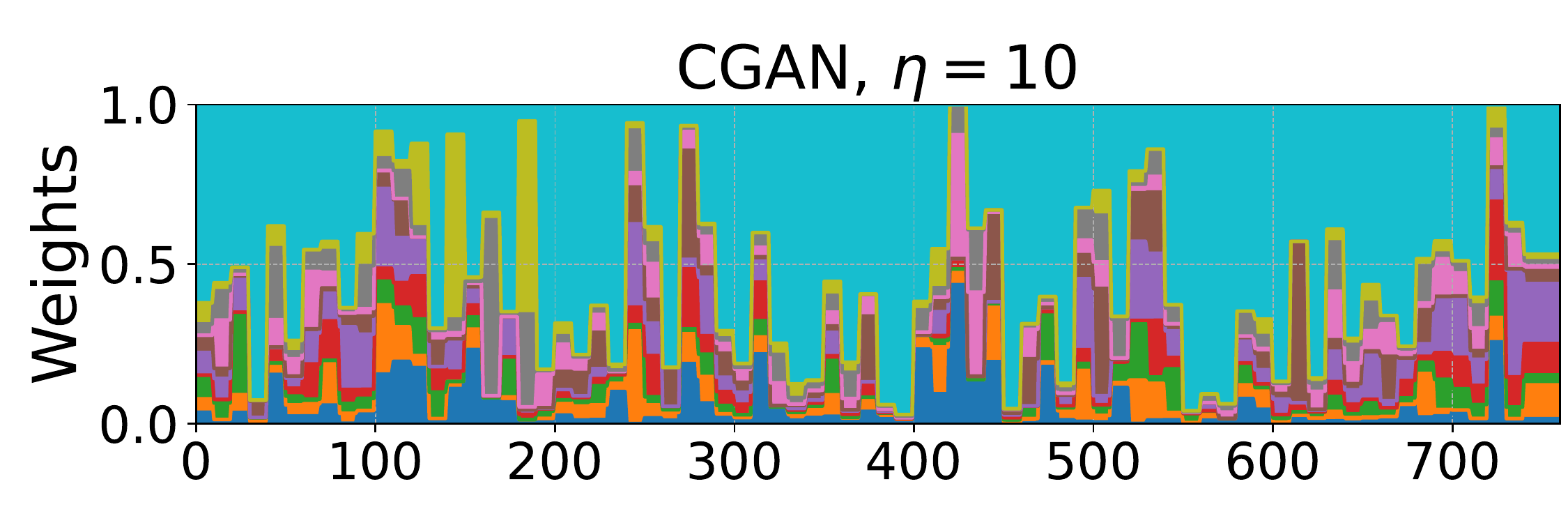} \label{fig:weight_diverse_us410}}\vspace{-0.3em}
\subfigure{\includegraphics[width=0.45\textwidth]{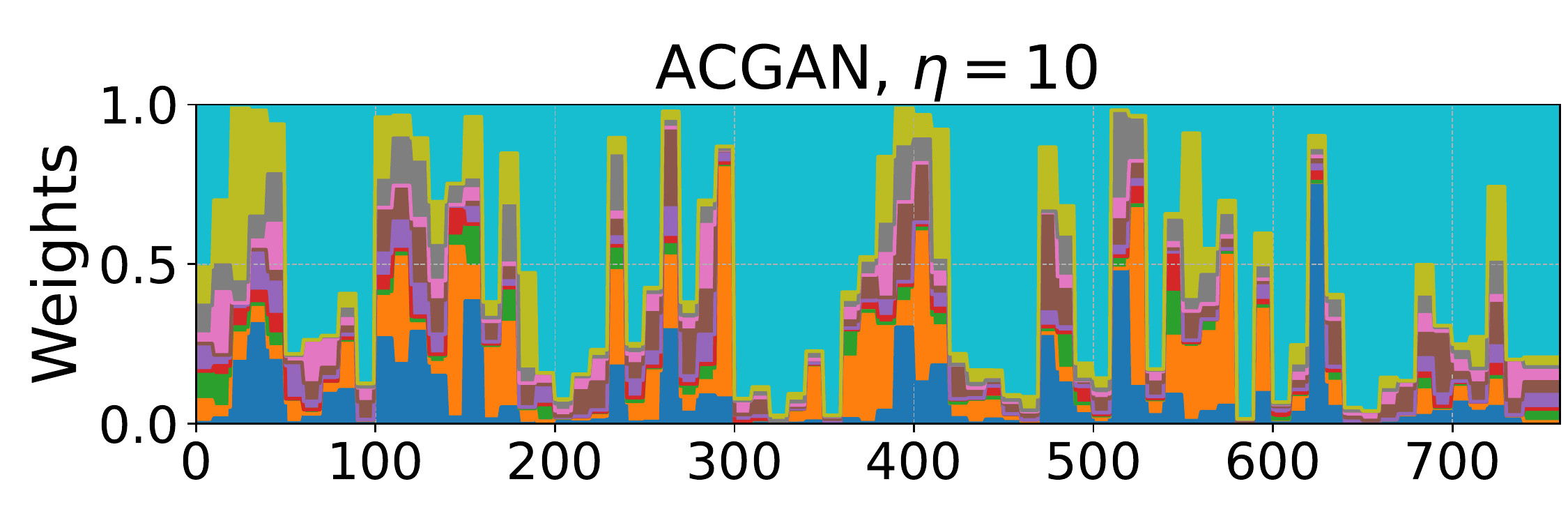} \label{fig:weight_diverse_us510}}\vspace{-0.3em}
\subfigure{\includegraphics[width=0.45\textwidth]{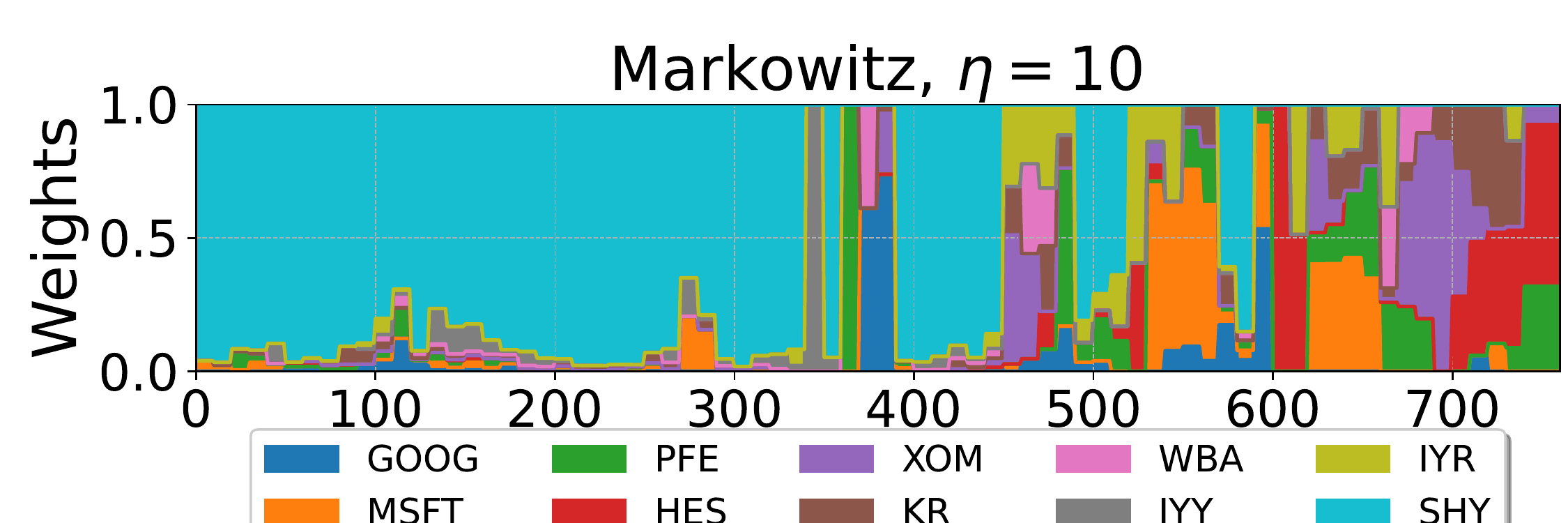} \label{fig:weight_diverse_us610}}
\caption{Weights distribution over time on the test period for HybridCGAN, HybridACGAN, CGAN, ACGAN, and Markowitz models for the US region with rebalancing every 10 days.}
\label{fig:weight_diverse_bydays_us_10}
\end{figure*}

\begin{figure*}[!htb]
\centering  
%\vspace{-0.35cm} 
\subfigtopskip=2pt 
\subfigbottomskip=2pt 
\subfigcapskip=10pt 
\subfigure{\includegraphics[width=0.45\textwidth]{imgs/acgan_dataset_us_test.pdf} \label{fig:weight_diverse_us113}}\vspace{-0.2em}
\subfigure{\includegraphics[width=0.45\textwidth]{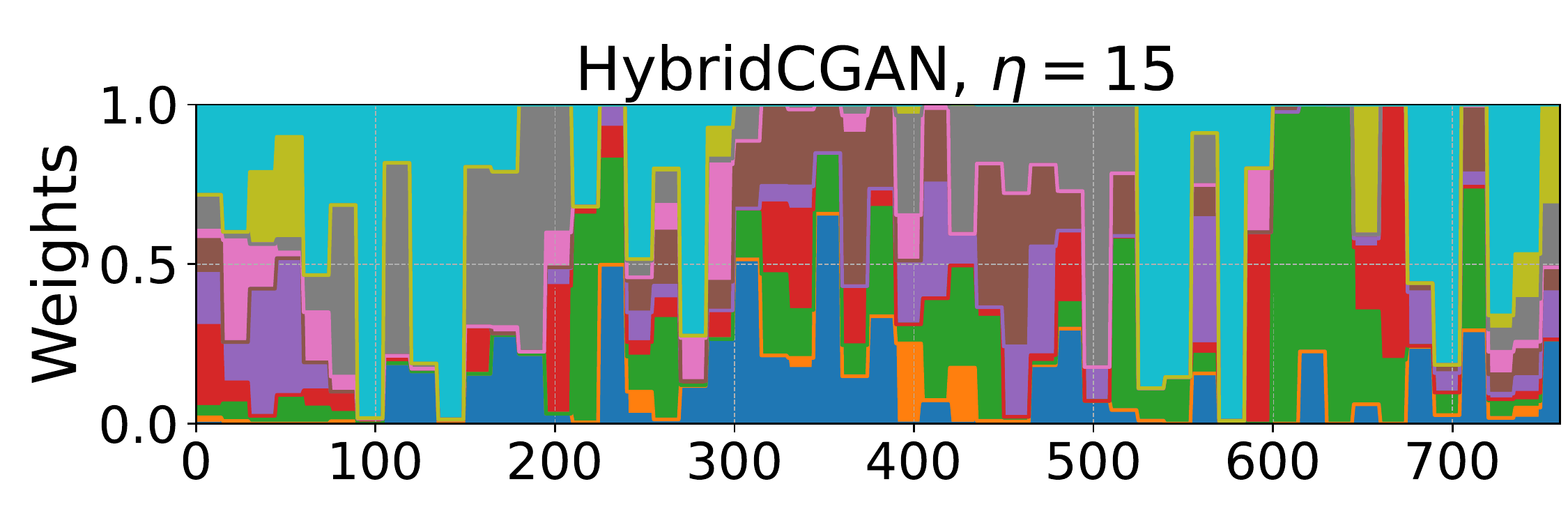} \label{fig:weight_diverse_us213}}\vspace{-0.2em}
\subfigure{\includegraphics[width=0.45\textwidth]{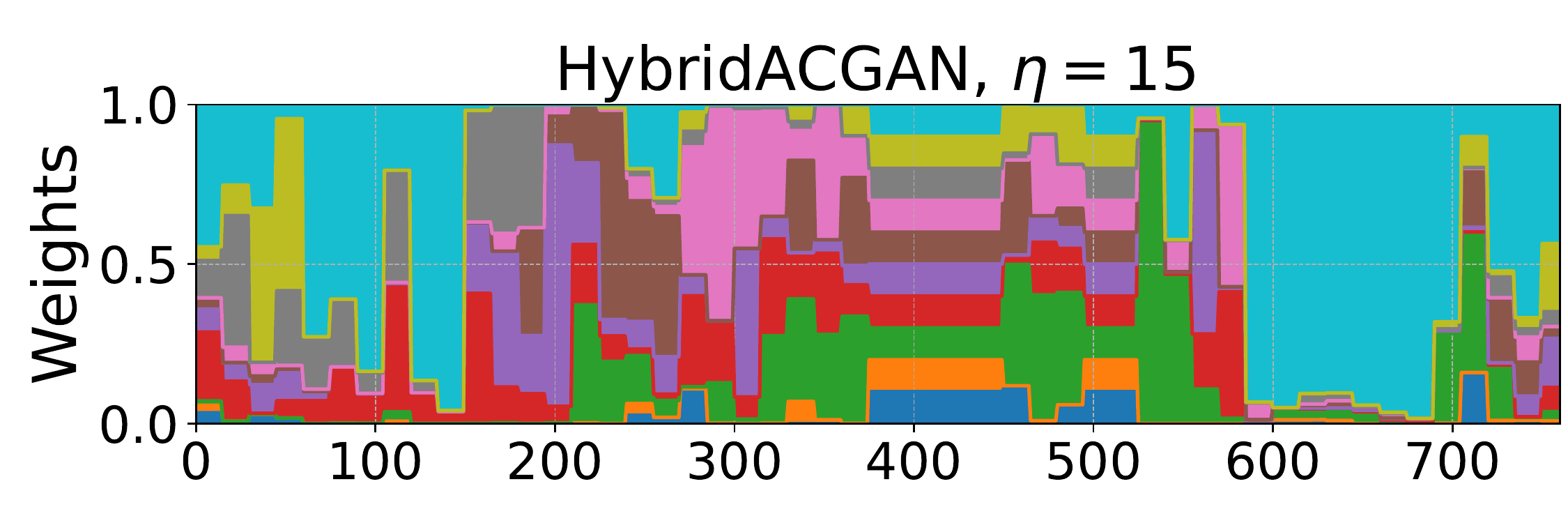} \label{fig:weight_diverse_us313}}\vspace{-0.3em}
\subfigure{\includegraphics[width=0.45\textwidth]{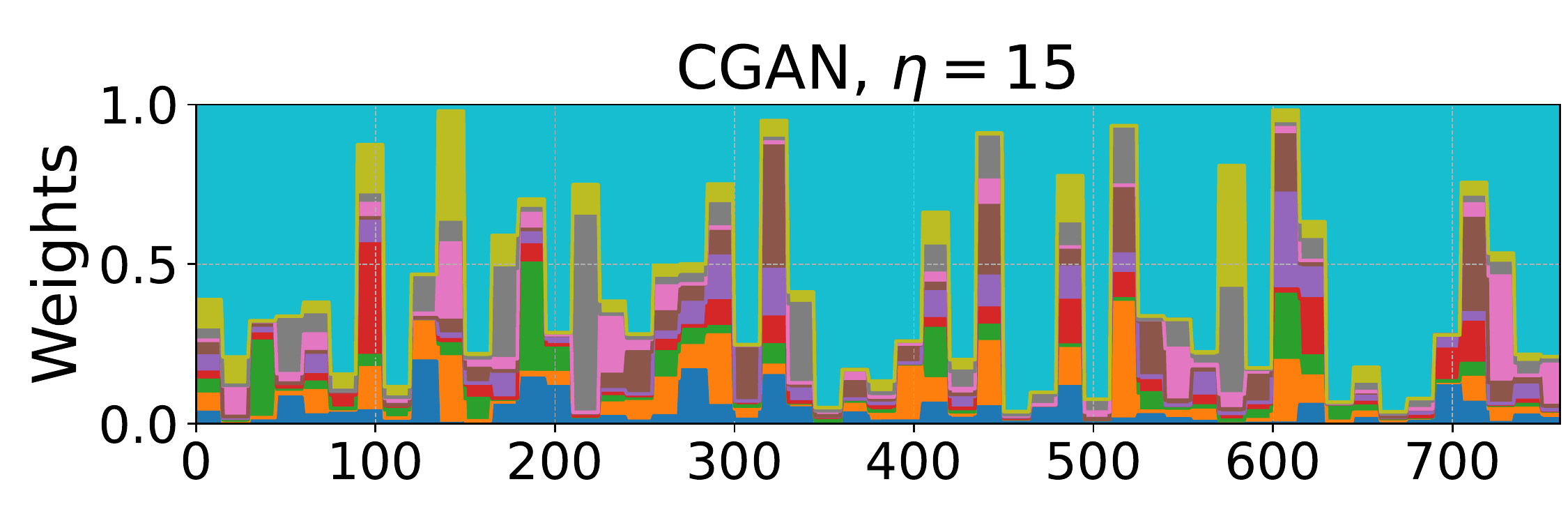} \label{fig:weight_diverse_us413}}\vspace{-0.3em}
\subfigure{\includegraphics[width=0.45\textwidth]{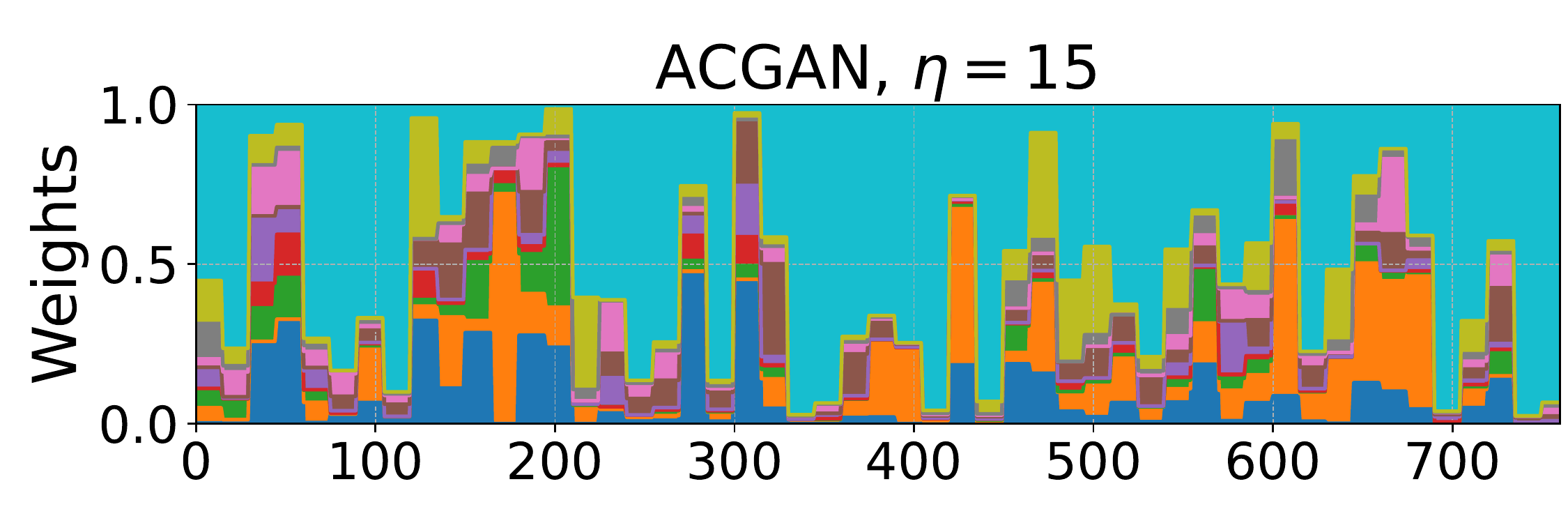} \label{fig:weight_diverse_us513}}\vspace{-0.3em}
\subfigure{\includegraphics[width=0.45\textwidth]{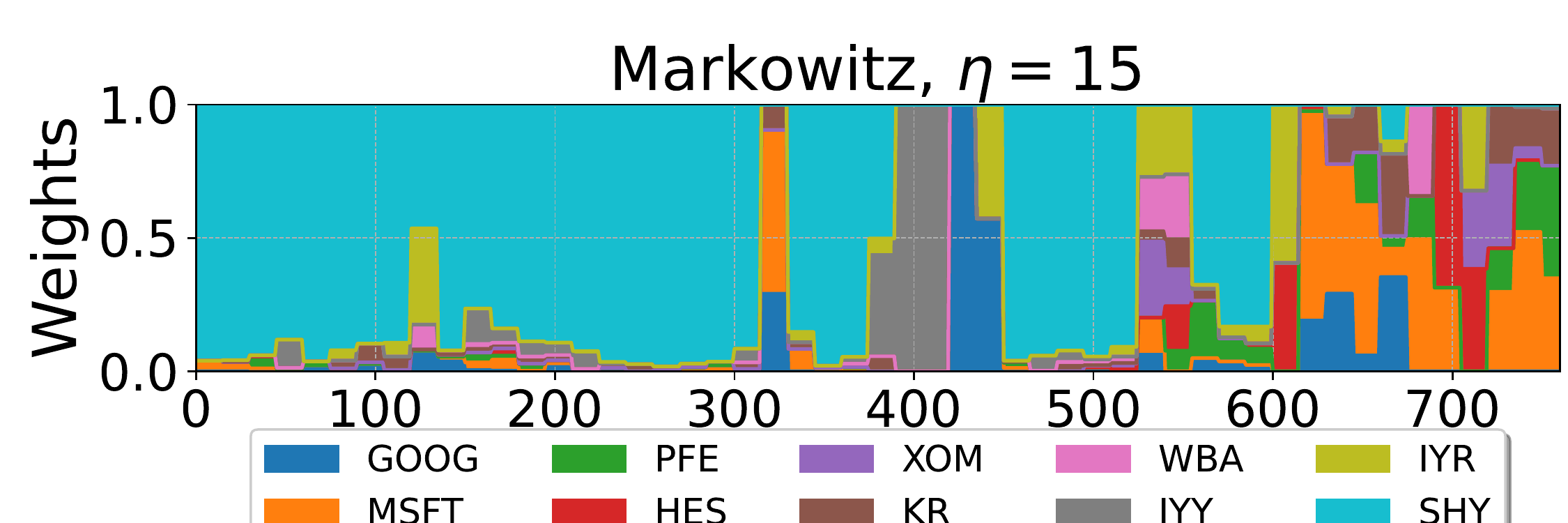} \label{fig:weight_diverse_us613}}
\caption{Weights distribution over time on the test period for HybridCGAN, HybridACGAN, CGAN, ACGAN, and Markowitz models for the US region with rebalancing every 15 days.}
\label{fig:weight_diverse_bydays_us_15}
\end{figure*}

\begin{figure*}[!htb]
	\centering  
	%\vspace{-0.35cm} 
	\subfigtopskip=2pt 
	\subfigbottomskip=2pt 
	\subfigcapskip=10pt 
	\subfigure{\includegraphics[width=0.45\textwidth]{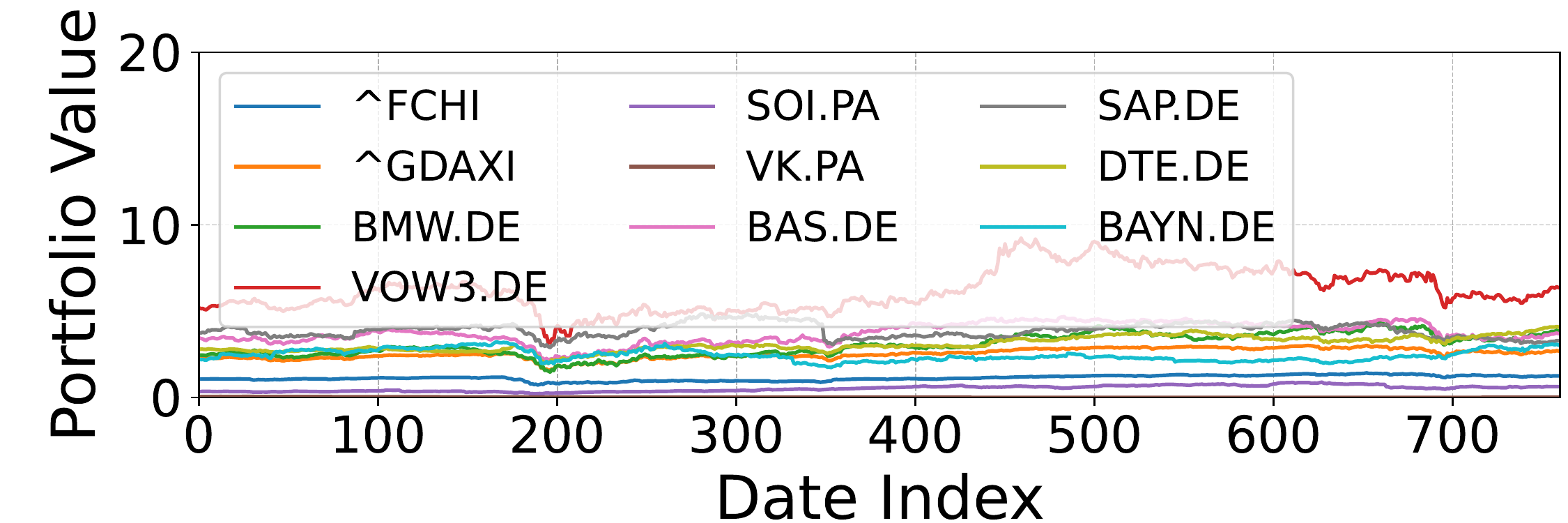} \label{fig:weight_diverse_eu110}}\vspace{-0.2em}
	\subfigure{\includegraphics[width=0.45\textwidth]{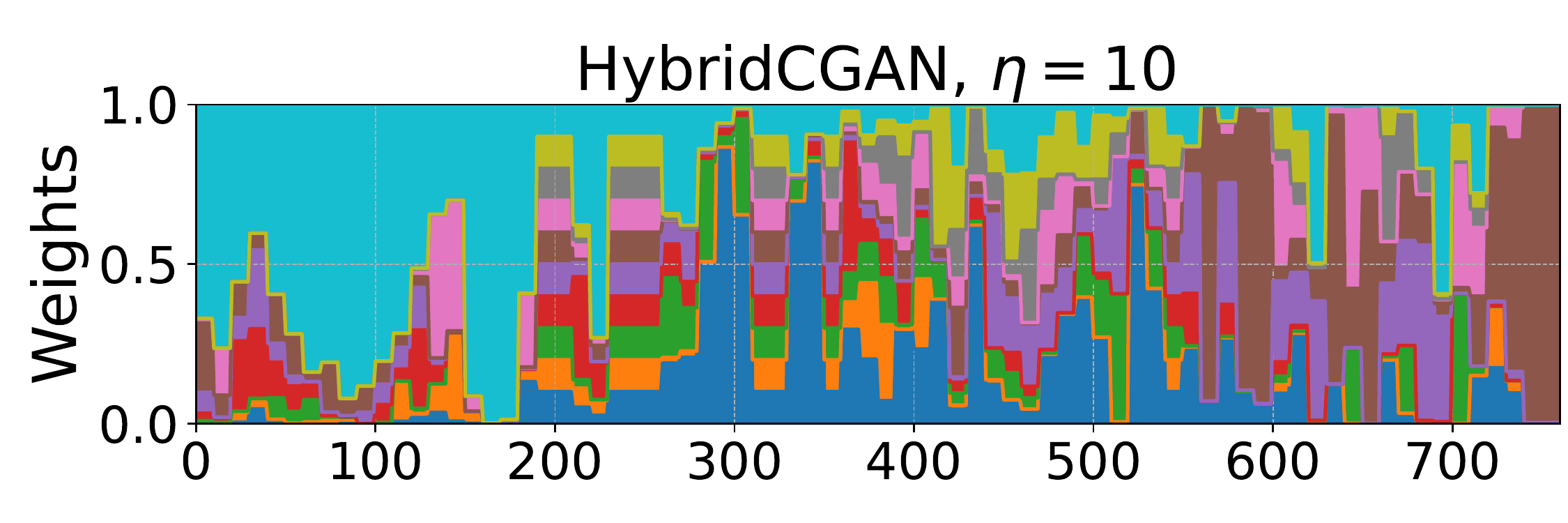} \label{fig:weight_diverse_eu210}}\vspace{-0.2em}
	\subfigure{\includegraphics[width=0.45\textwidth]{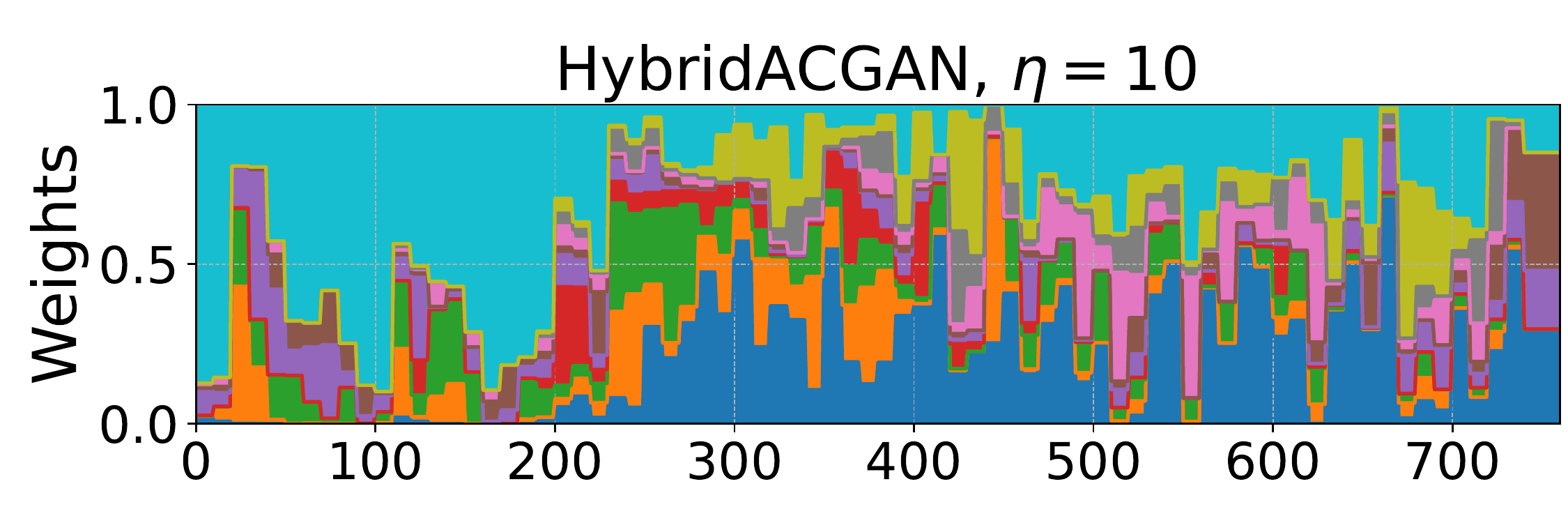} \label{fig:weight_diverse_eu310}}\vspace{-0.3em}
	\subfigure{\includegraphics[width=0.45\textwidth]{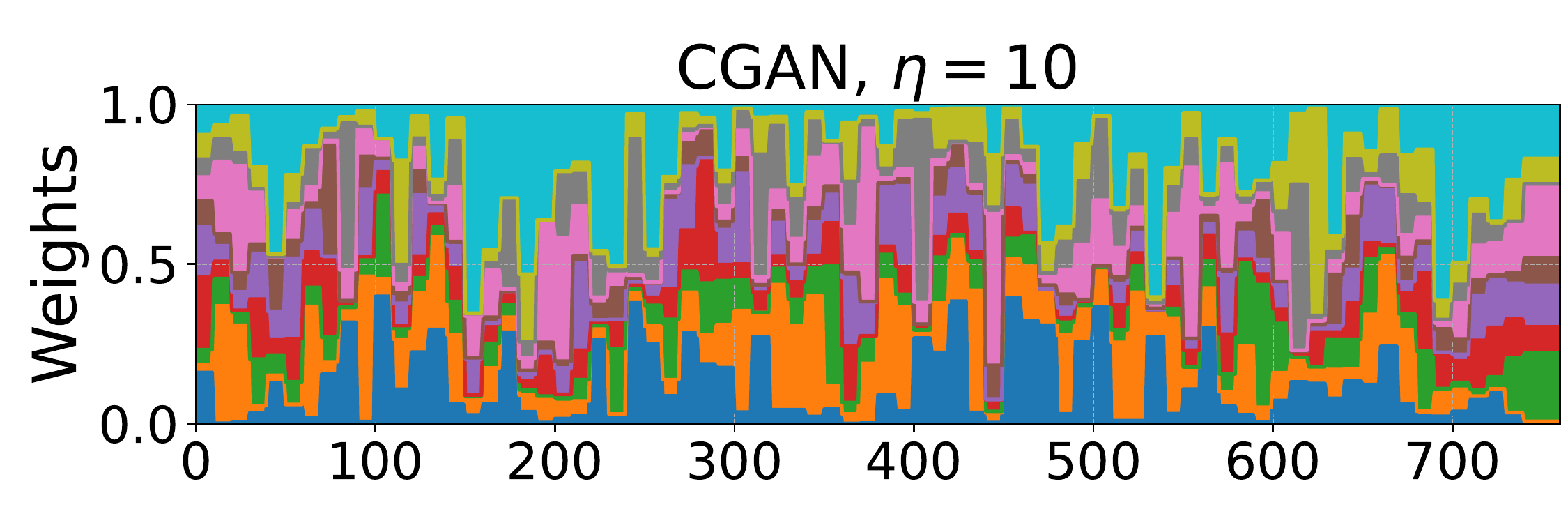} \label{fig:weight_diverse_eu410}}\vspace{-0.3em}
	\subfigure{\includegraphics[width=0.45\textwidth]{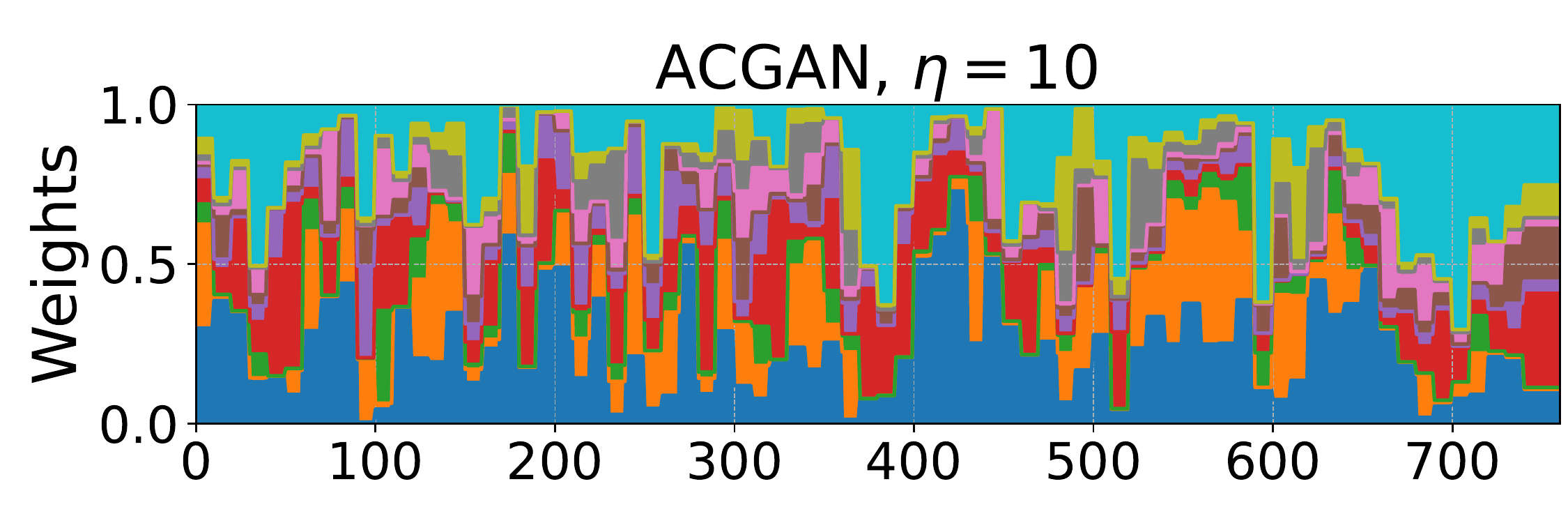} \label{fig:weight_diverse_eu510}}\vspace{-0.3em}
	\subfigure{\includegraphics[width=0.45\textwidth]{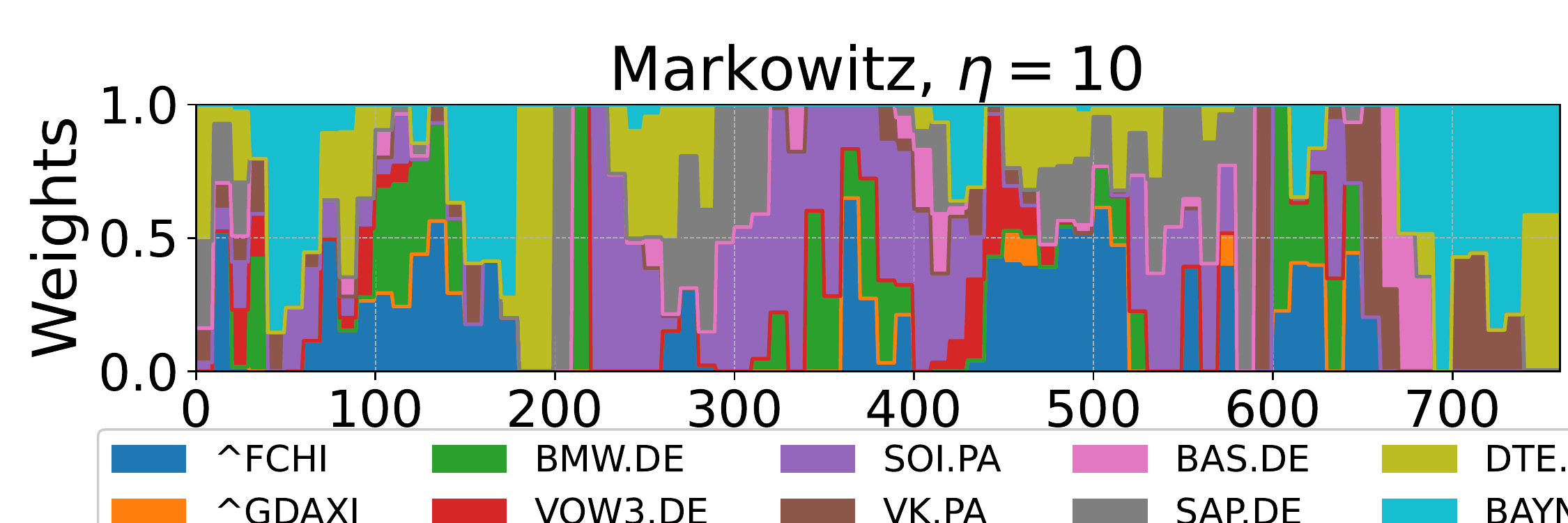} \label{fig:weight_diverse_eu610}}
	\caption{Weights distribution over time on the test period for HybridCGAN, HybridACGAN, CGAN, ACGAN, and Markowitz models for the EU region with rebalancing every 10 days.}
	\label{fig:weight_diverse_bydays_eu_10}
\end{figure*}

\begin{figure*}[!htb]
	\centering  
	%\vspace{-0.35cm} 
	\subfigtopskip=2pt 
	\subfigbottomskip=2pt 
	\subfigcapskip=10pt 
	\subfigure{\includegraphics[width=0.45\textwidth]{imgs/acgan_dataset_eu_test.pdf} \label{fig:weight_diverse_eu113}}\vspace{-0.2em}
	\subfigure{\includegraphics[width=0.45\textwidth]{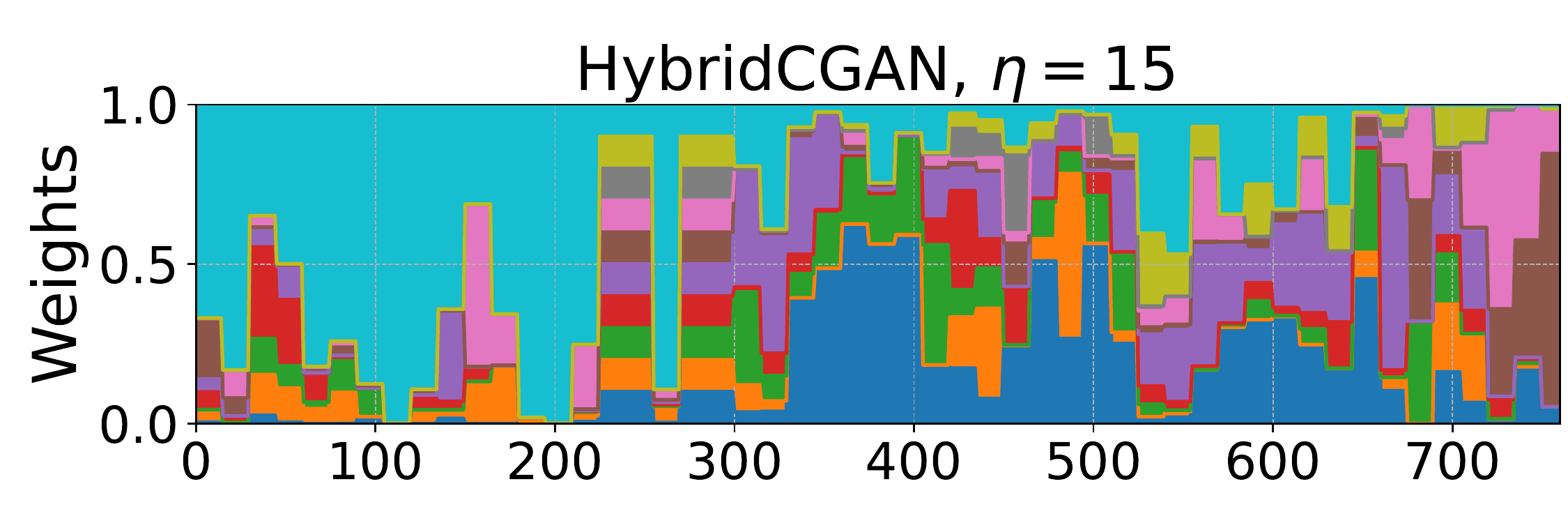} \label{fig:weight_diverse_eu213}}\vspace{-0.2em}
	\subfigure{\includegraphics[width=0.45\textwidth]{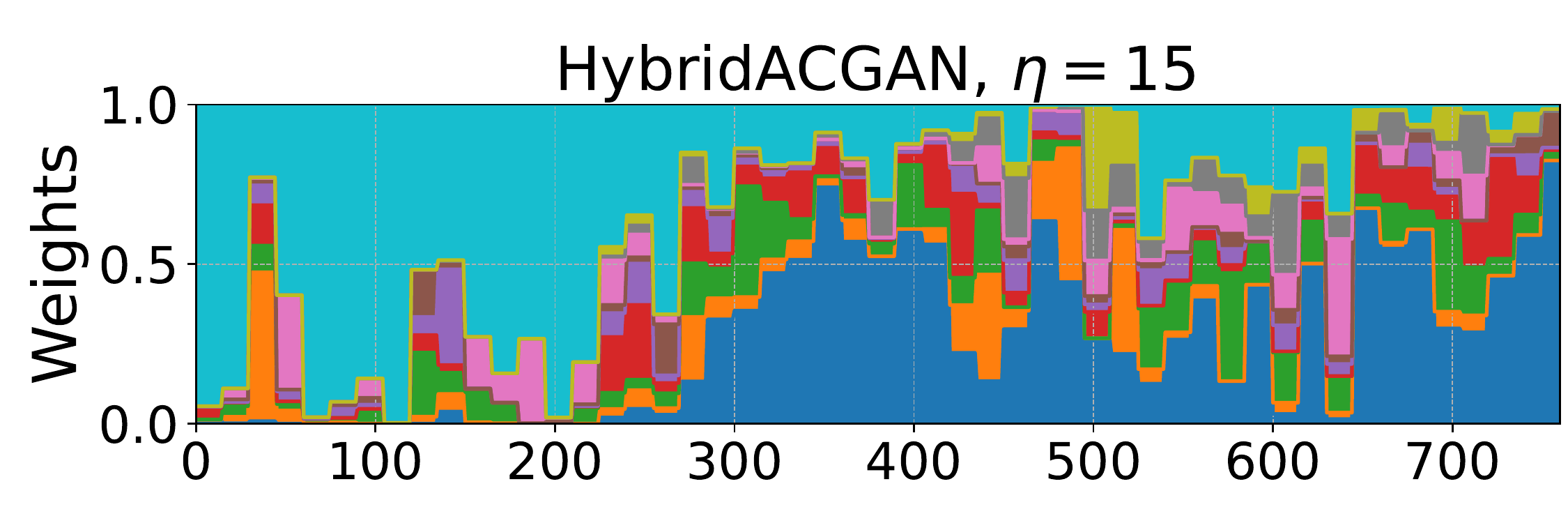} \label{fig:weight_diverse_eu313}}\vspace{-0.3em}
	\subfigure{\includegraphics[width=0.45\textwidth]{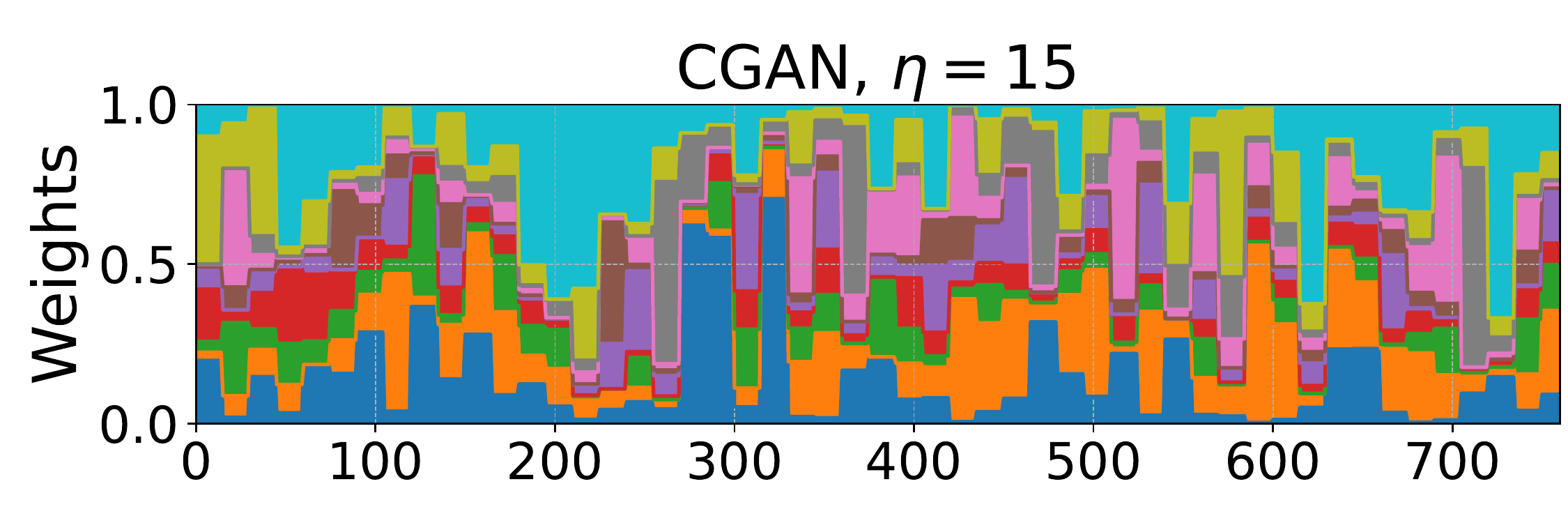} \label{fig:weight_diverse_eu413}}\vspace{-0.3em}
	\subfigure{\includegraphics[width=0.45\textwidth]{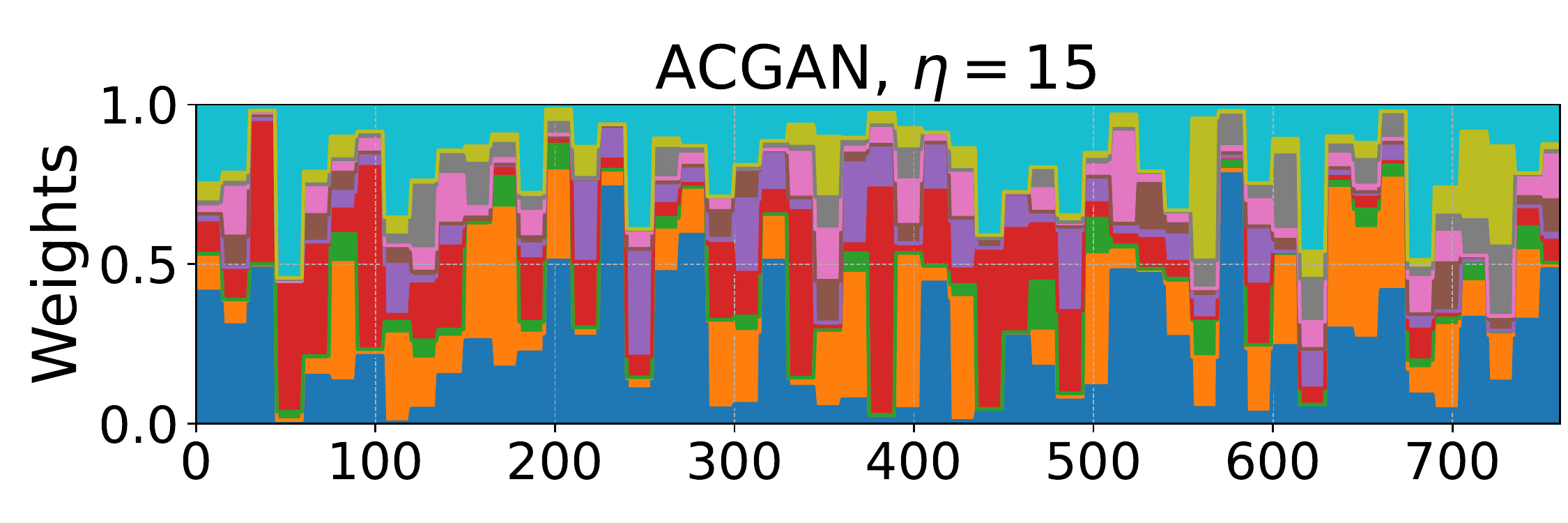} \label{fig:weight_diverse_eu513}}\vspace{-0.3em}
	\subfigure{\includegraphics[width=0.45\textwidth]{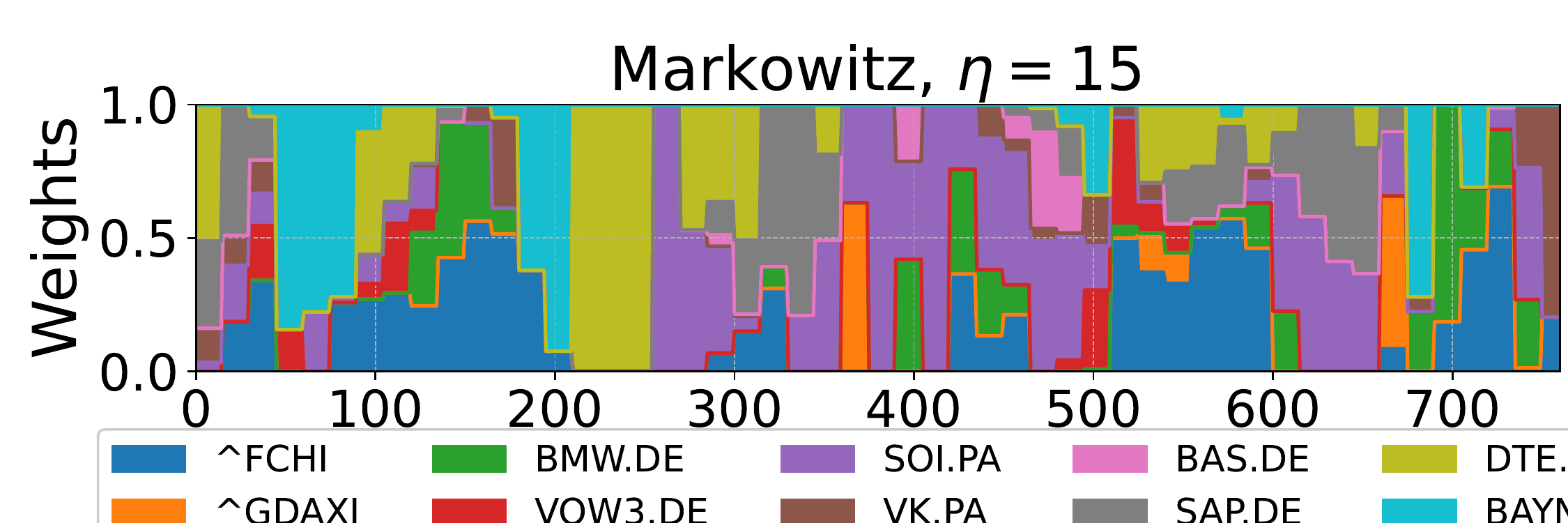} \label{fig:weight_diverse_eu613}}
	\caption{Weights distribution over time on the test period for HybridCGAN, HybridACGAN, CGAN, ACGAN, and Markowitz models for the EU region with rebalancing every 15 days.}
	\label{fig:weight_diverse_bydays_eu_15}
\end{figure*}

\begin{figure*}[!htb]
	\centering  
	%\vspace{-0.35cm} 
	\subfigtopskip=2pt 
	\subfigbottomskip=2pt 
	\subfigcapskip=10pt 
	\subfigure{\includegraphics[width=0.45\textwidth]{imgs/acgan_dataset_eu_test.pdf} \label{fig:weight_diverse_eu120}}\vspace{-0.2em}
	\subfigure{\includegraphics[width=0.45\textwidth]{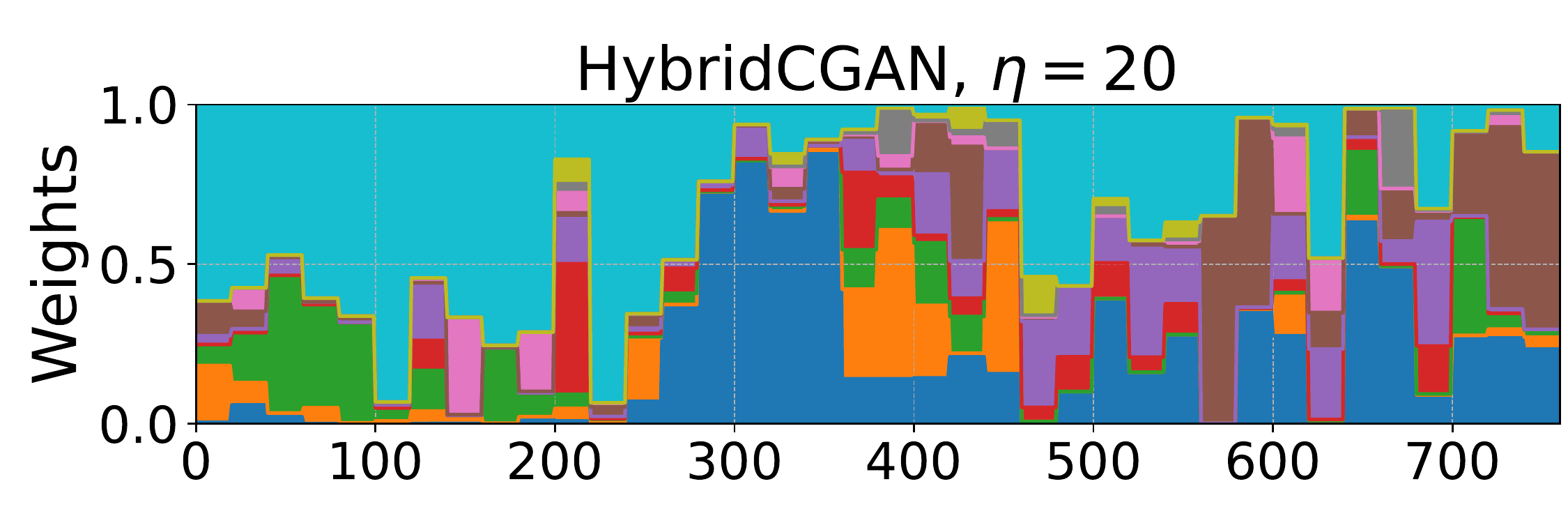} \label{fig:weight_diverse_eu220}}\vspace{-0.2em}
	\subfigure{\includegraphics[width=0.45\textwidth]{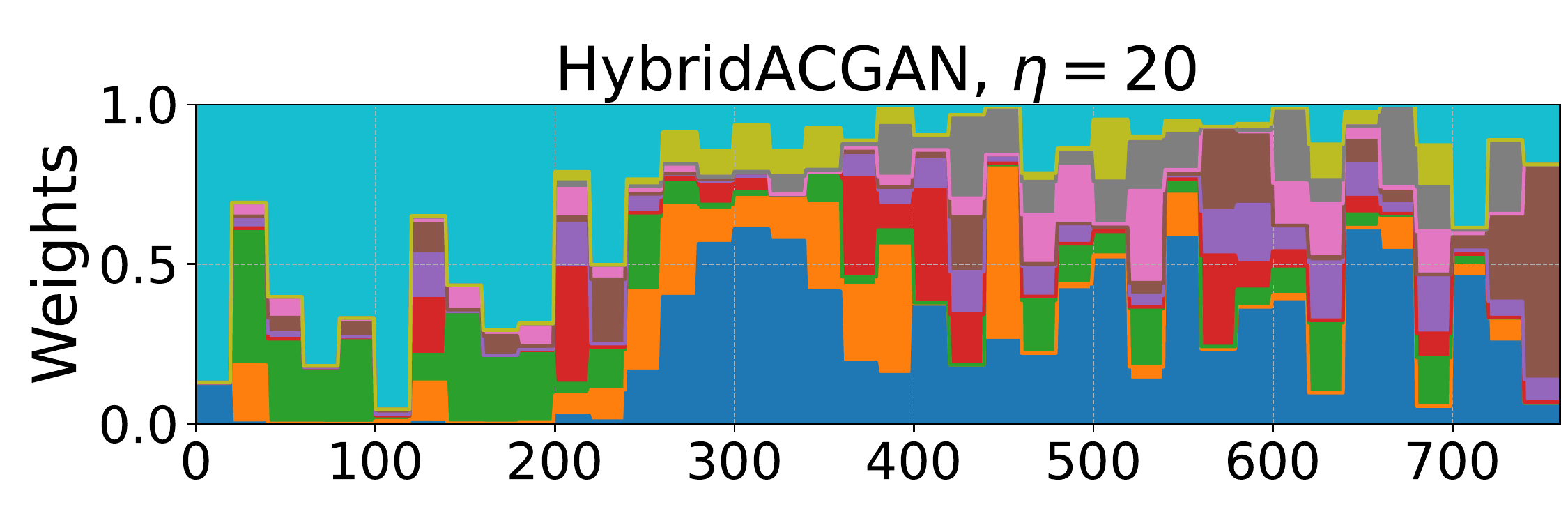} \label{fig:weight_diverse_eu320}}\vspace{-0.3em}
	\subfigure{\includegraphics[width=0.45\textwidth]{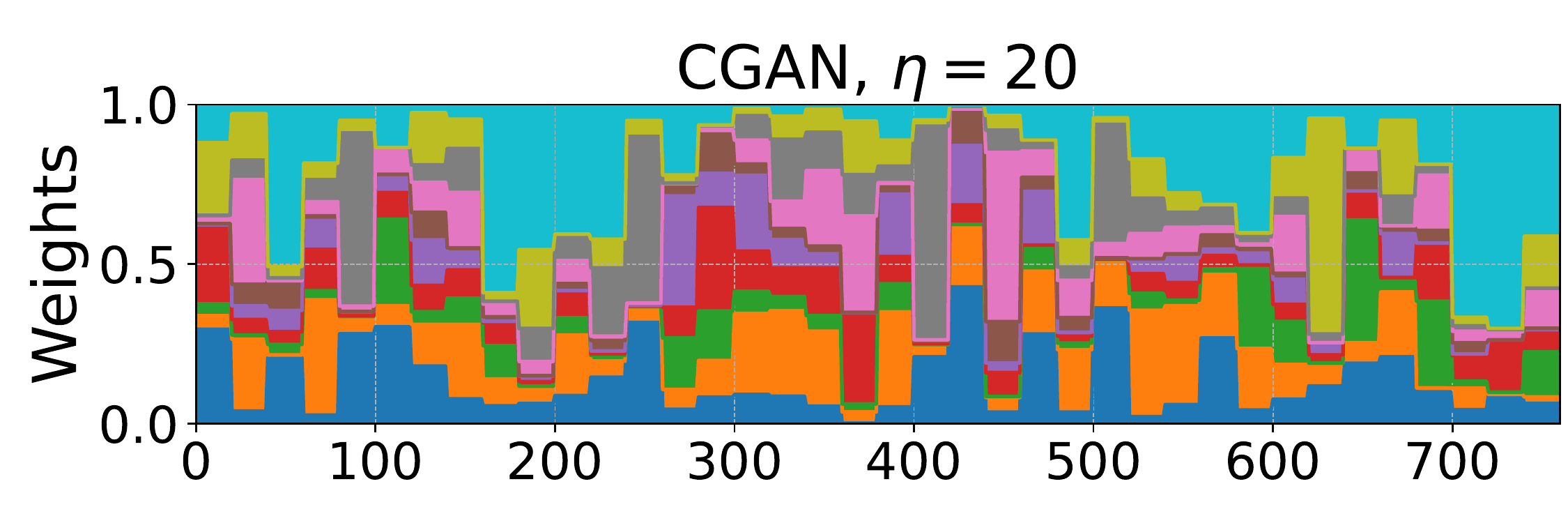} \label{fig:weight_diverse_eu420}}\vspace{-0.3em}
	\subfigure{\includegraphics[width=0.45\textwidth]{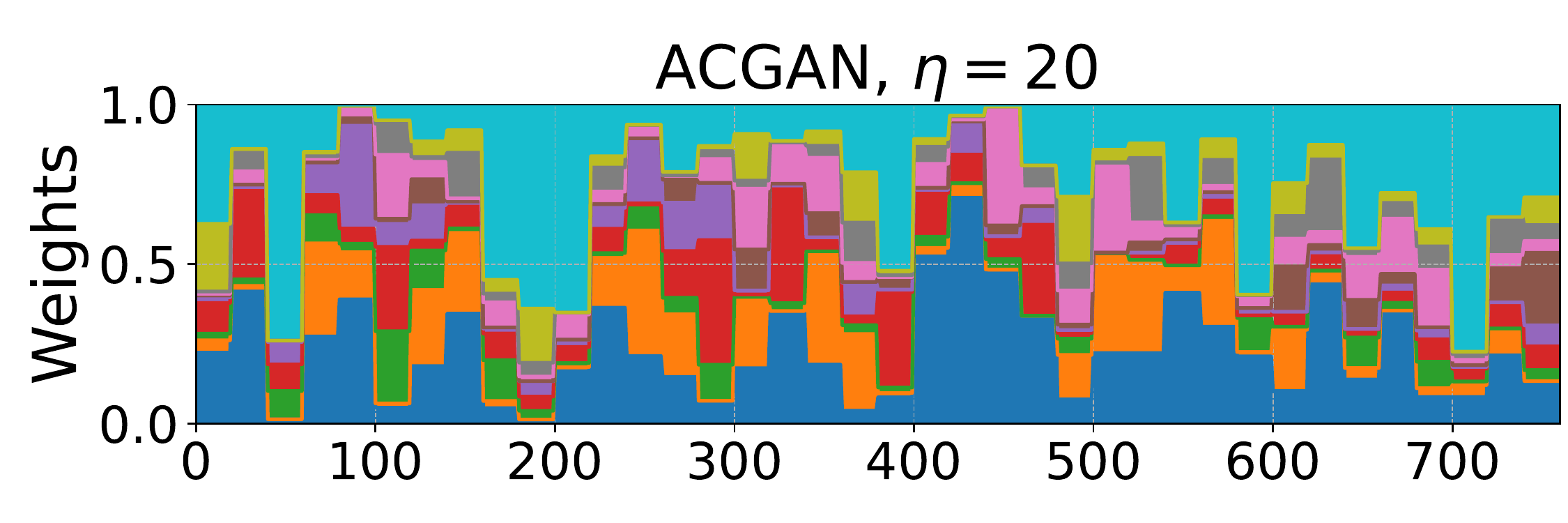} \label{fig:weight_diverse_eu520}}\vspace{-0.3em}
	\subfigure{\includegraphics[width=0.45\textwidth]{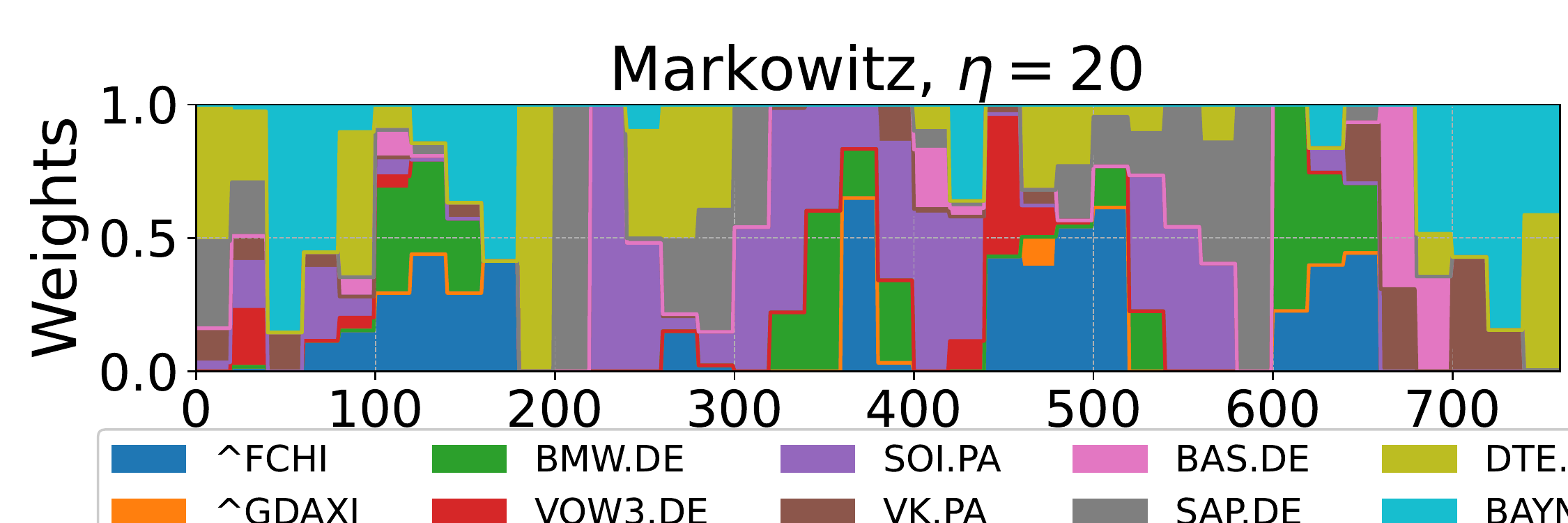} \label{fig:weight_diverse_eu620}}
	\caption{Weights distribution over time on the test period for HybridCGAN, HybridACGAN, CGAN, ACGAN, and Markowitz models for the EU region with rebalancing every 20 days.}
	\label{fig:weight_diverse_bydays_eu_20}
\end{figure*}

%In this section, we provide more weight distribution results for the comparative models. 
%We set window size $h=40, f=20$ and $w=60$ in all experiments.
As shown in the main paper, Figure~\ref{fig:weight_diverse_bydays_us20} presents the weight distribution over time on the test period for the US region with $\eta=20$.
Figure~\ref{fig:weight_diverse_bydays_us_10} and \ref{fig:weight_diverse_bydays_us_15} show the weight distributions for the US region with $\eta=10, 15$ respectively. Again, we observe that the HybridCGAN and HybridACGAN still have a more diverse portfolio allocation.
Figure~\ref{fig:weight_diverse_bydays_eu_10}, \ref{fig:weight_diverse_bydays_eu_15}, and \ref{fig:weight_diverse_bydays_eu_20} then present the weight distributions for the EU region with $\eta=10,15,20$ respectively.

\end{document}